\RequirePackage[british]{babel}
\pdfoutput=1
\documentclass[reqno,a4paper,12pt]{amsart}
\usepackage[a4paper,hmargin=2cm,tmargin=3cm,bmargin=3cm]{geometry}
\usepackage[utf8x]{inputenc}
\usepackage[T1]{fontenc}
\usepackage{tgschola}
\usepackage{amsmath,amssymb,amstext,amsthm,amscd,mathrsfs,eucal,bm,slashed}
\usepackage{graphicx,color}
\usepackage{array}
\usepackage{cite}
\usepackage{hyperref}
\hypersetup{%
  pdftitle   = {Conformal symmetry superalgebras for Lorentzian manifolds},
  pdfkeywords = {rigid supersymmetry, conformal superalgebra, twistor spinors, conformal Killing vectors},
  pdfauthor  = {Paul de Medeiros},
  pdfcreator = {\LaTeX\ with package \flqq hyperref\frqq}
}
\PrerenderUnicode{éÉ}
\newcommand{\half}{\tfrac12}
\newcommand{\cB}{\mathcal{B}}
\newcommand{\cF}{\mathcal{F}}

\newcommand{\cL}{\mathcal{L}}

\newcommand{\cP}{\mathcal{P}}
\newcommand{\cR}{\mathcal{R}}
\newcommand{\cS}{\mathcal{S}}

\newcommand{\ff}{\mathfrak{f}}
\newcommand{\fg}{\mathfrak{g}}

\newcommand{\fa}{\mathfrak{a}}

\newcommand{\fb}{\mathfrak{b}}
\newcommand{\fC}{\mathfrak{C}}
\newcommand{\fc}{\mathfrak{c}}

\newcommand{\fH}{\mathfrak{H}}
\newcommand{\fK}{\mathfrak{K}}

\newcommand{\fS}{\mathfrak{S}}

\newcommand{\fX}{\mathfrak{X}}
\newcommand{\fZ}{\mathfrak{Z}}

\newcommand{\fheis}{\mathfrak{heis}}

\newcommand{\fr}{\mathfrak{r}}

\newcommand{\fso}{\mathfrak{so}}
\newcommand{\fosp}{\mathfrak{osp}}

\newcommand{\fsl}{\mathfrak{sl}}
\newcommand{\fsp}{\mathfrak{sp}}
\newcommand{\fsu}{\mathfrak{su}}
\newcommand{\fu}{\mathfrak{u}}

\newcommand{\Cl}{\mathrm{C}\ell}

\newcommand{\rd}{\mathrm{d}}
\newcommand{\SO}{\mathrm{SO}}
\newcommand{\Pin}{\mathrm{Pin}}
\newcommand{\Spin}{\mathrm{Spin}}

\newcommand{\RR}{\mathbb{R}}
\newcommand{\CC}{\mathbb{C}}
\newcommand{\HH}{\mathbb{H}}
\newcommand{\KK}{\mathbb{K}}
\newcommand{\ZZ}{\mathbb{Z}}

\newcommand{\eN}{\mathscr{N}}

\newcommand{\be}{\boldsymbol{e}}

\newcommand{\te}{\mathrm{e}}
\renewcommand{\Im}{\mathrm{Im}}
\renewcommand{\Re}{\mathrm{Re}}

\DeclareMathOperator{\Mat}{Mat}

\newcommand{\rn}[1]{|\!|#1|\!|}
\theoremstyle{plain}

\theoremstyle{definition}

\newcommand{\MUNCH}[1]{\relax}

\allowdisplaybreaks
\begin{document}
\title[Submaximal conformal superalgebras for Lorentzian manifolds]{Submaximal conformal symmetry superalgebras for Lorentzian manifolds of low dimension}
\author[de Medeiros]{Paul de Medeiros}
\email{p.f.demedeiros@gmail.com}
\date{\today}
\begin{abstract}
We consider a class of smooth oriented Lorentzian manifolds in dimensions three and four which admit a nowhere vanishing conformal Killing vector and a closed two-form that is invariant under the Lie algebra of conformal Killing vectors. The invariant two-form is constrained in a particular way by the conformal geometry of the manifold. In three dimensions, the conformal Killing vector must be everywhere causal (or null if the invariant two-form vanishes identically). In four dimensions, the conformal Killing vector must be everywhere null and the invariant two-form vanishes identically if the geometry is everywhere of Petrov type N or O. To the conformal class of any such geometry, it is possible to assign a particular Lie superalgebra structure, called a conformal symmetry superalgebra. The even part of this superalgebra contains conformal Killing vectors and constant R-symmetries while the odd part contains (charged) twistor spinors. The largest possible dimension of a conformal symmetry superalgebra is realised only for geometries that are locally conformally flat. We determine precisely which non-trivial conformal classes of metrics admit a conformal symmetry superalgebra with the next largest possible dimension, and compute all the associated submaximal conformal symmetry superalgebras. In four dimensions, we also compute symmetry superalgebras for a class of Ricci-flat Lorentzian geometries not of Petrov type N or O which admit a null Killing vector.    
\end{abstract}
\maketitle
\clearpage
\vspace*{1.5cm}
\tableofcontents
\clearpage


\section{Introduction}
\label{sec:introduction}

The characterisation of non-trivial background geometries which support some amount of rigid (conformal) supersymmetry has attracted much attention in the recent literature \cite{Festuccia:2011ws,Jia:2011hw,Samtleben:2012gy,Klare:2012gn,Dumitrescu:2012he,Cassani:2012ri,Liu:2012bi,deMedeiros:2012sb,Dumitrescu:2012at,Kehagias:2012fh,Closset:2012ru,Martelli:2012sz,Samtleben:2012ua,Kuzenko:2012vd,Hristov:2013spa,deMedeiros:2013jja,deMedeiros:2013mca,Martelli:2013aqa,Cassani:2013dba,Alday:2013lba,Kuzenko:2013gva,Klare:2013dka,Pan:2013uoa,Closset:2013vra,Closset:2013sxa,Deger:2013yla,Kuzenko:2013uya,Cassani:2014zwa,DiPietro:2014moa,Kuzenko:2014mva,Anous:2014lia,Imamura:2014ima,Farquet:2014kma,Alday:2014rxa,Assel:2014paa,Alday:2014bta,Kuzenko:2014eqa,Farquet:2014bda}. The primary motivation being that it is often possible to obtain important exact results for quantum field theories defined on such backgrounds, with many novel holographic applications \cite{Klare:2012gn,Cassani:2012ri,Martelli:2012sz,Hristov:2013spa,Martelli:2013aqa,Cassani:2014zwa,Farquet:2014kma,Alday:2014rxa,Alday:2014bta,Farquet:2014bda}. Perhaps the most systematic strategy for generating admissible backgrounds is by taking a rigid limit of some local supergravity coupling, such that the dynamics of the gravity supermultiplet is effectively frozen out \cite{Festuccia:2011ws}. The resulting bosonic supergravity background supports rigid supersymmetry, with the supersymmetry parameter constrained by setting to zero the supersymmetry variation of the fermions in the gravity supermultiplet. For bosonic supersymmetric backgrounds of conformal supergravity \cite{Kaku:1978nz,Fradkin:1985am,vanNieuwenhuizen:1985cx,Rocek:1985bk}, the supersymmetry parameter typically obeys a particular conformally invariant first order PDE, known as a \lq twistor spinor equation', with respect to a certain superconnection whose precise form is dictated by the structure of the conformal gravity supermultiplet.

The Lie superalgebra which encodes the rigid (conformal) supersymmetry of a bosonic (conformal) supergravity background is known as the {\emph{(conformal) symmetry superalgebra}} of the background (\!\!\cite{deMedeiros:2013jja,deMedeiros:2013mca})\cite{FigueroaO'Farrill:2004mx,FigueroaO'Farrill:2007ar,FOF:2007F4E8,FigueroaO'Farrill:2008ka,FigueroaO'Farrill:2008if}. The even part of this superalgebra contains (conformal) Killing vectors, which generate (conformal) isometries of the background, together with R-symmetries of the associated rigid supermultiplet. The odd part contains (twistor) spinors valued in certain R-symmetry representations which generate rigid (conformal) supersymmetries of the background. The virtue of the (conformal) symmetry superalgebra construction is that it often reveals special geometrical properties of the background based on the type and amount of rigid (conformal) supersymmetry it supports. For example, in dimensions eleven, ten and six, this approach was used recently in \cite{FigueroaO'Farrill:2012fp,Figueroa-O'Farrill:2013aca} to prove that any bosonic supersymmetric supergravity background possessing more than half the maximal amount of supersymmetry is necessarily (locally) homogeneous.

The simplest class of conformal symmetry superalgebras contain odd elements which obey a \lq geometric' twistor spinor equation, with respect to the Levi-Civit\`{a} connection. Their generic structure was described in some detail in \cite{deMedeiros:2013jja}, where it was found that the inclusion of a non-trivial R-symmetry is crucial in solving the odd-odd-odd component of the Jacobi identity for the superalgebra. Indeed, this extra ingredient is what distinguishes the construction in \cite{deMedeiros:2013jja} from several earlier ones \cite{Hab:1990,Duval:1993hs,Klinker:2005,Rajaniemi:2006}. More general conformal symmetry superalgebras are further complicated by the presence of some assortment of non-trivial background fields (other than the metric). The details of these background fields depend on the composition of the conformal gravity supermultiplet but one common feature is the presence of R-symmetry gauge fields. In section~\ref{sec:ConformalSymmetrySuperalgebras} of this paper, we shall explore a natural generalisation of the construction in \cite{deMedeiros:2013jja} based on the gauging of R-symmetry. For Lorentzian geometries, we find that the resulting structure generically defines a Lie superalgebra only if the R-symmetry is one-dimensional and the background has dimension three or four. Indeed, these are precisely the cases where the bosonic sector of a conformal gravity supermultiplet contains only the metric and the R-symmetry gauge field.    

It is a well-known and useful fact that geometric twistor spinors \lq square' (in a sense which can be made precise) to conformal Killing vectors. More generally, for a conformal symmetry superalgebra, there is a similar (albeit somewhat more complicated) squaring map defined by the odd-odd bracket \cite{deMedeiros:2013jja,deMedeiros:2013mca}. Of course, if a pseudo-Riemannian spin manifold admits a conformal Killing vector, it need not admit a geometric twistor spinor.
\footnote{In Euclidean and Lorentzian signatures, up to local conformal equivalence, the classification of those geometries which do admit a nowhere vanishing geometric twistor spinor was established in \cite{Baum:2002,BL:2003,Baum:2012,Leitner:2005,Baum:2008}.}
However, as was shown in \cite{Cassani:2012ri,deMedeiros:2012sb,Hristov:2013spa}, at least for a certain class of Lorentzian geometries which need not admit a geometric twistor spinor, the existence of a nowhere vanishing conformal Killing vector with a particular causal character is in fact locally equivalent to the existence of a nowhere vanishing twistor spinor that is defined with respect to a particular connection with non-trivial intrinsic torsion. The precise form of this intrinsic torsion is dictated by the local isotropy of the twistor spinor. Moreover, in dimensions three and four, with one-dimensional R-symmetry, this data recovers precisely the defining conditions for a bosonic supersymmetric conformal supergravity background. 

If it is possible to define a quantum field theory on a background preserving a large amount of (conformal) supersymmetry, it is often the case that the theory is particularly well-behaved. Backgrounds which admit a conformal symmetry superalgebra with the largest possible dimension are necessarily locally conformally flat. In Lorentzian signature, any such conformal symmetry superalgebra has compact R-symmetry and is isomorphic to one of the well-known conformal superalgebras classified by Nahm in \cite{Nahm:1977tg}. However, the general structure of conformal symmetry superalgebras with the next largest possible, or {\emph{submaximal}}, dimension (for backgrounds that are not locally conformally flat) is much less clear. Our goal here will be to elucidate this structure for Lorentzian geometries in three and four dimensions which admit a conformal symmetry superalgebra with gauged one-dimensional R-symmetry. Our strategy will make use of some recent progress \cite{KruThe2013,KruMat2013,DouThe2013} which has determined the submaximal dimension of the Lie algebra of conformal Killing vectors for any Lorentzian manifold. We will also utilise some earlier results \cite{Kru:1954,DC:1975,ExactSolutions} on the classification of (conformal) Killing vectors for Lorentzian manifolds of low dimension. We then employ the results of \cite{Cassani:2012ri,Hristov:2013spa} to deconstruct a null (in four dimensions) or timelike (in three dimensions) conformal Killing vector in terms of the charged twistor spinors which form the odd part of the conformal symmetry superalgebra. Up to local conformal equivalence, we prove that there are precisely three types of Lorentzian three-manifold (see Table~\ref{tab:ClassIVData}) with a timelike conformal Killing vector and two types of Lorentzian four-manifold (see Table~\ref{tab:PlaneWaveExtraKV4d}) with a null conformal Killing vector which admit non-isomorphic submaximal conformal symmetry superalgebras (see Table~\ref{tab:ClassIVDataRhoKL} and Appendix~\ref{sec:PlaneWaves}). In each case, the submaximal conformal symmetry superalgebra can be assigned to the conformal class of a locally homogeneous Lorentzian metric. We also determine, up to local conformal equivalence, precisely which Lorentzian three-manifolds (see Table~\ref{tab:dim3PP}) with a null conformal Killing vector admit a conformal symmetry superalgebra with the next largest possible dimension. Finally, using the results of \cite{ExactSolutions}, we compute the symmetry superalgebras for a class of \lq physically admissible'
\footnote{In the sense that their energy-momentum tensor does not violate the dominant energy condition.}
Ricci-flat Lorentzian four-manifolds with a null Killing vector that are not of Petrov type N or O. We also compute the conformal symmetry superalgebra for the most symmetric geometry in this class, which is the unique representative of Petrov type D.

\section{Conformal Killing vectors}
\label{sec:CKV}

Let $M$ be a smooth oriented manifold equipped with a Lorentzian metric $g$ whose associated Levi-Civit\`{a} connection will be denoted by $\nabla$. We take $M$ to have dimension $d>2$.

Let $\fX ( M )$ denote the space of vector fields on $M$ (i.e. sections of the tangent bundle $TM$). Let $\rn{X}^2 = g(X,X)$ denote the norm squared of any $X \in \fX(M)$ with respect to $g$. At a point in $M$, $X$ may be either {\emph{spacelike}} (if $\rn{X}^2 >0$), {\emph{timelike}} (if $\rn{X}^2 <0$) or {\emph{null}} (if $\rn{X}^2 =0$). If $\rn{X}^2 \leq 0$ then $X$ is said to be {\emph{causal}}. At each point in $M$, clearly the sign of $\rn{X}^2$ with respect to any positive multiple of $g$ is the same, so the aforementioned causal properties of a vector field depend only on the conformal class $[g]$ of $g$. 

The Lie derivative $\cL_X$ along any $X \in \fX(M)$ defines an endomorphism of the space of tensor fields on $M$. The Lie bracket of vector fields is defined by $[X,Y] =\cL_X Y = \nabla_X Y - \nabla_Y X \in \fX(M)$, for all $X,Y \in \fX(M)$. This equips $\fX(M)$ with the structure of a Lie algebra. Furthermore
\begin{equation}\label{eq:LieDerHom}
\cL_X \cL_Y - \cL_Y \cL_X = \cL_{[X,Y]}~,
\end{equation}
for all $X,Y \in \fX ( M )$. Whence, the Lie derivative defines on the space of tensor fields a representation of the Lie algebra of vector fields.

The subspace of {\emph{conformal Killing vectors}} in $\fX ( M )$ is defined by
\begin{equation}\label{eq:CKV}
\fC (M,g) = \{ X \in \fX ( M ) \; |\; \cL_X g = -2 \sigma_X  g\}~,
\end{equation}
for some real function $\sigma_X$ on $M$. For any $X,Y \in \fC ( M,g )$, using \eqref{eq:LieDerHom}, it follows that $[X,Y] \in \fC ( M,g )$ with  
\begin{equation}\label{eq:sigma}
\sigma_{[X,Y]} = \nabla_X \sigma_Y - \nabla_Y \sigma_X~.
\end{equation}
Whence, restricting the Lie bracket to $\fC ( M,g )$ defines a (finite-dimensional) Lie subalgebra of conformal Killing vectors on $(M,g)$. 

Any $X \in \fC ( M,g )$ with $\sigma_X$ constant is called {\emph{homothetic}} and let $\fH (M,g)$ denote the subspace of homothetic conformal Killing vectors on $(M,g)$. Any $X \in \fH ( M,g )$ with $\sigma_X \neq 0$ is said to be {\emph{proper}}. Any $X \in \fH ( M,g )$ with $\sigma_X =0$ is called {\emph{isometric} and let $\fK (M,g)$ denote the subspace of isometric conformal Killing vectors (i.e. Killing vectors) on $(M,g)$. From \eqref{eq:sigma}, clearly $[ \fH(M,g) , \fH(M,g) ] < \fK(M,g)$ so restricting to the subspace of Killing vectors on $(M,g)$ defines the ideal $\fK ( M,g ) \lhd \, \fH (M,g)$. Furthermore, given any $X,Y \in \fH(M,g)$ with $\sigma_X \neq 0$, then $Y - \frac{\sigma_Y}{\sigma_X} X \in \fK(M,g)$. Whence, either $\fH(M,g) = \fK(M,g)$ or ${\mbox{dim}} ( \fH(M,g) / \fK(M,g) ) =1$.  

A real function $\phi$ on $M$ is called a {\emph{conformal scalar}} if $\nabla_X \phi = p_\phi \sigma_X \phi$, for all $X \in \fC ( M,g )$, in terms of some $p_\phi \in \RR$ ($\phi$ is said to be {\emph{proper}} if $p_\phi \neq 0$). A real one-form $\upsilon$ on $M$ is called a {\emph{conformal one-form}} if $\cL_X \upsilon = p_\upsilon \rd \sigma_X$, for all $X \in \fC ( M,g )$, in terms of some $p_\upsilon \in \RR$ ($\upsilon$ is said to be {\emph{proper}} if $p_\upsilon \neq 0$). If $\rd \upsilon =0$ then $\upsilon$ is called a {\emph{conformal gradient}}. For example, if $\phi$ is a (proper) conformal scalar, then $\rd ( \ln \phi )$ is a (proper) conformal gradient. If $\upsilon$ is a proper conformal gradient then, at least locally, $\upsilon = p_\upsilon  \rd \varphi$ for some real function $\varphi$ such that, for each $X \in \fC ( M,g )$, $\sigma_X - \nabla_X \varphi = s_X$ for some $s_X \in \RR$.  

Any metric ${\tilde g}$ in the same conformal class $[g]$ as $g$ is of the form ${\tilde g} = \te^{2\omega} g$, in terms of some real function $\omega$ on $M$. Each $X \in \fC ( M,g )$ (with conformal factor $\sigma_X$) is also in $\fC(M,{\tilde g})$ but with conformal factor ${\tilde \sigma}_X = \sigma_X - \nabla_X \omega$. Thus, we may assign the Lie algebra $\fC ( M,[g] )$ of conformal Killing vectors on $(M,g)$ to the conformal class $[g]$. Of course, there may be a preferred metric in $[g]$ with respect to which the conformal Killing vectors in $\fC ( M,[g] )$ are most conveniently represented (e.g. via a homothetic or isometric action). For example, if $(M,g)$ admits a proper conformal scalar $\phi$, then $\fC ( M,[g] ) = \fK(M , \te^{2\omega} g )$ for $\omega = \tfrac{1}{p_\phi} \ln \phi$. Alternatively, if $(M,g)$ admits a proper conformal gradient $\upsilon$, then $\fC ( M,[g] ) = \fH(M , \te^{2\omega} g )$ for $\omega = \varphi$ and ${\tilde \sigma}_X = s_X$. More generally, $\fC (M,[g])$ is said to be {\emph{conformally isometric}} if it can be represented by $\fK (M,  \te^{2\omega} g )$ or {\emph{conformally homothetic}} if it can be represented by $\fH (M, \te^{2\omega} g )$ (with ${\mbox{dim}} ( \fH(M,\te^{2\omega} g) / \fK(M,\te^{2\omega} g) ) =1$), for some choice of $\omega$. 

Important aspects of the conformal geometry of $(M,g)$ are characterised by its Weyl tensor $W$ and Cotton-York tensor $C$. If $d >3$, then $W=0$ only if $(M,g)$ is locally conformally flat. For any $X \in \fC(M,[g])$, $\cL_X W = -2 \sigma_X W$ which implies $\nabla_X \rn{W}^2 = 4 \sigma_X \rn{W}^2$, where $\rn{W}^2$ denotes the scalar norm-squared of $W$ with respect to $g$. Thus, if $d >3$, any $(M,g)$ with $\rn{W}^2$ nowhere vanishing is conformally isometric, with $\fC ( M,[g] ) = \fK(M , \rn{W} g )$ (i.e. $\phi = \rn{W}^2$ is a proper conformal scalar with $p_\phi =4$). If $d =3$, then $W$ vanishes identically and $C=0$ only if $(M,g)$ is locally conformally flat. In this case, for any $X \in \fC(M,[g])$, $\cL_X C = 0$ which implies $\nabla_X \rn{C}^2 = 6 \sigma_X \rn{C}^2$, where $\rn{C}^2$ denotes the scalar norm-squared of $C$ with respect to $g$. Thus, if $d =3$, any $(M,g)$ with $\rn{C}^2$ nowhere vanishing is conformally isometric, with $\fC ( M,[g] ) = \fK(M , \rn{C}^{2/3} g )$ (i.e. $\phi = \rn{C}^2$ is a proper conformal scalar with $p_\phi =6$). 

For any $X \in \fC(M,[g])$, $\nabla_X \rn{X}^2 = -2 \sigma_X \rn{X}^2$ so if ${\mbox{dim}} \, \fC(M,[g]) =1$ then $\phi = \rn{X}^{-2}$ defines a proper conformal scalar with $p_\phi =2$ provided $X$ is nowhere null. In this case, $\fC(M,[g])$ is therefore always conformally isometric. Alternatively, if $(M,g)$ is conformally flat then $\fC(M,[g]) \cong \fso(d,2)$, which is neither conformally homothetic nor conformally isometric.

In fact, the dimension of $\fC(M,[g])$ can never exceed ${d+2 \choose 2}$ and equals it only if $(M,g)$ is locally conformally flat. An important problem in conformal geometry is to determine the next largest value of ${\mbox{dim}}\, \fC(M,[g])$, or {\emph{submaximal dimension}}, which can be realised for some $(M,g)$ that is not conformally flat. In Lorentzian signature, this problem was recently solved in \cite{KruThe2013,KruMat2013,DouThe2013}. Any $(M,g)$ that is not locally conformally flat must have ${\mbox{dim}}\, \fC(M,[g]) \leq  4+{d-1 \choose 2}$ for any $d>3$ and ${\mbox{dim}}\, \fC(M,[g]) \leq  4$ for $d=3$. These upper bounds are sharp in that, for every $d>2$, there are explicit examples for which they are saturated.    

In $d = 4$, the conformal class of $(M,g)$ can be characterised locally as being of Petrov type I, II, D, III, N or O, depending on which components of the Weyl tensor vanish identically. Theorem 5.1.3 in \cite{KruThe2013} provides sharp upper bounds on ${\mbox{dim}}\, \fC(M,[g])$ for each Petrov type. Type O means $W=0$ so $(M,g)$ is locally conformally flat and ${\mbox{dim}}\, \fC(M,[g]) =15$. Type N must have ${\mbox{dim}}\, \fC(M,[g]) \leq 7$, type D must have ${\mbox{dim}}\, \fC(M,[g]) \leq 6$ while ${\mbox{dim}}\, \fC(M,[g]) \leq 4$ for types I, II and III. It was shown in \cite{DC:1975} that $\fC (M,[g])$ is conformally isometric only if $(M,g)$ admits a proper conformal scalar. Furthermore, from theorem 3 in \cite{DC:1975}, it follows that if $(M,g)$ does not admit a proper conformal scalar then it must be locally conformally equivalent to either Minkowski space (type O) or a plane wave (type N). Whence, any $(M,g)$ of type I, II, D or III must have $\fC (M,[g])$ conformally isometric.

\section{Twistor spinors}
\label{sec:SpinTS}

Let us now assume that $M$ has vanishing second Stiefel-Whitney class so the bundle $\SO(M)$ of oriented pseudo-orthonormal frames lifts to $\Spin(M)$ by the assignment of a {\emph{spin structure}}. For $d \leq 3$, this lift is always unobstructed.

The {\emph{Clifford bundle}} $\Cl (TM)$ over $(M,g)$ is defined by the relation 
\begin{equation}\label{eq:CliffordRelation}
{\bm X} {\bm Y} + {\bm X} {\bm Y} = 2 g(X,Y) {\bf 1}~,
\end{equation}
for all $X,Y \in \fX ( M )$, where each multi-vector field $\Phi$ on $M$ is associated with a section ${\bm{\Phi}}$ of $\Cl (TM)$. At each point $x \in M$, the exterior algebra of $T_x M \cong \RR^{d-1,1}$ is isomorphic, as a vector space, to the Clifford algebra $\Cl (T_x M)$ (the metric $g$ and its inverse provide a duality between multi-vector fields and differential forms on $M$). The canonical volume form for the metric $g$ on $M$ defines a unique idempotent section ${\bm \Gamma}$ of $\Cl (TM)$. If $d$ is odd, ${\bm \Gamma}$ is central in $\Cl (TM)$. If $d$ is even, ${\bm \Gamma} {\bm X} = - {\bm X} {\bm \Gamma}$, for all $X \in \fX ( M )$.

The Clifford algebra $\Cl (T_x M)$ is $\ZZ_2$-graded such that elements with even and odd degrees are assigned grades $0$ and $1$ respectively. The grade $0$ elements span an ungraded associative subalgebra $\Cl^0 (T_x M) < \Cl (T_x M)$. The degree two elements span a Lie subalgebra $\fso (T_xM) < \Cl^0 (T_x M)$, where $\Cl^0 (T_x M)$ is understood as a Lie algebra whose brackets are defined by commutators. 

At each point $x \in M$, the set of invertible elements in $\Cl (T_x M)$ forms a multiplicative group $\Cl^\times (T_x M)$. The vectors ${\bm X} \in \Cl (T_x M)$ with $\rn{X}^2 = \pm 1$ generate the subgroup $\Pin ( T_x M ) < \Cl^\times (T_x M)$. The group $\Spin ( T_x M ) = \Pin ( T_x M ) \cap \Cl^0 (T_x M)$, which also follows by exponentiating $\fso (T_xM) < \Cl^0 (T_x M)$.

The {\emph{pinor}} module is defined by the restriction to $\Pin ( T_x M )$ of an irreducible representation of $\Cl (T_x M)$. Every Clifford algebra is isomorphic, as an associative algebra with unit, to a matrix algebra and it is a simple matter to deduce their irreducible representations. The {\emph{spinor}} module is defined by the restriction to $\Spin ( T_x M )$ of an irreducible representation of $\Cl^0 (T_x M)$. Note that restricting to $\Spin ( T_x M )$ an irreducible representation of $\Cl (T_x M)$ need not define an irreducible spinor module. If $d$ is even, $\Cl (T_x M)$ has a unique irreducible representation which descends to a reducible representation when restricted to $\Spin ( T_x M )$, yielding a pair of inequivalent irreducible ({\emph{chiral}}) spinor modules associated with the two eigenspaces of ${\bm \Gamma}$ on which ${\bm \Gamma} = \pm {\bf 1}$. If $d$ is odd, $\Cl (T_x M)$ has two inequivalent irreducible representations which are isomorphic to each other when restricted to $\Spin ( T_x M )$. The isomorphism here is provided by the central element ${\bm \Gamma}$ and corresponds to Hodge duality in the exterior algebra. In either case, the spinor module defined at each point in $M$ defines a principle bundle $\Spin (M)$ and its associated vector bundle $\$ (M)$ is called the {\emph{spinor bundle}} over $M$.

Let $\fS ( M )$ denote the space of spinor fields on $M$ (i.e. sections of $\$(M)$). If $d$ is even, $\fS ( M ) = \fS_+ (M) \oplus \fS_- (M)$, where $\fS_\pm ( M )$ denote the subspaces of chiral spinor fields (defined via projection operators ${\bf P}_\pm = \half ( {\bf 1} \pm {\bm \Gamma})$) on which ${\bm \Gamma} = \pm {\bf 1}$. The action of $\nabla$ induced on $\fS ( M )$ is compatible with the Clifford action, i.e.
\begin{equation}\label{eq:LCCliffComp}
\nabla_X ( {\bm Y} \psi ) = ( \nabla_X {\bm Y} )  \psi + {\bm Y} \nabla_X \psi~,
\end{equation}  
for all $X,Y \in \fX ( M )$ and $\psi \in \fS ( M )$. Furthermore,
\begin{equation}\label{eq:LCBracket}
( \nabla_X \nabla_Y - \nabla_Y \nabla_X ) \psi = \nabla_{[X,Y]} \psi + \half {\bm R}(X,Y) \psi~,
\end{equation}
for all $X,Y \in \fX ( M )$ and $\psi \in \fS ( M )$, where ${\bm R}(X,Y) = \half X^\mu Y^\nu R_{\mu\nu\rho\sigma} {\bm \Gamma}^{\rho\sigma}$, in terms of the basis conventions described in Appendix~\ref{sec:CoordinateBasisConventions}.

There always exists on $\fS(M)$ a non-degenerate bilinear form $\langle -,- \rangle$ with the properties
\begin{align}
\label{eq:SpinorInnerProd}
\langle \psi , \varphi \rangle &= \sigma \langle \varphi , \psi \rangle \nonumber \\
\langle {\bm X} \psi , \varphi \rangle &= \tau \langle \psi ,  {\bm X} \varphi \rangle \\
X \langle \psi , \varphi \rangle &= \langle \nabla_X \psi , \varphi \rangle + \langle \psi , \nabla_X \varphi \rangle~, \nonumber
\end{align}
for all $\psi , \varphi \in \fS (M)$ and $X \in \fX(M)$, with respect to a pair of fixed signs $\sigma$ and $\tau$ (see \cite{deMedeiros:2013jja,AlekCort1995math,Alekseevsky:2003vw} for more details). The possible choices for $\sigma$ and $\tau$ depend critically on both $d$ and the signature of $g$. The sign $\sigma = \pm 1$ indicates whether $\langle -,- \rangle$ is symmetric or skewsymmetric. The third line in \eqref{eq:SpinorInnerProd} says that $\langle -,- \rangle$ is spin-invariant. For $d$ even, this implies $\langle {\bm \Gamma} \psi , \varphi \rangle = (-1)^{d/2} \langle \psi , {\bm \Gamma} \varphi \rangle$. Whence,
\begin{align}\label{eq:ChiralSpinorProd}
\langle \psi_\pm , \varphi_\mp \rangle &= 0 \quad {\mbox{if $d=0$ mod $4$}} \nonumber \\
\langle \psi_\pm , \varphi_\pm \rangle &= 0 \quad {\mbox{if $d=2$ mod $4$}}~,
\end{align}
for all $\psi_\pm , \varphi_\pm \in \fS_\pm (M)$. To any pair $\psi , \varphi \in \fS(M)$, let us assign a vector field $\xi_{\psi , \varphi}$ defined such that
\begin{equation}\label{eq:Bilinear}
g(X, \xi_{\psi , \varphi} ) = \langle \psi , {\bm X} \varphi \rangle~,
\end{equation}
for all $X \in \fX (M)$. From the first two properties in \eqref{eq:SpinorInnerProd}, it follows that $\xi_{\psi,\varphi} = \sigma\tau \xi_{\varphi,\psi}$, for all $\psi , \varphi \in \fS (M)$ and $X \in \fX(M)$. 

Now let the dual ${\overline \psi}$ of any $\psi \in \fS (M)$ with respect to $\langle -,- \rangle$ be defined such that ${\overline \psi} \varphi = \langle \psi , \varphi \rangle$, for all $\varphi \in \fS (M)$. From any $\psi , \varphi \in \fS (M)$, one can define $\psi {\overline \varphi}$ as an endomorphism of $\fS (M)$. Whence, it can be expressed relative to the basis \eqref{eq:ClBasis} in Appendix~\ref{sec:CoordinateBasisConventions}, with coefficients proportional to multi-vectors of the form ${\overline \varphi} {\bm \Gamma}^{\mu_1 ... \mu_k} \psi$. Such expressions are known as {\emph{Fierz identities}}, full details of which can be found in section 4 of~\cite{deMedeiros:2013jja}. 

The {\emph{spinorial Lie derivative}} \cite{Lic:1963,Kosmann:1972,BG:1992,Habermann:1996} along any $X \in \fC(M,[g])$ is defined by
\begin{equation}\label{eq:SpinLieDer}
\cL_X = \nabla_X + \tfrac{1}{4} {\bm{\rd X}}~,
\end{equation}
where ${\bm{\rd X}} = ( \nabla_{\mu} X_\nu ) {\bm \Gamma}^{\mu\nu}$. 
\footnote{In a slight abuse of notation, we use the same symbol for a vector field and its dual one-form with respect to $g$.}
The spinorial Lie derivative \eqref{eq:SpinLieDer} obeys
\begin{align}\label{eq:LieDerBracketSpinors}
\cL_X ( \nabla_Y \psi ) -  \nabla_Y ( \cL_X \psi ) &= \nabla_{[X,Y]} \psi + \tfrac{1}{4} {\bm{\rd \sigma_X \!\wedge Y}} \psi \nonumber \\
\cL_X  ( {\bm \nabla} \psi ) -  {\bm \nabla} ( \cL_X \psi ) & = \sigma_X {\bm \nabla} \psi - \half (d-1) ( {\bm{\nabla}} \sigma_X ) \psi~,
\end{align}
for all $X \in \fC (M,[g])$, $Y \in \fX ( M )$ and $\psi \in \fS ( M )$. Moreover, for all $X,Y \in \fC (M,[g])$ and any $w\in \RR$, using \eqref{eq:SpinLieDer} and \eqref{eq:sigma}, it follows that
\begin{equation}\label{eq:LieDerSpinHomAlpha}
( \cL_X + w \sigma_X {\bf 1} ) ( \cL_Y + w \sigma_Y {\bf 1} ) - ( \cL_Y + w \sigma_Y {\bf 1} ) ( \cL_X + w \sigma_X {\bf 1} ) = \cL_{[X,Y]} + w \sigma_{[X,Y]} {\bf 1}~.
\end{equation}
Whence, the map $X \mapsto \cL_X + w \sigma_X {\bf 1}$ defines on $\fS(M)$ a representation of $\fC (M,[g])$. 

With respect to a metric ${\tilde g} = \te^{2\omega} g$ in $[g]$, compatibility with the Clifford relation \eqref{eq:CliffordRelation} requires that ${\tilde {\bm X}} = \te^{\omega} {\bm X}$, for all $X \in \fX(M)$. Moreover, given any $\psi \in \fS(M)$, ${\tilde \psi} =  \te^{\omega/2} \psi$ defines the corresponding spinor field with respect to ${\tilde g}$. For $w=\half$, the representation of $\fC (M,[g])$ on $\fS(M)$ defined by
\begin{equation}\label{eq:KSLieDer}
{\hat \cL}_X = \cL_X + \half  \sigma_X {\bf 1}~,
\end{equation}
for all $X \in \fC (M,[g])$, is known as the {\emph{Kosmann-Schwarzbach Lie derivative}}. It is worthy of note because only for this particular value of $w$ does \eqref{eq:KSLieDer} define a conformally equivariant operator on $\fS(M)$, i.e. if $g \mapsto \te^{2\omega} g$ then
\begin{equation}\label{eq:KSLieDerConf}
{\hat \cL}_X \mapsto \te^{\omega/2} {\hat \cL}_X \te^{-\omega/2}~,
\end{equation} 
for all $X \in \fC (M,[g])$. The {\emph{Penrose operator}} 
\begin{equation}\label{eq:PenroseOperator}
\cP_X = \nabla_X -  \tfrac{1}{d} {\bm X} {\bm \nabla}~,
\end{equation}
acts on $\fS ( M )$ along any $X \in \fX (M)$. It is also conformally equivariant on $\fS(M)$ and, using \eqref{eq:LieDerBracketSpinors}, obeys
\begin{equation}\label{eq:LieDerBracketPenroseSpinors}
{\hat \cL}_X ( \cP_Y \psi ) - \cP_Y ( {\hat \cL}_X  \psi ) = \cP_{[X,Y]} \psi~,
\end{equation}
for all $X \in \fC (M,[g])$, $Y \in \fX ( M )$ and $\psi \in \fS ( M )$.

The subspace of {\emph{conformal Killing (or twistor) spinors}} in $\fS ( M )$ is defined by 
\begin{equation}\label{eq:CKS}
\fZ ( M,[g] ) = \{ \psi \in \fS ( M ) \; |\; \nabla_X \psi = \tfrac{1}{d} {\bm X} {\bm \nabla} \psi \; , \; \forall \, X \in \fX (M) \}~.
\end{equation}
By construction, $\fZ(M,[g]) = \ker \cP$ and conformal equivariance of the Penrose operator explains why the subspace \eqref{eq:CKS} is assigned to the conformal class $[g]$ rather than to the particular metric $g$ on $M$. Furthermore, \eqref{eq:LieDerBracketPenroseSpinors} shows that $\fZ(M,[g])$ is preserved by the action of the Kosmann-Schwarzbach Lie derivative \eqref{eq:KSLieDer}. A key property of twistor spinors is that they \lq square' to conformal Killing vectors, in the sense that the vector field defined by \eqref{eq:Bilinear} is $\xi_{\psi,\varphi} \in \fC(M,[g])$, for any $\psi,\varphi \in \fZ ( M,[g] )$. 

With respect to a particular metric in $[g]$, any $\psi \in \fZ (M,[g])$ with $\tfrac{1}{d} {\bm \nabla} \psi = \lambda \psi$, for some $\lambda \in \CC$, is said to be {\emph{Killing}} if $\lambda \neq 0$ or {\emph{parallel}} if $\lambda =0$. The non-zero constant $\lambda$ is called the {\emph{Killing constant}} of a Killing spinor $\psi$.

Taking a derivative of the defining equation for any $\psi \in \fZ(M,[g])$ yields the following conditions 
\begin{equation}\label{eq:2DerTwistorSpinors}
\nabla_X {\bm \nabla} \psi  = \tfrac{d}{2} {\bm K}(X) \psi \;\; , \quad\quad {\bm \nabla}^2 \psi = - \tfrac{d}{4(d-1)} R\, \psi~,
\end{equation}
and combining \eqref{eq:2DerTwistorSpinors} with \eqref{eq:LCBracket} implies the important integrability conditions
\begin{equation}\label{eq:TwistorSpinorIntegrability}
{\bm W}(X,Y) \psi = 0 \;\; , \quad\quad {\bm C}(X,Y) \psi = \tfrac{1}{d} {\bm W}(X,Y) {\bm \nabla}\psi~,
\end{equation}
for all $X,Y \in \fX (M)$, where ${\bm K}(X) = X^\mu K_{\mu\nu} {\bm \Gamma}^\nu$, ${\bm W}(X,Y) = \half X^\mu Y^\nu W_{\mu\nu\rho\sigma} {\bm \Gamma}^{\rho\sigma}$ and ${\bm C}(X,Y) = X^\mu Y^\nu C_{\mu\nu\rho} {\bm \Gamma}^{\rho}$, in terms of the basis conventions in Appendix~\ref{sec:CoordinateBasisConventions}.

There are a number of classification results concerning the existence of twistor spinors (with and without zeros) on $(M,g)$ in different dimensions and signatures. It can be shown that ${\mbox{dim}}\, \fZ(M,[g]) \leq 2\, {\mbox{dim}}\, \fS(M)$ and, from \eqref{eq:TwistorSpinorIntegrability}, it follows that this bound is saturated only if $(M,g)$ is locally conformally flat. The classification in $d\geq 3$ of all local conformal equivalence classes of Lorentzian spin manifolds which admit a twistor spinor without zeros was established by Baum and Leitner \cite{Baum:2002,BL:2003,Baum:2012,Leitner:2005,Baum:2008}. Their results generalise the classification in $d=4$ obtained earlier by Lewandowski in \cite{Lewandowski:1991bx} which established that any Lorentzian spin manifold admitting a twistor spinor must be locally conformally equivalent to either a pp-wave, a Fefferman space or $\RR^{3,1}$. In $d=3$, any Lorentzian spin manifold admitting a twistor spinor must be locally conformally equivalent to either a pp-wave or $\RR^{2,1}$. In $d>4$, there are a few more distinct classes of Lorentzian manifolds which admit a non-vanishing twistor spinor (see \cite{deMedeiros:2013jja} for more details). Any such geometry is locally conformally equivalent to either a Lorentzian Einstein-Sasaki manifold (if $d$ is odd) or the direct product of a Lorentzian Einstein-Sasaki manifold with a Riemannian manifold admitting Killing spinors. 

\section{Conformal symmetry superalgebras and gauged R-symmetry}
\label{sec:ConformalSymmetrySuperalgebras}

A certain class of pseudo-Riemannian spin manifolds which admit a twistor spinor may be equipped with a real Lie superalgebra structure, that we refer to as a {\emph{conformal symmetry superalgebra}} \cite{deMedeiros:2013jja,deMedeiros:2013mca}. A conformal symmetry superalgebra $\cS = \cB \oplus \cF$ contains conformal Killing vectors and constant R-symmetries in its even part $\cB$ and twistor spinors valued in certain R-symmetry representations in its odd part $\cF$. A brief review of the general structure of Lie superalgebras is provided at the beginning of Appendix~\ref{sec:AppendixF}. 

Let $\cR$ denote the real Lie algebra of R-symmetries. On any background $(M,g)$ that admits a conformal symmetry superalgebra $\cS$, the even part $\cB = \fC(M,[g]) \oplus \cR$, as a real Lie algebra. The action of $\cB$ on $\cF$ which defines the $[\cB,\cF]$ bracket of $\cS$ involves the action of $\fC(M,[g])$ on $\fZ(M,[g])$ defined by the Kosmann-Schwarzbach Lie derivative \eqref{eq:KSLieDer} and the action of $\cR$ defined by the R-symmetry representation of $\cF$. The $\fC(M,[g])$ part of the $[\cF,\cF]$ bracket of $\cS$ involves pairing twistor spinors, using the spinorial bilinear form in \eqref{eq:SpinorInnerProd} to make a conformal Killing vector \eqref{eq:Bilinear}, and projecting onto the $\cR$-invariant part. 

The type of spinor representation must be compatible with the type of R-symmetry representation in order to define $\cF$ as a real $\cB$-module. This puts restrictions on $\cR$ according to the dimension $d$ of $M$ and the signature of $g$. In Lorentzian signature, the critical data is summarised in Table~\ref{tab:GenericLSAData}. Entries in the \lq Type' row in Table~\ref{tab:GenericLSAData} denote the ground field $\KK$ over which the representation of $\cR$ is defined. The dimension over $\KK$ of this representation is denoted by $\eN$. Entries in the \lq $\cS_{\mathrm{max}}$' row of Table~\ref{tab:GenericLSAData} denote the conformal symmetry superalgebra $\cS \cong \cS_{\mathrm{max}}$ that is realised only when $(M,g)$ is locally conformally equivalent to Minkowski space. 
\begin{table}
\begin{center}
\begin{tabular}{|c||c|c|c|c|c|}
  \hline
  &&&&& \\ [-.4cm]
  $d$ & $3$ & $4$ & $4$ & $5$ & $6$ \\ [.05cm] 
  \hline
  &&&&& \\ [-.4cm]
  Type & $\RR$ & $\CC$ & $\CC$ & $\HH$ & $\HH$ \\ [.05cm] 
  \hline
  &&&&& \\ [-.4cm]
  $\cR$ & $\fso(\eN)$ & $\fu(\eN{\neq} 4)$ & $\fsu(4)$ & $\fsp(1)$ & $\fsp(\eN)$ \\ [.05cm] 
  \hline
  &&&&& \\ [-.4cm]
  $\cS_{\mathrm{max}}$ & $\fosp ( \eN | 4)$ & $\fsu(2,2 |\eN{\neq}4)$ & ${\mathfrak{psu}}(2,2 |4)$ & $\ff(4)$ & $\fosp(6,2|\eN)$ \\ [.05cm] 
  \hline
  \end{tabular} \vspace*{.2cm}
\caption{Generic conformal symmetry superalgebra data for Lorentzian $(M,g)$.}
\label{tab:GenericLSAData}
\end{center}
\end{table}

If there exists a $\fK(M,g)$-invariant subspace $\cF_\circ \subset \cF$ such that the $\fC(M,[g])$ part of $[\cF_\circ,\cF_\circ]$ is in $\fK(M,g) < \fC(M,[g])$ then the background $(M,g)$ can be assigned the {\emph{symmetry superalgebra}} $\cS_\circ = \cB_\circ \oplus \cF_\circ$, where $\cB_\circ = \fK(M,g) \oplus \cR$.

Consider now the gauging of R-symmetry in the construction above. This amounts to promoting a given $\cR$-module to a non-trivial vector bundle that is equipped with a conformally invariant connection $A$. Locally, elements in $\cF$ now correspond to spinors on $(M,g)$ that are valued in sections of this vector bundle. By replacing all occurrences of the Levi-Civit\`{a} connection $\nabla$ with the gauged connection $D=\nabla +A$ (i.e. in the Kosmann-Schwarzbach Lie derivative \eqref{eq:KSLieDer} and the Penrose operator \eqref{eq:PenroseOperator}), the construction is made manifestly equivariant with respect to the gauged R-symmetry. 

Following this prescription for a given conformal symmetry superalgebra $\cS$ implies that each $\epsilon \in \cF$ obeys a twistor spinor equation 
\begin{equation}\label{eq:TSGauged}
D_X \epsilon =  \tfrac{1}{d} {\bm X} {\bm D} \epsilon~,
\end{equation}
with respect to the gauged connection $D$, for all $X \in \fX(M)$. For any $\epsilon \in \cF$, let $\Xi_\epsilon$ denote the component of $[\epsilon,\epsilon]$ in $\fC(M,[g])$. That $\Xi_\epsilon$ indeed remains a conformal Killing vector after gauging the R-symmetry follows directly from  \eqref{eq:TSGauged}, using the fact that $\Xi_\epsilon$ is $\cR$-invariant. 

The action of $\fC(M,[g])$ on $\cF$ is of the form
\begin{equation}\label{eq:BFGauged}
[X,\epsilon] = {\hat \cL}_X \epsilon + ( A_X + \rho_X ) \cdot \epsilon~,
\end{equation}
for all $X \in \fC(M,[g])$ and $\epsilon \in \cF$, in terms of some $\cR$-valued function $\rho_X$ on $M$. Generically, \eqref{eq:BFGauged} is not in $\cF$ since $[X,\epsilon]$ does not obey the twistor spinor equation \eqref{eq:TSGauged}. However, this property does follow if 
\begin{equation}\label{eq:iXF}
\iota_X F = D \rho_X~,
\end{equation}
for all $X \in \fC(M,[g])$, where $F =\rd A + A \wedge A$ denotes the curvature of $D$. 
\footnote{In fact, it is sufficient that $\Phi_X \cdot \epsilon = \tfrac{1}{d} {\bm X} {\bm \Phi} \cdot \epsilon$, for all $X \in \fC(M,[g])$ and $\epsilon \in \cF$, where $\Phi_X = \iota_X F - D \rho_X$. Generically this condition is weaker than \eqref{eq:iXF} but is equivalent to it for all the cases of interest here.}
For any $X,Y \in \fC(M,[g])$ which obey \eqref{eq:iXF}, $[X,Y] \in \fC(M,[g])$ also obeys \eqref{eq:iXF} with
\begin{equation}\label{eq:iXFConsistency}
\rho_{[X,Y]} = F(X,Y) + D_X \rho_Y - D_Y \rho_X + [ \rho_X , \rho_Y ]~.
\end{equation}
The condition \eqref{eq:iXF} ensures not only that $[\cB,\cF] \subset \cF$ but also that the $[\cB\cB\cF]$  and $[\cB\cF\cF]$ components of the Jacobi identity \eqref{eq:LSAJacobi} for $\cS$ remain satisfied. Furthermore, $[\cF,\cF] \subset \cB$ and the $[\cF\cF\cF]$ component of the Jacobi identity for $\cS$ remains satisfied as a consequence of manifest equivariance with respect to the gauged R-symmetry. Thus, $\cS$ remains a Lie superalgebra after gauging the R-symmetry provided \eqref{eq:iXF} is satisfied. 

Taking $X =\Xi_\epsilon$ in \eqref{eq:iXF}, for all $\epsilon \in \cF$, typically constrains the form of $F$. If $F=0$, one can choose a gauge such that $\cS$ recovers the form it took before gauging the R-symmetry. In Lorentzian signature, one finds that \eqref{eq:iXF} is satisfied identically (with $F\neq 0$) for all conformal Killing vectors in $[\cF,\cF]$ only if $d=3$ with $\cR = \fso(2)$ or $d=4$ with $\cR = \fu(1)$. Notice that these are the only two cases in Table~\ref{tab:GenericLSAData} where $\cR$ is abelian. It follows that $F = \rd A$ and $D \rho_X = \rd \rho_X$ in \eqref{eq:iXF}, which is the local characterisation of  
\begin{equation}\label{eq:iXFAbelian}
\cL_X F = 0~,
\end{equation}
for all $X \in \fC(M,[g])$. Furthermore, it is precisely for these two cases that the data $(g,A)$ describes the full set of bosonic fields in an off-shell conformal gravity supermultiplet and \eqref{eq:TSGauged} is the defining condition for bosonic supersymmetric vacua. The structure of field theories with rigid supersymmetry on such backgrounds has been explored recently in \cite{Cassani:2012ri,deMedeiros:2012sb,Hristov:2013spa,Kuzenko:2013uya}.

\section{$d=3$}
\label{sec:dequals3}

\subsection{Null triads}
\label{sec:NullTriad3d}

Let $(\xi , \theta , \chi )$ denote a {\emph{null triad}} of real vector fields on $(M,g)$, subject to the defining relations
\begin{equation}\label{eq:NullTriadRelations}
\rn{\xi}^2 = \rn{\theta}^2 = g(\xi,\chi) = g(\theta,\chi) =0   \; , \quad\quad g(\xi,\theta) = \rn{\chi}^2 =1~.
\end{equation}
The relations \eqref{eq:NullTriadRelations} are preserved under the following transformations (which collectively generate ${\mathrm{O}}(2,1)$):
\begin{itemize}
\item $( \xi , \theta , \chi ) \mapsto ( \theta , \xi , \chi )$.
\item $( \xi , \theta , \chi ) \mapsto ( \xi , \theta - \alpha \chi - \half\alpha^2 \xi , \chi + \alpha \xi )$, for any $\alpha \in \RR$.
\item $( \xi , \theta , \chi ) \mapsto ( \beta \xi , \beta^{-1} \theta , \pm \chi )$, for any $\beta \in \RR \backslash \{ 0 \}$.
\end{itemize}

It is convenient to use the null triad to express the metric and volume form on $M$ as
\begin{equation}\label{eq:NullTriadMetric}
g = \xi \otimes \theta + \theta \otimes \xi + \chi \otimes \chi \; , \quad\quad \varepsilon = \xi \wedge \theta \wedge \chi~.
\end{equation}

\subsection{Majorana spinors}
\label{sec:MajoranaSpinors3d}

The Clifford algebra $\Cl (2,1) \cong \Mat_2 ( \RR ) \oplus \Mat_2 ( \RR )$ has two inequivalent irreducible representations, each isomorphic to $\RR^2$, which are both identified with the unique irreducible representation of $\Cl^0 (2,1) \cong \Mat_2 ( \RR )$ after restricting to $\Spin(2,1)$. This restriction defines the Majorana spinor representation.  

Relative to the basis conventions \eqref{eq:ClBasis} in Appendix~\ref{sec:CoordinateBasisConventions}, Hodge duality in the exterior algebra implies
\begin{equation}\label{eq:3dgammadual}
{\bm \Gamma}_{\mu\nu} = \varepsilon_{\mu\nu\rho} {\bm \Gamma}^{\rho} \; , \quad\quad {\bm \Gamma}_{\mu\nu\rho} = \varepsilon_{\mu\nu\rho} {\bf 1}~,
\end{equation}
on Majorana spinors.

The bilinear form \eqref{eq:SpinorInnerProd} on $\fS(M)$ is unique and skewsymmetric. For all $\psi , \varphi \in \fS ( M )$, it follows that 
\begin{equation}\label{eq:3dbilnears}
{\overline \psi} \varphi  = - {\overline \varphi} \psi \; , \quad\quad {\overline \psi} {\bm \Gamma}_{\mu} \varphi  = {\overline \varphi} {\bm \Gamma}_\mu \psi~,
\end{equation} 
and the associated Fierz identity is given by
\begin{equation}\label{eq:3dfierz2}
\psi {\overline \varphi} = \half ( ( {\overline \varphi} \psi ) {\bf 1} + ( {\overline \varphi} {\bm \Gamma}^\mu \psi ) {\bm \Gamma}_\mu )~.
\end{equation} 
In terms of a unitary basis for $\Cl (2,1)$, all the quantities above are manifestly real. 

For any $\epsilon \in \fS(M)$, let us define the real vector field
\begin{equation}\label{eq:3dbilinearsEpsilon}
\xi_\epsilon^\mu = {\overline \epsilon} {\bm \Gamma}^\mu \epsilon~.
\end{equation}
We shall assume henceforth that $\xi_\epsilon$ is nowhere vanishing, which is so only if $\epsilon$ is nowhere vanishing. Furthermore, using \eqref{eq:3dfierz2}, it follows that ${\bm \xi_\epsilon} \epsilon =0$. Consequently, $\xi_\epsilon$ is null.

For any $\epsilon , \epsilon^\prime \in \fS(M)$, let us also define
\begin{equation}\label{eq:3dbilinearsEpsilon2}
\zeta^\mu = {\overline \epsilon} {\bm \Gamma}^\mu \epsilon^\prime \; , \quad\quad \kappa = {\overline \epsilon} \epsilon^\prime ~.
\end{equation}
The scalar $\kappa$ vanishes identically only if $\epsilon$ and $\epsilon^\prime$ are linearly dependent. We shall assume henceforth that $\kappa \neq 0$. Using \eqref{eq:3dfierz2}, it follows that  
\begin{equation}\label{eq:3dbilinearsEpsilonIdentities}
{\bm \xi_\epsilon} \epsilon^\prime = -2 {\bm \zeta} \epsilon = 2 \kappa \epsilon~.
\end{equation}
Thus, $\xi_\epsilon$ and $\xi_{\epsilon^\prime}$ are both null while $\zeta$ is spacelike, with
\begin{equation}\label{eq:3dbilinearsEpsilonIdentities2}
\rn{\zeta}^2 = \kappa^2 = -\half g( \xi_\epsilon , \xi_{\epsilon^\prime} ) \;\; , \quad\quad g( \xi_\epsilon , \zeta ) = g( \xi_{\epsilon^\prime} , \zeta ) =0~.
\end{equation}

It is sometimes convenient to identify \eqref{eq:3dbilinearsEpsilon} and \eqref{eq:3dbilinearsEpsilon2} in terms of the null triad introduced in section~\ref{sec:NullTriad3d}, such that 
\begin{equation}\label{eq:3dNullTriadBilinears}
\xi_\epsilon = \xi \; , \quad\quad -\half \kappa^{-2} \xi_{\epsilon^\prime} = \theta \; , \quad\quad \kappa^{-1} \zeta = \chi~.
\end{equation}
In terms of this identification, $\epsilon^\prime = \kappa {\bm \theta} \epsilon$. Since $( \epsilon , \epsilon^\prime )$ define a basis of Majorana spinors, it follows that any $\psi \in \fS(M)$ can be written 
\begin{equation}\label{eq:3dfierz3}
\psi = \alpha \epsilon + \beta {\bm \theta} \epsilon~,
\end{equation} 
where $\alpha = {\overline \epsilon} {\bm \theta} \psi$ and $\beta = {\overline \epsilon} \psi$. 

\subsection{Causal conformal Killing vectors}
\label{sec:CausalCKV3d}

The action of the Levi-Civit\`{a} connection $\nabla$ on the null triad one-forms is constrained by the relations \eqref{eq:NullTriadRelations} such that
\begin{align}\label{eq:NablaTriad}
\nabla_\mu \xi_\nu &= p_\mu \xi_\nu - q_\mu \chi_\nu \nonumber \\ 
\nabla_\mu \theta_\nu &= - p_\mu \theta_\nu + r_\mu \chi_\nu \\
\nabla_\mu \chi_\nu &= - r_\mu \xi_\nu + q_\mu \theta_\nu~, \nonumber
\end{align}
in terms of three real one-forms $(p,q,r)$.

Given any $X \in \fX(M)$ with $\rn{X}^2 =0$, using the transformations below \eqref{eq:NullTriadRelations}, one can always define a null triad such that $X=\xi$. From the first line in \eqref{eq:NablaTriad}, it follows that $\xi$ is a conformal Killing vector only if
\begin{equation}\label{eq:CKVcondition3d}
p_\theta = q_\xi =0 \; , \quad\quad p_\xi = -2 q_\chi \; , \quad\quad p_\chi = q_\theta~. 
\end{equation}
The conformal factor \eqref{eq:CKV} for $\xi \in \fC(M,[g])$ is $\sigma_\xi = q_\chi$. The condition $q_\xi =0$ in \eqref{eq:CKVcondition3d} is equivalent to $\xi \wedge \rd \xi =0$. If $\xi$ is a Killing vector then \eqref{eq:CKVcondition3d} are satisfied with $\sigma_\xi = q_\chi = 0 = p_\xi$. If $\xi$ is $\nabla$-parallel then $p=q=0$.

Identifying $\xi = \xi_\epsilon$ as in \eqref{eq:3dNullTriadBilinears} implies
\begin{equation}\label{eq:NablaTriadSpinors}
\nabla_\mu \epsilon = \half ( p_\mu \epsilon - q_\mu {\bm \theta} \epsilon )~. 
\end{equation}
It follows that \eqref{eq:CKVcondition3d} are in fact necessary and sufficient for $\epsilon \in \fZ (M,[g])$. Since $\epsilon \in \fZ (M,[g])$ implies $\xi_\epsilon = \xi \in \fC(M,[g])$, clearly \eqref{eq:CKVcondition3d} are necessary for any $\epsilon \in \fZ (M,[g])$. The point is that, at least locally, the existence of a nowhere vanishing null conformal Killing vector is actually equivalent to the existence of a nowhere vanishing Majorana twistor spinor \cite{deMedeiros:2012sb}. Furthermore, if $\xi_\epsilon$ is a Killing vector then $\epsilon$ is necessarily a Killing spinor (with Killing constant $-\half q_\theta$). If $\xi_\epsilon$ is $\nabla$-parallel then $\epsilon$ is necessarily also $\nabla$-parallel.  

Given any $X \in \fX(M)$ with $\rn{X}^2 \neq 0$, using the transformations below \eqref{eq:NullTriadRelations}, one can always define a null triad such that $X=\xi + \half \rn{X}^2 \, \theta$. If $\rn{X}^2 < 0$, let $\rn{X}^2 = -4 \kappa^2$ so that $X = \xi - 2\kappa^2 \theta$. From the first two lines in \eqref{eq:NablaTriad}, it follows that $X$ is a conformal Killing vector only if
\begin{align}\label{eq:TimelikeCKVcondition3d}
p_\theta &=0 \; , \quad\quad q_\xi + 2\kappa^2 r_\xi = 2( \kappa^2 p_\chi - \partial_\chi \kappa^2 ) \; , \quad\quad p_\xi -2 \partial_\theta \kappa^2 = -2 ( q_\chi +2 \kappa^2 r_\chi ) \; ,  \nonumber \\ 
p_\chi &= q_\theta + 2\kappa^2 r_\theta \; , \quad\quad \kappa^2 p_\xi = \partial_\xi \kappa^2~. 
\end{align}
The conformal factor $\sigma_X = -\partial_X (\ln \kappa )$. If $X$ is a Killing vector then \eqref{eq:TimelikeCKVcondition3d} are satisfied with $p_\xi = 2 \partial_\theta \kappa^2$. 

Let us now identify the non-zero scalar $\kappa$ above with its namesake in \eqref{eq:3dbilinearsEpsilon2}. Identifying $\xi = \xi_\epsilon$ and $\theta = -\half \kappa^{-2} \xi_{\epsilon^\prime}$ as in \eqref{eq:3dNullTriadBilinears} then implies $X = \xi - 2\kappa^2 \theta = \xi_\epsilon + \xi_{\epsilon^\prime}$. Let us name this timelike vector field $\Xi = \xi_\epsilon + \xi_{\epsilon^\prime}$, with $\rn{\Xi}^2 = -4\kappa^2$. With respect to the aforementioned identifications, it follows that
\begin{align}\label{eq:TimelikeNablaTriadSpinors}
\nabla_\mu \epsilon &= \half ( p_\mu \epsilon - \kappa^{-1} q_\mu \epsilon^\prime ) \nonumber \\
\nabla_\mu \epsilon^\prime &= - \kappa r_\mu \epsilon - \half ( p_\mu + \partial_\mu ( \ln \kappa ) ) \epsilon^\prime~. 
\end{align}

If $\Xi$ is a conformal Killing vector then
\begin{equation}\label{eq:dXi}
\rd ( \kappa^{-2} \Xi ) = - \kappa^{-3} \rho_\Xi \, {*\Xi}~, 
\end{equation}
where 
\begin{equation}\label{eq:RhoXi}
\rho_\Xi = 2( \partial_\chi \kappa - \kappa p_\chi )~.
\end{equation}
Thus $\Xi \wedge \rd \Xi = 4\kappa \rho_\Xi \, \xi\wedge \theta\wedge\chi$ is zero only if $\rho_\Xi =0$. 

\subsection{Charged twistor spinors}
\label{sec:CTS3d}

Consider now the implications of the existence of a charged twistor spinor, which can be thought of locally as a pair of Majorana spinors $(\epsilon,\epsilon^\prime)$ both obeying \eqref{eq:TSGauged} with respect to the action of the $\cR = \fso(2)$ gauged connection
\begin{align}\label{eq:GaugedD3d}
D_\mu \epsilon &= \nabla_\mu \epsilon + A_\mu \epsilon^\prime \nonumber \\
D_\mu \epsilon^\prime &= \nabla_\mu \epsilon^\prime - A_\mu \epsilon~. 
\end{align}
An equivalent version of \eqref{eq:GaugedD3d} follows by taking the real and imaginary parts of $D_\mu \epsilon_\CC = \nabla_\mu \epsilon_\CC - i A_\mu \epsilon_\CC$, where $\epsilon_\CC = \epsilon + i \epsilon^\prime$. Because $(\epsilon,\epsilon^\prime)$ transform as a $2$-vector under $\fso(2)$, $\epsilon_\CC = \epsilon + i \epsilon^\prime$ has unit charge under $\fu(1) \cong \fso(2)$.

The first consequence of \eqref{eq:GaugedD3d} is that the timelike vector field $\Xi = \xi_\epsilon + \xi_{\epsilon^\prime} \in \fC(M,[g])$. Thus, using \eqref{eq:TimelikeNablaTriadSpinors}, one finds that the five conditions in \eqref{eq:TimelikeCKVcondition3d} are necessary in order for $(\epsilon,\epsilon^\prime)$ to define a charged twistor spinor with respect to $D$. The remaining three conditions which come from the twistor spinor equation for $(\epsilon,\epsilon^\prime)$ are precisely sufficient to fix all three components of $A$, such that
\begin{equation}\label{eq:3dA}
A_\mu = -\kappa r_\theta \xi_\mu + \half \kappa^{-1} q_\xi \theta_\mu + ( \partial_\theta \kappa - \kappa r_\chi ) \chi_\mu~, 
\end{equation}
in terms of the identifications in \eqref{eq:3dNullTriadBilinears}.

Thus we conclude that if $(M,g)$ admits a nowhere vanishing everywhere timelike conformal Killing vector $X$, then there must exist a nowhere vanishing charged twistor spinor  pair $( \epsilon , \epsilon^\prime )$ on $(M,g)$, with $X = \Xi$, that is defined with respect to the gauged connection $A$ in \eqref{eq:3dA}. This characterisation of charged twistor spinors on a smooth orientable Lorentzian three-manifold was first obtained in \cite{Hristov:2013spa}. 

Using \eqref{eq:3dA}, the twistor spinor equation for $(\epsilon,\epsilon^\prime)$ implies that the real function 
\begin{equation}\label{eq:rhoeps3d}
\rho_{\Xi} = \tfrac{2}{3} ( {\overline \epsilon} {\bm D} \epsilon^\prime - {\overline \epsilon^\prime} {\bm D} \epsilon )~,
\end{equation}
is identical to its namesake in \eqref{eq:RhoXi}. Moreover, it follows that
\begin{equation}\label{eq:iXiF}
\iota_\Xi F = \rd \rho_\Xi~,
\end{equation}
so \eqref{eq:iXF} is satisfied identically for $X=\Xi$. 

Differentiating the defining condition \eqref{eq:TSGauged} with \eqref{eq:GaugedD3d} gives  
\begin{equation}\label{eq:2DerTwistorSpinorsGauged3d}
\tfrac{2}{3} D_\mu {\bm D} \epsilon_\CC  = K_{\mu\nu} {\bm \Gamma}^\nu \epsilon_\CC -i( F_{\mu\nu} {\bm \Gamma}^\nu - {\tilde F}_\mu {\bf 1} ) \epsilon_\CC~,
\end{equation}
where ${\tilde F}_\mu = \half \varepsilon_{\mu\nu\rho} F^{\nu\rho}$. Combining \eqref{eq:2DerTwistorSpinorsGauged3d} with \eqref{eq:LCBracket} implies the integrability condition
\begin{equation}\label{eq:TwistorSpinorIntegrabilityGauged3d}
i C_{\mu\nu\rho} {\bm \Gamma}^\rho \epsilon_\CC  = ( 2\nabla_{[\mu} {\tilde F}_{\nu]} {\bf 1} + {\bm \nabla} F_{\mu\nu} ) \epsilon_\CC + \kappa^{-1} ( F_{\mu\nu} {\bf 1} - {\tilde F}_{[\mu} {\bm \Gamma}_{\nu]} ) ( \rho_\Xi {\bf 1} + 2 {\bm \nabla} \kappa ) \epsilon_\CC~,
\end{equation}
using the identity
\begin{equation}\label{eq:EDE3d}
\tfrac{4}{3} {\bm D} \epsilon_\CC  = \kappa^{-1} ( \rho_\Xi {\bf 1} + 2 {\bm \nabla} \kappa ) \epsilon_\CC~.
\end{equation} 

It is convenient to define ${\tilde C}_{\mu\nu} = \half \varepsilon_{\mu}{}^{\rho\sigma} C_{\rho\sigma\nu}$ in $d=3$. From the definition of the Cotton-York tensor \eqref{eq:CottonTensor}, it follows that ${\tilde C}_{\mu\nu} = {\tilde C}_{\nu\mu}$, ${\tilde C}_{\mu\nu} g^{\mu\nu} =0$ and $\nabla^\mu {\tilde C}_{\mu\nu} =0$ identically. An equivalent, but somewhat more wieldy, form of \eqref{eq:TwistorSpinorIntegrabilityGauged3d} is given by
\begin{equation}\label{eq:TwistorSpinorIntegrabilityGauged3dCtilde}
i {\tilde C}_{\mu\nu} {\bm \Gamma}^\nu \epsilon_\CC  = \kappa^{-1} ( ( \rho_\Xi  {\tilde F}_{\mu} - \nabla^{\nu} ( \kappa F_{\mu\nu} ) ) {\bf 1} + \kappa {\bm \nabla} {\tilde F}_\mu +2 {\tilde F}_{\mu} {\bm \nabla} \kappa + 2 {\tilde F}^\nu ( \partial_{[\mu} \kappa ) {\bm \Gamma}_{\nu]} - \half \rho_\Xi   {\tilde F}^\nu {\bm \Gamma}_{\mu\nu} ) \epsilon_\CC~.
\end{equation}
From \eqref{eq:TwistorSpinorIntegrabilityGauged3dCtilde}, it follows that 
\begin{equation}\label{eq:TwistorSpinorIntegrabilityGauged3dCtildeXi}
\half {\tilde C}_{\mu\nu} \Xi^\nu = \rho_\Xi  {\tilde F}_{\mu} - \nabla^\nu ( \kappa F_{\mu\nu} )~.
\end{equation}

\subsection{Conformal symmetry superalgebras}
\label{sec:CKS3d}

Let $(M,g)$ be any smooth orientable Lorentzian three-manifold equipped with a nowhere vanishing causal conformal Killing vector $X$ and a $\fC(M,[g])$-invariant closed two-form $F$. As we have explained, any null $X \in \fC(M,[g])$ defines a nowhere vanishing twistor spinor $\epsilon$ (with $X = \xi = \xi_\epsilon$) while any timelike $X \in \fC(M,[g])$ defines a nowhere vanishing charged twistor spinor pair $(\epsilon,\epsilon^\prime) $ (with $X = \Xi = \xi_\epsilon + \xi_{\epsilon^\prime}$) that is charged with respect to the connection $A$ in \eqref{eq:3dA} whose curvature is $F = \rd A$. From section~\ref{sec:ConformalSymmetrySuperalgebras}, we recall that this data is sufficient to assign to $(M,[g])$ a conformal symmetry superalgebra $\cS$ with trivial $\cR$ for null $X$ or gauged $\cR = \fso(2)$ for timelike $X$.

\subsubsection{Null case}
\label{sec:nullcase}
If $(M,g)$ admits a nowhere vanishing conformal Killing vector $\xi$ that is everywhere null then it is locally conformally equivalent to a pp-wave or $\RR^{2,1}$. We shall assume henceforth that $(M,g)$ is not locally conformally flat. The structure of conformal Killing vectors of three-dimensional pp-waves is discussed in some detail in Appendix~\ref{sec:3dppwaves}. In the generic case, the explicit form of the pp-wave metric $g_{\mathrm{pp}}$ is given by \eqref{eq:ppwave}, in terms of Brinkmann coordinates $(u,v,x)$ on $M$. In these coordinates, one can take $\xi = \partial_v$. 

If $2 \leq {\mbox{dim}}\, \fC(M,[g_{\mathrm{pp}}]) \leq  4$, it was shown in Appendix~\ref{sec:3dppwaves} that the structure of $\fC ( M,[g_{\mathrm{pp}}] )$ depends critically on whether or not $\xi \in Z( \fC ( M,[g_{\mathrm{pp}}] ))$. 

If $\xi \in Z( \fC ( M,[g_{\mathrm{pp}}] ))$ then $\fC ( M,[g_{\mathrm{pp}}] ) = \fK (M,g)$ (i.e. conformally isometric). This requires the existence of a Lorentzian three-manifold $(M,g)$ with $2 \leq {\mbox{dim}}\, \fK(M,g) \leq  4$ which admits a non-zero null $\xi \in Z( \fK(M,g) )$. From the second paragraph in Appendix~\ref{sec:TheKruckovicClassification}, it transpires that the only option is to have $\fC(M,[g_{\mathrm{pp}}]) \cong \RR^2$ with isometric action of Killing vectors $(\partial_u , \xi = \partial_v )$ on a metric in $[g_{\mathrm{pp}}]$ of the form 
\begin{equation}\label{eq:ppwave2}
2 \rd u \rd v + H(x) {\rd u}^2 + {\rd x}^2~,
\end{equation}
with $\partial_x^3 H \neq0$. 

If $\xi \notin Z( \fC ( M,[g_{\mathrm{pp}}] ))$ then $\fC ( M,[g_{\mathrm{pp}}] ) =  \fH(M,g)$ (i.e. conformally homothetic) and $\xi \in Z( \fK(M,g) )$. Just as in the previous case, $\fK(M,g) \cong \RR^2$ is the only option if ${\mbox{dim}}\, \fC(M,[g_{\mathrm{pp}}]) >2$ (the ${\mbox{dim}}\, \fC(M,[g_{\mathrm{pp}}]) =2$ case is covered in Appendix~\ref{sec:3dppwaves}). Clearly ${\mbox{dim}}\, \fC(M,[g_{\mathrm{pp}}]) =4$ is impossible since ${\mbox{dim}}\, \fC(M,[g_{\mathrm{pp}}]) = {\mbox{dim}}\, \fH(M,g) = {\mbox{dim}}\, \fK(M,g) +1$. The only remaining option is ${\mbox{dim}}\, \fC(M,[g_{\mathrm{pp}}]) =3$. As shown in Appendix~\ref{sec:3dppwaves}, the homothetic action of $\fC ( M,[g_{\mathrm{pp}}] )$ is with respect to $g = ( \partial_x^3 H )^{2/5}  g_{\mathrm{pp}}$, where $g_{\mathrm{pp}}$ is of the form \eqref{eq:ppwave2}. The proper homothetic $X \in \fH ( M,g ) / \fK (M,g)$ obeys $\cL_X g = \tfrac{8}{5} c_X g$ and $[ \xi , X] = 2 c_X \xi$ with $c_X \neq 0$. The non-zero value of $c_X$ is irrelevant and it is convenient to work in terms of $\vartheta := -\tfrac{1}{2c_X} X \in \fH ( M,g ) / \fK (M,g)$ which has $c_\vartheta = -\half$. Solving $\cL_\vartheta g = -\tfrac{4}{5} g$ and $[ \vartheta , \xi ] =  \xi$ yields three inequivalent classes of solutions for $H$ and $\vartheta$ which are displayed in Table~\ref{tab:dim3PP}. 
\begin{table}
\begin{center}
\begin{tabular}{|c|c|c|}
  \hline
  $H$ & $\vartheta$ & $ \fC(M,[g_{\mathrm{pp}}])$ \\ [.05cm]
  \hline\hline
  && \\ [-.4cm]
  $b + c\, {\mathrm{e}}^x$ & $u \partial_u - ( v + b u) \partial_v - 2\partial_x$ & $\fb$[VI$_0$] \\ [.05cm]
  && \\ [-.4cm]
  $b - 2c \ln x$ & $-u \partial_u - ( v + c u) \partial_v - x\partial_x$ & $\fb$[IV] \\ [.05cm]
  && \\ [-.4cm]
  $b + c\, x^{-2\left(\frac{a -1}{a +1}\right)}$ & $-a u \partial_u - ( v -\half(a -1) b u) \partial_v - \half (a +1) x\partial_x$ & $\fb$[VI] \\ [.15cm]
  \hline  
\end{tabular} \vspace*{.2cm}
\caption{Data for geometries with ${\mbox{dim}}\, \fC(M,[g_{\mathrm{pp}}]) =3$.}
\label{tab:dim3PP}
\end{center}
\end{table}
The parameters $a \in \RR \backslash \{ 0 , \pm 1 , \tfrac{1}{3} \}$, $b \in \RR$ and $c \in \RR \backslash \{ 0\}$. Entries in the third column of Table~\ref{tab:dim3PP} denote three-dimensional indecomposable real Lie algebras in the Bianchi classification, which is summarised in Appendix~\ref{sec:LALow}. For the functions $H$ in Table~\ref{tab:dim3PP}, $\vartheta$ is homothetic not only with respect to $g = ( \partial_x^3 H )^{2/5}  g_{\mathrm{pp}}$ but also with respect to $g_{\mathrm{pp}}$ (albeit with a different constant conformal factor). For $H = b + c\, {\mathrm{e}}^x$, $\vartheta$ is a Killing vector with respect to $g_{\mathrm{pp}}$ and $(M,g_{\mathrm{pp}})$ is homogeneous. In the other two cases, $\vartheta$ is a Killing vector with respect to $x^{-2} g_{\mathrm{pp}}$. For each geometry in Table~\ref{tab:dim3PP}, it follows by direct calculation that the associated conformal symmetry superalgebra $\cS \cong \cS^\lhd ( \fC(M,[g_{\mathrm{pp}}]))$, in terms of the notation introduced in Appendix~\ref{sec:AppendixF} (data for the three Lie superalgebras of interest appears in Table~\ref{tab:41LSAUncentered}). 

\subsubsection{Timelike case}
\label{sec:timelike}
Let us now assume that $(M,g)$ admits a nowhere vanishing conformal Killing vector $\Xi$ that is everywhere timelike. As we have explained, this data defines a nowhere vanishing charged twistor spinor pair $( \epsilon , \epsilon^\prime )$ such that $\Xi = \xi_\epsilon + \xi_{\epsilon^\prime}$. We shall also insist that the closed two-form $F =\rd A$ defined by \eqref{eq:3dA} is $\fC(M,[g])$-invariant so that we may assign to $(M,[g])$ a conformal symmetry superalgebra $\cS$.

If $F=0$ then locally $A=\rd \lambda$ and the charged twistor spinor pair $( \epsilon , \epsilon^\prime )$ is equivalent to a pair of ordinary twistor spinors $( \cos \lambda \, \epsilon + \sin \lambda \, \epsilon^\prime , -\sin \lambda \, \epsilon + \cos \lambda \, \epsilon^\prime )$. Since $\epsilon$ and $\epsilon^\prime$ are linearly independent, clearly $F=0$ implies $(M,g)$ must be locally conformally flat. 
\footnote{Conversely, if $(M,g)$ is locally conformally flat, the integrability condition \eqref{eq:TwistorSpinorIntegrabilityGauged3dCtilde} does not imply $F=0$. In this case, in terms of the conformally equivalent metric ${\hat g} = \kappa^{-2} g$, \eqref{eq:TwistorSpinorIntegrabilityGauged3dCtilde} just states that the one-form ${\hat *} F$ defines a Killing vector with respect to ${\hat g}$ and obeys $\rd  {\hat *} F = - \rho_\Xi F$. However, since $\fC(M,[g]) \cong \fso(3,2)$, it is the condition \eqref{eq:iXFAbelian} that fixes $F=0$. Of course, if $(M,g)$ is locally conformally flat and admits a non-zero $\fK(M,g)$-invariant $F$ which solves  \eqref{eq:TwistorSpinorIntegrabilityGauged3dCtilde}, one can define an associated symmetry superalgebra by restricting to $\fK(M,g) < \fso(3,2)$, as was done for a number of examples in \cite{Hristov:2013spa}.} 
In that case, $\fC(M,[g]) \cong \fso(3,2) \cong \fsp (4,\RR)$ and $\cS \cong \fosp (2|4)$ (i.e. the $\eN =2$, $d=3$ case in Table~\ref{tab:GenericLSAData}). We shall assume henceforth that $(M,g)$ is not locally conformally flat so both the Cotton-York tensor $C$ and the two-form $F$ are not identically zero.   

For any $X \in \fC(M,[g])$, $\cL_X C =0$ and $\cL_X F =0$ while $\cL_X {\tilde C} = \sigma_X {\tilde C}$ and $\cL_X {\tilde F} = \sigma_X {\tilde F}$ since $\cL_X g = -2 \sigma_X g$. Any non-zero scalar built from ${\tilde C}$, ${\tilde F}$ and $g$ therefore defines a conformal scalar $\phi$ (with weight $p_\phi$). Five different options are displayed in Table~\ref{tab:3dConformalScalars}. 
\begin{table}
\begin{center}
\begin{tabular}{|c||c|c|c|c|c|}
  \hline
  &&&&& \\ [-.4cm]
  $\phi$ & ${\tilde C}_{\mu\nu} {\tilde C}^{\mu\nu}$ & ${\tilde C}_{\mu\nu} {\tilde C}^{\nu}{}_\rho {\tilde C}^{\rho\mu}$ & ${\tilde F}_\mu {\tilde F}^\mu$  & ${\tilde C}_{\mu\nu} {\tilde F}^\mu {\tilde F}^\nu$ & ${\tilde F}^\mu {\tilde C}_{\mu\nu} {\tilde C}^{\nu\rho} {\tilde F}_\rho$ \\ [.05cm] 
  \hline
  &&&&& \\ [-.4cm]
  $p_\phi$ & $6$ & $9$ & $4$ & $7$ & $10$ \\ [.05cm] 
  \hline
   \end{tabular} \vspace*{.2cm}
\caption{Some proper conformal scalars.}
\label{tab:3dConformalScalars}
\end{center}
\end{table}
If just one $\phi$ in Table~\ref{tab:3dConformalScalars} is nowhere vanishing then $\fC(M,[g])$ is conformally isometric. Conversely, let us examine what happens if all $\phi$ in Table~\ref{tab:3dConformalScalars} are identically zero. 

In this case, ${\tilde F}$ defines a non-zero vector field that is everywhere null with respect to $g$. As such, at least locally, there must exist a null triad $({\tilde F},G,H)$ such that 
\begin{equation}\label{eq:NullTriadFTilde}
g_{\mu\nu} = 2 {\tilde F}_{(\mu} G_{\nu)} + H_\mu H_\nu~.
\end{equation}
Using ${\tilde C}_{\mu\nu} {\tilde F}^\mu {\tilde F}^\nu =0$, it follows that ${\tilde F}^\mu {\tilde C}_{\mu\nu} {\tilde C}^{\nu\rho} {\tilde F}_\rho = 0$ implies ${\tilde C}_{\mu\nu} {\tilde F}^\mu H^\nu =0$. Moreover, ${\tilde C}_{\mu\nu} {\tilde C}^{\mu\nu} =0$ and ${\tilde C}_{\mu\nu} g^{\mu\nu} =0$ then imply ${\tilde C}_{\mu\nu} {\tilde F}^\nu =0$ and ${\tilde C}_{\mu\nu} H^\mu H^\nu =0$. Therefore
\begin{equation}\label{eq:CTildeNoScalars}
{\tilde C}_{\mu\nu} = \alpha {\tilde F}_\mu {\tilde F}_\nu + 2\beta {\tilde F}_{(\mu} H_{\nu)}~,
\end{equation}
in terms of a pair of real functions $\alpha$ and $\beta$ which are not both zero. The form of the Cotton-York tensor in \eqref{eq:CTildeNoScalars} further implies that ${\tilde C}_{\mu\nu} {\tilde C}^{\nu}{}_\rho {\tilde C}^{\rho\mu}$ vanishes identically. Now taking $\cL_X$ of \eqref{eq:NullTriadFTilde} and using $\cL_X g_{\mu\nu} = -2 \sigma_X g_{\mu\nu}$ and $\cL_X {\tilde F}_\mu = \sigma_X {\tilde F}_\mu$ implies that 
\begin{equation}\label{eq:LXGH}
\cL_X G_\mu = -3 \sigma_X G_\mu + \gamma_X H_\mu \; , \quad\quad \cL_X H_\mu = -\gamma_X {\tilde F}_\mu - \sigma_X H_\mu~,
\end{equation}
in terms of some real function $\gamma_X$, for all $X \in \fC(M,[g])$. Taking $\cL_X$ of \eqref{eq:CTildeNoScalars} and using $\cL_X {\tilde C}_{\mu\nu} = \sigma_X {\tilde C}_{\mu\nu}$ together with the expressions above yields
\begin{equation}\label{eq:LXGH}
\partial_X \alpha = - \sigma_X \alpha + 2\gamma_X \beta \; , \quad\quad \partial_X \beta = \sigma_X \beta~,
\end{equation}
for all $X \in \fC(M,[g])$. Whence, if $\beta \neq 0$ then $\phi = \beta$ defines a proper conformal scalar with $p_\phi =1$. If $\beta =0$ then $\phi = \alpha$ defines a proper conformal scalar with $p_\phi =-1$. Thus we have proved that $\fC(M,[g])$ is always conformally isometric in the timelike case.

This fact means that we may choose a representative metric $g$ in $[g]$ for any admissible geometry such that $\fC(M,[g]) = \fK(M,g)$. By definition, the geometry $(M,g)$ is therefore {\emph{stationary}} because it is equipped with a timelike Killing vector $\Xi$. It is convenient to express the stationary metric
\begin{equation}\label{eq:StationaryMetric}
g = - 4\kappa^2 \varpi \otimes \varpi + h~,
\end{equation}
in terms of the one-form $\varpi$ defined by $\varpi (X) = -\tfrac{1}{4\kappa^2} g(\Xi,X)$, for all $X\in \fX(M)$. By construction, $\iota_\Xi \varpi =1$ and $\iota_\Xi  h =0$ since $\rn{\Xi}^2 = -4\kappa^2$. It follows that $\cL_\Xi \varpi =0$ while $\partial_\Xi \kappa = - \sigma_\Xi \kappa =0$ and $\cL_\Xi h = -2\sigma_\Xi h=0$ because $\sigma_\Xi =0$. If $\rd \varpi =0$ then $(M,g)$ is {\emph{static}}, which occurs only if $\Xi \wedge \rd \Xi =0$. Thus, from the observation below \eqref{eq:RhoXi}, $(M,g)$ is static only if $\rho_\Xi =0$.

Any admissible geometry with ${\mbox{dim}}\, \fC(M,[g]) =4$ must be locally conformally equivalent to one of three Lorentzian geometries of class IV in \cite{Kru:1954} that is not conformally flat (see Appendix~\ref{sec:TheKruckovicClassification} for a brief summary of \cite{Kru:1954}). All three of these class IV geometries are stationary and, when expressed in the form \eqref{eq:StationaryMetric}, have
\begin{equation}\label{eq:ClassIV}
2 \kappa \varpi = \rd t + \omega(x) \rd y \; , \quad\quad h = \rd x^2 + f(x)^2 \rd y^2~,
\end{equation}
in terms of local coordinates $(t,x,y)$ on $M$, for particular pairs of functions $\omega$ and $f$ that are related such that $\partial_x \omega = a f$, for some non-zero real number $a$ (the precise data is displayed in Table~\ref{tab:ClassIVData}).
\begin{table}
\begin{center}
\hspace*{-.5cm}
\begin{tabular}{|c|c|c|c|c|c|}
  \hline
  Class & $f$ & $a$ & $k$ & $l$ & $ \fK(M,g)$ \\ [.05cm]
  \hline\hline
  &&&&& \\ [-.4cm]
  IV.4 & $1$ & $1$ & $\partial_x - y \partial_t$ & $y \partial_x - x \partial_y +\half ( x^2 - y^2 ) \partial_t$ & $\fc$[X] \\ [.05cm]
  &&&&& \\ [-.4cm]
  IV.5 & ${\mathrm{e}}^x$ & $\neq \pm 1$ & $\partial_x - y \partial_y$ & $y \partial_x +\half ( {\mathrm{e}}^{-2x} - y^2 ) \partial_y - a {\mathrm{e}}^{-x} \partial_t$ & $\RR \oplus \fb$[VIII] \\ [.05cm]
  &&&&& \\ [-.4cm]
  IV.6 & $\sin x$ & - & $\cos y \, \partial_x - \frac{\sin y}{\sin x} ( \cos x \, \partial_y + a \partial_t )$ & $-\sin y \, \partial_x - \frac{\cos y}{\sin x} ( \cos x \, \partial_y + a \partial_t )$ & $\RR \oplus \fb$[IX] \\ [.15cm]
  \hline  
\end{tabular} \vspace*{.2cm}
\caption{Data for geometries with ${\mbox{dim}}\, \fC(M,[g]) =4$.}
\label{tab:ClassIVData}
\end{center}
\end{table}
This identification fixes $\Xi = -2\kappa \partial_t$ with $\kappa$ constant. The associated null triad one-forms are 
\begin{equation}\label{eq:ClassIVNullTriadOneForms}
\xi = \kappa ( \rd t + ( \omega + f ) \rd y ) \; , \quad\quad \theta = -\frac{1}{2\kappa} ( \rd t + ( \omega - f ) \rd y ) \; , \quad\quad \chi = \rd x~.
\end{equation}
The volume form \eqref{eq:NullTriadMetric} is $\varepsilon = -f \rd t \wedge \rd x \wedge \rd y$. The action of $\nabla$ on \eqref{eq:ClassIVNullTriadOneForms} can be defined via the triple of one-forms $(p,q,r)$ in \eqref{eq:NablaTriad}, which are given by
\begin{align}\label{eq:ClassIVNablaTriad}
p_\xi &= p_\theta = q_\chi = r_\chi = 0 \; , \quad\quad q_\theta = - r_\xi = \half f^{-1} \partial_x f~, \nonumber \\ 
p_\chi &= \half a \; , \quad\quad q_\xi = \kappa^2 ( a+ f^{-1} \partial_x f ) \; , \quad\quad r_\theta = \frac{1}{4\kappa^2} ( a - f^{-1} \partial_x f )~.
\end{align}
Comparison with \eqref{eq:TimelikeCKVcondition3d} confirms that $\Xi$ is indeed a Killing vector. Moreover, from \eqref{eq:RhoXi}, $\rho_\Xi = -2\kappa p_\chi = -\kappa a$. Whence, none of the three class IV geometries is static.

Substituting \eqref{eq:ClassIVNullTriadOneForms} and \eqref{eq:ClassIVNablaTriad} into \eqref{eq:3dA} yields
\begin{equation}\label{eq:ClassIVA}
A = -\half a ( \rd t + \omega \rd y ) + \half \partial_x f \rd y~.
\end{equation}
Whence, $F = \rd A = \half ( \partial_x^2 f - a^2 f ) \rd x \wedge \rd y$. Furthermore, substituting \eqref{eq:ClassIVNullTriadOneForms} and \eqref{eq:ClassIVNablaTriad} into \eqref{eq:TimelikeNablaTriadSpinors}, and using \eqref{eq:ClassIVA}, implies the gauged connection in \eqref{eq:GaugedD3d} is given by  
\begin{equation}\label{eq:ClassIVD}
D_\mu = \partial_\mu - \tfrac{1}{4} a {\bm \Gamma}_\mu~.
\end{equation}
The identification $\Xi = \xi_\epsilon + \xi_{\epsilon^\prime} = -2\kappa \partial_t$ and $\kappa = {\overline \epsilon} \epsilon^\prime$ implies $\epsilon^\prime = {\bm \Gamma}_t \epsilon$. Moreover, $\epsilon_\CC = \epsilon + i {\bm \Gamma}_t \epsilon$ is a charged twistor spinor with respect to \eqref{eq:ClassIVD} only if $\epsilon$ is constant.

For each of the three Lorentzian geometries of class IV defined by Table~\ref{tab:ClassIVData}, there are four Killing vectors $( \partial_t , \partial_y , k , l) \in \fK(M,g)$, with $\partial_t \in Z ( \fK(M,g) )$. In each case, using \eqref{eq:ClassIVA}, it is straightforward to check that the condition $\cL_X F =0$ from \eqref{eq:iXFAbelian} is satisfied identically, for all $X \in \fK(M,g)$. Consequently, the locally equivalent condition $\iota_X F = \rd \rho_X$ from \eqref{eq:iXF} defines each function $\rho_X$ in \eqref{eq:BFGauged}, up to the addition of an arbitrary constant.
\footnote{Any such constant term in \eqref{eq:BFGauged} can be set to zero in the $[\cB,\cF]$ bracket for the conformal symmetry superalgebra $\cS$ via an appropriate compensating constant R-symmetry contribution.}
We have already noted that $\rho_\Xi = -\kappa a$ while, up to an arbitrary constant, $\rho_{\partial_y} = \half ( a \omega - \partial_x f )$. For the remaining Killing vectors $k$ and $l$, the non-constant parts of $\rho_k$ and $\rho_l$ are displayed in Table~\ref{tab:ClassIVDataRhoKL}.
\begin{table}
\begin{center}
\begin{tabular}{|c|c|c|c|}
  \hline
  Class & $\rho_k$ & $\rho_l$ & $\cS$ \\ [.05cm]
  \hline\hline
  &&& \\ [-.4cm]
  IV.4 & $-\half y$ & $-\tfrac{1}{4} ( x^2 + y^2 )$ & $\cS^\circ ( \fc$[X]$ |\cR )$ \\ [.05cm]
  &&& \\ [-.4cm]
  IV.5 & $\half (1- a^2) y {\mathrm{e}}^x$ & $\tfrac{1}{4} (1- a^2) ( y^2 {\mathrm{e}}^x + {\mathrm{e}}^{-x} )$ & $\cS^\circ ( \RR | \cR ) \oplus \fb$[VIII] \\ [.05cm]
  &&& \\ [-.4cm]
  IV.6 & $-\half ( 1+a^2 ) \sin x \sin y$ & $-\half ( 1+a^2 ) \sin x \cos y$ & $\cS^\circ ( \RR | \cR ) \oplus \fb$[IX] \\ [.15cm]
  \hline  
\end{tabular} \vspace*{.2cm}
\caption{Conformal superalgebra data for geometries with ${\mbox{dim}}\, \fC(M,[g]) =4$.}
\label{tab:ClassIVDataRhoKL}
\end{center}
\end{table}
Using this data, it is straightforward to compute the associated submaximal conformal symmetry superalgebras $\cS$, which are displayed in the rightmost column of Table~\ref{tab:ClassIVDataRhoKL}, in terms of the notation introduced at the end of Appendix~\ref{sec:AppendixF} (data for the three Lie superalgebras of interest appears in Table~\ref{tab:41LSACentered}). In each case, $\cR \cong \fu(1)$ and we can take $[ \fK(M,g), \cF ]=0$ in $\cS$. For some non-zero $R\in \cR$ and all $\epsilon_\CC \in \cF$, we can take $[R,\epsilon_\CC] = i \epsilon_\CC$ and $[\epsilon_\CC , \epsilon_\CC^* ] = \Xi$.  

\section{$d=4$}
\label{sec:dequals4}

\subsection{Null tetrads}
\label{sec:NullTetrad4d}

Let $(\xi , \theta , \chi )$ denote a {\emph{null tetrad}} of vector fields on $(M,g)$, subject to the defining relations
\begin{equation}\label{eq:NullTetradRelations}
\rn{\xi}^2 = \rn{\theta}^2 = 0 = g(\xi,\chi) = g(\theta,\chi) = g(\chi,\chi)  \; , \quad\quad g(\xi,\theta) = g(\chi,\chi^*) =1~.
\end{equation}
The elements $\xi$ and $\theta$ are real while $\chi$ is complex. The relations \eqref{eq:NullTetradRelations} are preserved under the following transformations (which collectively generate ${\mathrm{O}}(3,1)$):
\begin{itemize}
\item $(\xi,\theta,\chi) \mapsto (\theta,\xi,\chi)$.
\item $(\xi,\theta,\chi) \mapsto (\xi, \theta - {\alpha^*} \chi - \alpha \chi^* - |\alpha|^2 \xi , \chi + \alpha \xi )$, for any $\alpha \in \CC$. 
\item $(\xi,\theta,\chi) \mapsto (\xi, \theta , \te^{i\beta} \chi )$, for any $\beta \in \RR$. 
\item $(\xi,\theta,\chi) \mapsto (\gamma \xi, \gamma^{-1} \theta , \chi )$, for any $\gamma \in \RR \backslash \{ 0\}$. 
\end{itemize}

It is convenient to express the metric and orientation tensor on $M$ as
\begin{equation}\label{eq:NullTetradMetric}
g = \xi \otimes \theta + \theta \otimes \xi + \chi \otimes {\chi^*} + {\chi^*} \otimes \chi \; , \quad\quad \varepsilon = i\;\! \xi \wedge \theta \wedge \chi \wedge \chi^*~,
\end{equation}
in terms of the null tetrad.

\subsection{Majorana spinors}
\label{sec:MajoranaSpinors4d}

The Clifford algebra $\Cl (3,1) \cong \Mat_4 ( \RR )$ has a unique irreducible Majorana spinor representation that is isomorphic to $\RR^4$. On the other hand, its complexification (the Dirac spinor representation) decomposes into a pair of inequivalent irreducible chiral spinor representations, each isomorphic to $\CC^2$, associated with the two eigenspaces of ${\bm \Gamma}$ on which ${\bm \Gamma} = \pm {\bf 1}$. The action of a subalgebra $\Mat_2 (\CC) < \Mat_4 ( \RR )$ on $\CC^2$ which commutes with the complex structure $i{\bm \Gamma}$ defines the action of $\Cl (3,1)$ on each chiral projection (the two chiral projections transform in complex conjugate representations).

Relative to the Clifford algebra basis \eqref{eq:ClBasis} in Appendix~\ref{sec:CoordinateBasisConventions}, taking ${\bm \Gamma} = \frac{i}{4!} \varepsilon^{\mu\nu\rho\sigma} {\bm \Gamma}_{\mu\nu\rho\sigma}$, it follows that
\begin{equation}\label{eq:4dgammadual}
{\bm \Gamma}_{\mu\nu} = -\tfrac{i}{2} \varepsilon_{\mu\nu\rho\sigma} {\bm \Gamma}^{\rho\sigma} {\bm \Gamma} \; , \quad\quad {\bm \Gamma}_{\mu\nu\rho} = -i \, \varepsilon_{\mu\nu\rho\sigma} {\bm \Gamma}^{\sigma} {\bm \Gamma} \; , \quad\quad {\bm \Gamma}_{\mu\nu\rho\sigma}  = i \, \varepsilon_{\mu\nu\rho\sigma} {\bm \Gamma}~.
\end{equation}

There exists on $\fS (M)$ a skewsymmetric bilinear form \eqref{eq:SpinorInnerProd}, with respect to which
\begin{equation}\label{eq:4dbilnears}
{\overline \psi}_\pm \varphi_\pm  = - {\overline \varphi}_\pm \psi_\pm \; , \quad {\overline \psi}_\pm {\bm \Gamma}_{\mu} \varphi_\mp  = {\overline \varphi}_\mp {\bm \Gamma}_\mu \psi_\pm  \; , \quad {\overline \psi}_\pm {\bm \Gamma}_{\mu\nu} \varphi_\pm  = {\overline \varphi}_\pm {\bm \Gamma}_{\mu\nu} \psi_\pm = \mp \tfrac{i}{2} \varepsilon_{\mu\nu\rho\sigma} \, {\overline \varphi_\pm} {\bm \Gamma}^{\rho\sigma} \psi_\pm~,
\end{equation} 
for all $\psi_\pm , \varphi_\pm \in \fS_\pm ( M )$. The associated Fierz identities are given by
\begin{align}\label{eq:4dfierz2}
\psi_\pm \, {\overline \varphi_\pm} &= \tfrac{1}{2} \! \left( ( {\overline \varphi_\pm} \psi_\pm ) -\tfrac{1}{4}  ( {\overline \varphi_\pm} {\bm \Gamma}^{\mu\nu} \psi_\pm ) {\bm \Gamma}_{\mu\nu} \right) {\bf{P}}_\pm \nonumber \\
\psi_\pm \, {\overline \varphi_\mp} &= \half ( {\overline \varphi_\mp} {\bm \Gamma}^\mu \psi_\pm ) {\bm \Gamma}_\mu {\bf{P}}_\mp~.
\end{align} 

In terms of a unitary basis for $\Cl (3,1)$, it follows that under complex conjugation
\begin{equation}\label{eq:4dbilnearscc}
({\overline \psi}_+ \varphi_+ )^* = {\overline \psi}_- \varphi_- \; , \quad ( {\overline \psi}_+ {\bm \Gamma}_{\mu} \varphi_- )^* = {\overline \psi}_- {\bm \Gamma}_{\mu} \varphi_+  \; , \quad ({\overline \psi}_+ {\bm \Gamma}_{\mu\nu} \varphi_+ )^* = {\overline \psi}_- {\bm \Gamma}_{\mu\nu}  \varphi_-~.
\end{equation} 

For a given $\epsilon \in \fS(M)$, we define 
\begin{equation}\label{eq:4dbilinearsEpsilon}
\xi_\epsilon^\mu = {\overline \epsilon} \Gamma^\mu \epsilon = 2 \, {\overline \epsilon}_- \Gamma^\mu \epsilon_+ \; , \quad\quad \zeta_{\mu\nu} = {\overline \epsilon}_+ \Gamma_{\mu\nu} \epsilon_+~.
\end{equation}
From \eqref{eq:4dbilnearscc}, it follows that the vector field $\xi_\epsilon$ is real while the two-form $\zeta$ is complex and obeys $\zeta_{\mu\nu} = -\tfrac{i}{2} \varepsilon_{\mu\nu\rho\sigma} \zeta^{\rho\sigma}$. The vector field $\xi_\epsilon$ is nowhere vanishing only if $\epsilon$ is nowhere vanishing, which we shall assume henceforth. Furthermore, using \eqref{eq:4dfierz2}, one obtains
\begin{equation}\label{eq:4dbilinearsEpsilonIdentities}
\rn{\xi_\epsilon}^2 = 0 \; , \quad\quad \xi_\epsilon^\mu \zeta_{\mu\nu} =0 \; , \quad\quad  \zeta_{\mu\rho} \zeta^{\nu\rho} =0 \; , \quad\quad \zeta^*_{\mu\rho} \zeta^{\nu\rho} =\half g_{\mu\rho} \xi_\epsilon^\rho \xi_\epsilon^\nu~.
\end{equation}
It is convenient to identify \eqref{eq:4dbilinearsEpsilon} in terms of the null tetrad introduced in section~\ref{sec:NullTetrad4d}, such that $\xi_\epsilon = \xi$. The second identity in \eqref{eq:4dbilinearsEpsilonIdentities} then implies $\xi \wedge \zeta =0$, whence $\zeta = \xi \wedge \tau$, in terms of the complex one-form $\tau = \iota_\theta \zeta$. The remaining identities in \eqref{eq:4dbilinearsEpsilonIdentities} fix $\tau = \tfrac{1}{\sqrt{2}} \chi^*$, i.e. $\zeta = \tfrac{1}{\sqrt{2}} \xi \wedge \chi^*$ in terms of the null tetrad one-forms.

Using \eqref{eq:4dfierz2}, it is possible to express any $\psi \in \fS (M)$ in terms of the null tetrad and the reference spinor $\epsilon$ that defines $\xi$. In particular,  
\begin{align}\label{eq:4dfierz3}
\psi_+ &= \alpha \epsilon_+ + \beta {\bm \theta} \epsilon_- \nonumber \\
\psi_- &= \alpha^* \epsilon_- + \beta^* {\bm \theta} \epsilon_+~,
\end{align} 
where $\alpha = 2 {\overline \epsilon_-} {\bm \theta} \psi_+ $ and $\beta = 2 {\overline \epsilon_+} \psi_+$. Moreover, it is easily verified that ${\bm \chi} \epsilon_- =0$ and ${\bm \chi^*} \epsilon_- = \sqrt{2} \epsilon_+$.

\subsection{Petrov types}
\label{sec:PetrovTypes}

The Weyl tensor $W$ of $g$ may also be expressed in terms of the null tetrad and its non-trivial components are characterised by five complex functions 
\begin{align}\label{eq:WeylScalars}
\Psi_0 &= W(\xi,\chi,\xi,\chi) \; , \quad \Psi_1 = W(\xi,\theta,\xi,\chi) \; , \quad \Psi_2 = W(\xi,\chi,\theta,\chi^*) \; , \nonumber \\
\Psi_3 &= W(\xi,\theta,\theta,\chi^*) \; , \quad \Psi_4 = W(\theta,\chi^*,\theta,\chi^*)~,
\end{align}
called {\emph{Weyl scalars}}. The conformal class of $g$ at each point in $M$ may be classified as being one of the following six {\emph{Petrov types}}
\begin{itemize}
\item {\emph{Type I.}} $\Psi_0 =0$.
\item {\emph{Type II.}} $\Psi_0 = \Psi_1 = 0$.
\item {\emph{Type D.}} $\Psi_0 = \Psi_1 = \Psi_3 = \Psi_4 = 0$. 
\item {\emph{Type III.}} $\Psi_0 = \Psi_1 = \Psi_2 = 0$. 
\item {\emph{Type N.}} $\Psi_0 = \Psi_1 = \Psi_2 = \Psi_3 = 0$. 
\item {\emph{Type O.}} $\Psi_0 = \Psi_1 = \Psi_2 = \Psi_3 = \Psi_4 = 0$. 
\end{itemize}
We shall only be concerned with geometries which have the same Petrov type at each point, and thus refer to the Petrov type of $(M,g)$. If $(M,g)$ is type I then it is called {\emph{algebraically general}}, otherwise it is called {\emph{algebraically special}}. If $(M,g)$ is type O then it is locally conformally flat (i.e. $W=0$).

\subsection{Null conformal Killing vectors}
\label{sec:NullCKV4d}

The relations \eqref{eq:NullTetradRelations} constrain the action of the Levi-Civit\`{a} connection $\nabla$ on the null tetrad one-forms such that
\begin{align}\label{eq:NablaTetrad}
\nabla_\mu \xi_\nu &= 2\Re (p)_\mu \xi_\nu - q_\mu \chi_\nu - q^*_\mu \chi^*_\nu \nonumber \\ 
\nabla_\mu \theta_\nu &= -2 \Re (p)_\mu \theta_\nu + r_\mu \chi_\nu + r^*_\mu  \chi^*_\nu \\
\nabla_\mu \chi_\nu &= - r^*_\mu \xi_\nu + q^*_\mu \theta_\nu - 2i \Im (p)_\mu \chi_\nu~, \nonumber
\end{align}
in terms of data that it is convenient to assemble into three complex one-forms $(p,q,r)$ on $M$.

Given any null vector field $X$ on $(M,g)$, it is always possible to define a null tetrad with respect to which $X=\xi$. From the first line in \eqref{eq:NablaTetrad}, it follows that $\xi$ is a conformal Killing vector only if
\begin{equation}\label{eq:CKVcondition}
\Re(p)_\theta = 0 \; , \quad q_\xi =0 = q_{\chi^*} \; , \quad 2 \Re(p)_{\chi^*} = q_\theta \; , \quad 2 \Re(p)_\xi + q_\chi + q^*_{\chi^*} =0~. 
\end{equation}
Furthermore, $\xi$ is a Killing vector only if \eqref{eq:CKVcondition} are satisfied with $\Re(p)_\xi =0$ (whence, $q_\chi$ must be pure imaginary). In that case, 
\begin{equation}\label{eq:dKV}
\rd \xi = 2( q_\theta \chi + q^*_\theta \chi^*) \wedge \xi + 2 q_\chi \, \chi \wedge \chi^*~. 
\end{equation}

Any Killing vector $X$ on $(M,g)$ obeys 
\begin{equation}\label{eq:nablasquaredKV}
\nabla_\mu \nabla_\nu X_\rho = - R_{\nu\rho\mu\sigma} X^\sigma~, 
\end{equation}
in terms of the components $R_{\mu\nu\rho\sigma}$ of the Riemann tensor of $g$. For a null Killing vector $X = \xi$, \eqref{eq:nablasquaredKV} and \eqref{eq:dKV} imply
\begin{equation}\label{eq:nullKVRicci}
R_{\mu\nu} \xi^\mu \xi^\nu = \nabla_\mu \xi_\nu \nabla^\mu \xi^\nu = 2 | q_\chi |^2 \geq 0~, 
\end{equation}
where $R_{\mu\nu} = g^{\rho\sigma} R_{\mu\rho\nu\sigma}$ is the Ricci tensor. The geometry $(M,g)$ is {\emph{Einstein}} if it obeys $R_{\mu\nu} = \Lambda g_{\mu\nu}$, for some $\Lambda \in \RR$ (if $\Lambda =0$ then $(M,g)$ is {\emph{Ricci-flat}}). Whence, if $(M,g)$ is Einstein and admits a null Killing vector $\xi$ then $q_\chi =0$ and $\rd \xi = k \wedge \xi$ (where $k = 2( q_\theta \chi + q^*_\theta \chi^*)$), which implies $\xi \wedge \rd \xi =0$. Indeed $\xi \wedge \rd \xi =0$ only if $q_\chi =0$ so any $(M,g)$ with a null Killing vector $\xi$ for which $\xi \wedge \rd \xi \neq 0$ cannot be Einstein.

The three-form $\xi \wedge \rd \xi$ is called the {\emph{twist}} of $\xi$ and $\xi$ is said to be {\emph{twisting}} if $\xi \wedge \rd \xi \neq 0$ or {\emph{non-twisting}} if $\xi \wedge \rd \xi =0$. Restricting to the subspace $\xi^\perp = \{ X \in \fK(M,g) \, |\, g(X,\xi)=0 \}$ defines a Lie subalgebra of $\fK(M,g)$ only if $\xi$ is non-twisting, thus ensuring that the associated hypersurface in $(M,g)$ is integrable. Furthermore, if $\xi$ is non-twisting then substituting $\rd \xi = k \wedge \xi$ into \eqref{eq:nablasquaredKV} with $X=\xi$ leads to several more useful properties. Contracting the resulting expression with $\xi$ on different indices and with the inverse metric implies
\begin{equation}\label{eq:nullKVTwistId1}
\nabla_\xi k^\mu = f \xi^\mu \; , \quad R_{\mu\rho\nu\sigma} \xi^\rho \xi^\sigma = \tfrac{1}{4} \rn{k}^2 \xi_\mu \xi_\nu \; , \quad R_{\mu\nu} \xi^\nu = - \half \left( \nabla_\nu k^\nu + \half \rn{k}^2 - f \right) \xi_\mu~, 
\end{equation}
where $f = \theta_\mu \nabla_\xi k^\mu$ and $\rn{k}^2  > 0$ if $k \neq 0$. Substituting \eqref{eq:nullKVTwistId1} into the definition of the Weyl tensor \eqref{eq:WeylTensor} then implies
\begin{equation}\label{eq:nullKVTwistId2}
W_{\mu\nu\rho\sigma} \xi^\rho \xi^\sigma = -\half \left( \tfrac{1}{3} R + \nabla_\rho k^\rho - f \right) \xi_\mu \xi_\nu~. 
\end{equation}
Thus it follows that any $(M,g)$ with a non-twisting null Killing vector has $\Psi_0 = \Psi_1 =0$ and is therefore algebraically special. Moreover, if any such geometry is Ricci-flat with $\Psi_2 =0$ then $k=0$ so $\xi$ must be parallel and $(M,g)$ is necessarily of Petrov type N or O. Whence, any Ricci-flat $(M,g)$ with a null Killing vector must be of Petrov type II, D, N or O. 

Let us conclude this section by illustrating some of the properties above via the introduction of local coordinates $(u,v,x,y)$ on $(M,g)$. If $(M,g)$ admits a nowhere vanishing null Killing vector $\xi$ then we can take $\xi = \partial_v$ tangent to a family of null geodesics, with geodesic distance parameterised by the affine coordinate $v$. A convenient local form of the metric in these adapted coordinates is 
\begin{equation}\label{eq:NullCoordMetric}
g = 2 G ( \rd u + \alpha ) ( \rd v + \beta + \half H ( \rd u + \alpha )) + E^2 ( \rd x^2 + \rd y^2 )~,
\end{equation}   
in terms of three real functions $G$, $H$ and $E$ and two real one-forms $\alpha = \alpha_x \rd x + \alpha_y \rd y$ and $\beta = \beta_x \rd x + \beta_y \rd y$. All of these components are functions only of $(u,x,y)$. The null tetrad one-forms are identified such that 
\begin{equation}\label{eq:NullTetradCoord}
\xi = G ( \rd u + \alpha ) \; , \quad\quad \theta = \rd v + \beta + \half G^{-1} H\, \xi \; , \quad\quad \chi = \tfrac{1}{\sqrt{2}} E ( \rd x + i\rd y )~.
\end{equation}
If $\xi$ is non-twisting then integrability of its associated hypersurface implies $\xi = f \rd u$, for some function $f$ of $(u,x,y)$.  Up to a redefinition of $G$, this allows us to fix $\alpha =0$ in \eqref{eq:NullCoordMetric}. Furthermore, if $\xi$ is parallel, then $\rd \xi =0$ so $f$ must be a function only of $u$ and we can fix $f=1$, $\beta =0$ and $E=1$. Indeed, the existence of a parallel null vector on $(M,g)$ characterises a four-dimensional pp-wave, with local metric as in \eqref{eq:ppwaveGen} of Appendix~\ref{sec:PlaneWaves}. In $d=4$, the pp-wave metric \eqref{eq:ppwaveGen} is of Petrov type N or O. It is type O only if the real function $H$ of coordinates $(u,x,y)$ obeys $\partial_x^2 H = \partial_y^2 H$ and $\partial_x \partial_y H=0$. It is a plane wave only if $H$ is a quadratic function of $(x,y)$. 

Up to local isometry, there exists a classification of \lq physically admissible' Ricci-flat $(M,g)$ with a null Killing vector. Chapter 24.4-5 in \cite{ExactSolutions} contains a detailed summary of the local metrics and their Killing vectors. From table 24.2 in \cite{ExactSolutions}, if $(M,g)$ is a Ricci-flat type N pp-wave then $1\leq {\mbox{dim}} \, \fK(M,g) \leq 6$ (within this class, ${\mbox{dim}} \, \fK(M,g)$ never equals $4$ and equals $5$ or $6$ only for plane waves). The remaining solutions of type II and D are summarised in table 24.1 of \cite{ExactSolutions}. The local metrics in this subclass correspond to \eqref{eq:NullCoordMetric} with $\alpha = \beta =0$, $G =-x$, $E^2 = x^{-1/2}$ and $\partial_x ( x \partial_x H) + x \partial_y^2 H =0$. These geometries all have  $1\leq {\mbox{dim}} \, \fK(M,g) \leq 4$. Contained within this subclass are the type II \lq van Stockum' solutions (for $\partial_u H =0$) and a static type D solution with ${\mbox{dim}} \, \fK(M,g) = 4$ (for $H$ constant). We shall return to compute symmetry superalgebras for these geometries in section~\ref{sec:CKS4d}.

\subsection{Charged twistor spinors}
\label{sec:CTS4d}

Let us now examine the implications of the existence of a charged twistor spinor $\epsilon$, which can be thought of locally as a Majorana spinor obeying \eqref{eq:TSGauged} with respect to the action of the $\cR = \fu(1)$ gauged connection 
\begin{equation}\label{eq:4dD}
D_\mu \epsilon = \nabla_\mu \epsilon + i A_\mu {\bm \Gamma} \epsilon~,
\end{equation}
i.e. $D_\mu \epsilon_\pm = \nabla_\mu \epsilon_\pm \pm i A_\mu \epsilon_\pm$. 

The first consequence is that $\xi_\epsilon$ defined in \eqref{eq:4dbilinearsEpsilon} is a conformal Killing vector. Let us now deconstruct the defining condition $D_\mu \epsilon = \tfrac{1}{4} {\bm \Gamma}_\mu {\bm D} \epsilon$ for charged twistor spinor $\epsilon$ in terms of the null tetrad. Identifying $\xi_\epsilon = \xi$ then using \eqref{eq:4dfierz3} and \eqref{eq:NablaTetrad} implies 
\begin{align}\label{eq:DepsilonTorsion}
\nabla_\mu \epsilon_+ &= p_\mu \epsilon_+ + \tfrac{1}{\sqrt{2}} q_\mu {\bm \theta} \epsilon_- \nonumber \\ 
\nabla_\mu \epsilon_- &= p_\mu^* \epsilon_- + \tfrac{1}{\sqrt{2}} q_\mu^* {\bm \theta} \epsilon_+~.
\end{align}
It follows that the defining condition for charged twistor spinor $\epsilon$ is equivalent to the following conditions on $p$, $q$ and $A$:
\begin{equation}\label{eq:TScondition}
p_\theta + i A_\theta = 0 = p_\chi + i A_\chi \; , \quad q_\xi =0 = q_{\chi^*} \; , \quad\quad p_{\chi^*} + iA_{\chi^*} = q_\theta \; , \quad p_\xi +i A_\xi = - q_\chi~. 
\end{equation}
Since $\xi_\epsilon \in \fC(M,[g])$ for any charged twistor spinor $\epsilon$, it is straightforward to identify a subset of conditions in \eqref{eq:TScondition} with precisely the conditions in \eqref{eq:CKVcondition} required for $\xi$ to be a conformal Killing vector. The conditions in \eqref{eq:CKVcondition} describe eight real constraints on $\Re(p)$ and $q$ which are contained in the twelve real constraints on $p$, $q$ and $A$ in \eqref{eq:TScondition}. The four remaining constraints in \eqref{eq:TScondition} are precisely sufficient to fix all four components of $A$, such that   
\begin{equation}\label{eq:CTSa}
A_\mu = \tfrac{i}{2} ( p_\mu + q_\chi \theta_\mu - q_\theta \chi_\mu - p_\mu^* -  q^*_{\chi^*} \theta_\mu + q^*_\theta \chi^*_\mu )~. 
\end{equation}

Thus we conclude that if $(M,g)$ admits a nowhere vanishing null conformal Killing vector $X$, then there must exist a nowhere vanishing charged twistor spinor $\epsilon$ on $(M,g)$, with $X = \xi_\epsilon$, that is defined with respect to the gauged connection $A$ in \eqref{eq:CTSa}. This characterisation of charged twistor spinors on a smooth orientable Lorentzian four-manifold was first obtained in \cite{Cassani:2012ri} (see also \cite{deMedeiros:2012sb}).

For any charged twistor spinor $\epsilon$, the twist of $\xi_\epsilon = \xi$ can be written 
\begin{equation}\label{eq:twistxiepsilon}
\xi \wedge \rd \xi = \tfrac{4}{3} \rho_{\xi} \, {*\xi}~.
\end{equation}
in terms of the real function
\begin{equation}\label{eq:rhoeps}
\rho_{\xi} = -\tfrac{3i}{4} {\overline \epsilon} {\bm \Gamma} {\bm D} \epsilon = -\tfrac{3i}{4} ( q_\chi - q^*_{\chi^*} )~.
\end{equation}
It follows that
\begin{equation}\label{eq:ixiF4d}
\iota_\xi F = \rd \rho_\xi~,
\end{equation}
using $F = \rd A$ from \eqref{eq:CTSa}. Whence, the condition \eqref{eq:iXF} is satisfied identically for $X = \xi$. Furthermore, as noted below \eqref{eq:CKVcondition}, if $\xi$ is a Killing vector then $q_\chi$ is pure imaginary. In that case, $\xi$ is non-twisting only if $q_\chi =0$ which, from \eqref{eq:rhoeps}, occurs only if $\rho_\xi =0$.  

Differentiating the defining condition for $\epsilon$ gives  
\begin{equation}\label{eq:2DerTwistorSpinorsGauged}
D_\mu {\bm D} \epsilon_\pm  = 2 K_{\mu\nu} {\bm \Gamma}^\nu \epsilon_\pm \pm \tfrac{4i}{3} F_{\mu\nu} {\bm \Gamma}^\nu \epsilon_\pm - \tfrac{1}{3} \varepsilon_{\mu\nu\rho\sigma} F^{\rho\sigma} {\bm \Gamma}^\nu \epsilon_\pm~.
\end{equation}
Combining \eqref{eq:2DerTwistorSpinorsGauged} with \eqref{eq:LCBracket} implies the integrability condition
\begin{equation}\label{eq:TwistorSpinorIntegrabilityGauged}
\tfrac{1}{4} W_{\mu\nu\rho\sigma} {\bm \Gamma}^{\rho\sigma} \epsilon_\pm = \mp \tfrac{i}{3} F^{\rho}{}_{[\mu} {\bm \Gamma}_{\nu ]\rho} \epsilon_\pm \mp \tfrac{i}{3} F_{\mu\nu} \epsilon_\pm - \tfrac{1}{6} \varepsilon_{\mu\nu\rho\sigma} F^{\rho\sigma} \epsilon_\pm~.
\end{equation}
Identifying $\xi_\epsilon = \xi$, in terms of the null tetrad, one finds that \eqref{eq:TwistorSpinorIntegrabilityGauged} is equivalent to the following conditions on the Weyl scalars \eqref{eq:WeylScalars}:
\begin{equation}\label{eq:TwistorSpinorIntegrability2}
\Psi_0 = 0 \; , \quad \Psi_1 = -\tfrac{i}{3} F (\xi,\chi)  \; , \quad \Psi_2 = -\tfrac{i}{3} ( F(\xi,\theta) + F(\chi,\chi^*) ) \; , \quad \Psi_3 = -i F(\theta,\chi^*)~.  
\end{equation}

The Petrov type of $(M,g)$ therefore determines which components of $F$ must vanish identically. For the algebraically special geometries, we have
\begin{itemize}
\item Type II $\;\,\Longleftrightarrow$ $F(\xi,\chi) =0$.
\item Type D $\;\,\Longleftrightarrow$ $F(\xi,\chi) = F(\theta,\chi) = 0$ and $\Psi_4 = 0$. 
\item Type III $\Longleftrightarrow$ $F(\xi,\chi) = 0$, $F(\xi,\theta) = 0 = F(\chi,\chi^*)$. 
\item Type N $\;\,\Longleftrightarrow$ $F=0$ and $W \neq 0$.
\item Type O $\;\,\Longleftrightarrow$ $F=0$ and $W = 0$. 
\end{itemize}
These equivalences were first obtained in \cite{Cassani:2013dba}. Notice that if $F=0$ then, locally, $A = \rd \lambda$ and the charged twistor spinor $\epsilon_\pm$ is equivalent to an ordinary twistor spinor $\te^{\pm i\lambda}\epsilon_\pm$. 

Up to local conformal equivalence, the classification of Lorentzian four-manifolds which admit a nowhere vanishing twistor spinor $\epsilon$ is due to Lewandowski \cite{Lewandowski:1991bx}. Any such $(M,g)$ must be of type N or O. The admissible type N geometries are distinguished by the twist of the conformal Killing vector $\xi_\epsilon$. If the twist of $\xi_\epsilon$ vanishes then $(M,g)$ is in the conformal class of a pp-wave with $\epsilon$ parallel. If the twist of $\xi_\epsilon$ does not vanish then $(M,g)$ is in the conformal class of a Fefferman space \cite{Feff:1976}.

Thus, at least locally, any algebraically special $(M,g)$ which admits a charged twistor spinor $\epsilon$ with $F\neq 0$ is of Petrov type II, D or III with null conformal Killing vector $\xi_\epsilon$. 

\subsection{Conformal symmetry superalgebras}
\label{sec:CKS4d}

Let $(M,g)$ be any smooth orientable Lorentzian four-manifold equipped with a nowhere vanishing null conformal Killing vector $X$ and a $\fC(M,[g])$-invariant closed two-form $F$. As we have explained, $X$ defines a nowhere vanishing charged twistor spinor $\epsilon$ (with $X = \xi_\epsilon$) that is charged with respect to the connection $A$ in \eqref{eq:CTSa} whose curvature is $F = \rd A$. From section~\ref{sec:ConformalSymmetrySuperalgebras}, we recall that this data is precisely what is needed to assign to $(M,[g])$ a conformal symmetry superalgebra $\cS$ with gauged $\cR = \fu(1)$. 

From the discussion in section~\ref{sec:CTS4d}, it follows that if $F=0$ then $(M,g)$ must have Petrov type N (if $W\neq 0$) or type O (if $W=0$).  If $(M,g)$ is of type O (i.e. locally conformally flat) then $\fC(M,[g]) \cong \fso(4,2) \cong \fsu(2,2)$ and $\cS \cong \fsu(2,2|1)$ (i.e. the $\eN =1$, $d=4$ case in Table~\ref{tab:GenericLSAData}). If $(M,g)$ is of type N, with $\cF$ non-trivial, then it is locally conformally equivalent to a pp-wave or a Fefferman space. For a pp-wave, the twistor spinors in $\cF$ are all parallel. In $d=4$, any $(M,g)$ that is not of type O must have ${\mbox{dim}}\, \fC(M,[g]) \leq 7$ and, from theorem 5.1.3 in \cite{KruThe2013}, ${\mbox{dim}}\, \fC(M,[g]) = 7$ requires $(M,g)$ to be of type N. More precisely, it is known \cite{Keane:2013fta} that ${\mbox{dim}}\, \fC(M,[g]) = 7$ only if $(M,g)$ is locally conformally equivalent to a homogeneous plane wave geometry defined by either the first or third entry in Table~\ref{tab:PlaneWaveExtraKV4d} of Appendix~\ref{sec:PlaneWaves}. In both cases, the homogeneous plane wave $(M,g_{\mathrm{pw}})$ has $\fC(M,[g_{\mathrm{pw}}]) = \fH(M,g_{\mathrm{pw}})$ and the null Killing vector $\xi \in Z( \fK(M,g_{\mathrm{pw}}) )$. The associated submaximal conformal symmetry superalgebras $\cS = \cS_{\mathrm{pw}} \oplus \cR$ are described explicitly at the end of Appendix~\ref{sec:PlaneWaves}.  

From the discussion in section~\ref{sec:NullCKV4d}, we recall that if $(M,g)$ is equipped with a null Killing vector $\xi$ (and $F \neq 0$) then it must be of Petrov type II or D with $\xi$ non-twisting in order to be Ricci-flat. Let us now take advantage of the classification in \cite{ExactSolutions} of \lq physically admissible' Ricci-flat geometries which admit a null Killing vector and compute their associated symmetry superalgebras. We shall also compute the conformal symmetry superalgebra for the unique admissible Ricci-flat geometry of Petrov type D. In general, the symmetry superalgebra $\cS_\circ$ and conformal symmetry superalgebra $\cS$ associated with any admissible geometry $(M,g)$ are isomorphic only if $\fK(M,g) = \fC(M,[g])$ and $\cF_\circ = \cF$. It is worth recalling from theorem 3 in \cite{DC:1975} that any $(M,g)$ not of Petrov type N or O has $\fC(M,[g])$ conformally isometric. Of course, if $(M,g)$ is Ricci-flat with $\fC(M,[g]) \neq \fK(M,g)$, the particular geometry $(M,{\tilde g})$ for which $\fC(M,[g]) = \fK(M,{\tilde g})$ need not be Ricci-flat. 

In terms of the local coordinates introduced at the end of section~\ref{sec:NullCKV4d}, each geometry $(M,g)$ within this class has a null Killing vector $\xi = \partial_v$ with respect to a metric $g$ of the form \eqref{eq:NullCoordMetric} with $\alpha = \beta =0$, $G = -x$ and $E = x^{-1/4}$. The only non-trivial component of the Ricci tensor of $g$ is $R_{uu}$, which vanishes only if $\partial_x ( x \partial_x H) + x \partial_y^2 H =0$. By explicit calculation of the Weyl tensor of $g$, one finds that the Weyl scalars $\Psi_0 = \Psi_1 = \Psi_3 =0$ while $\Psi_2 = \tfrac{1}{8} x^{-3/2}$. The remaining Weyl scalar $\Psi_4$ vanishes only if 
\begin{equation}\label{eq:Psi4}
\partial_x ( x^{3/2} \partial_x H ) = x^{3/2} \partial_y^2 H \; , \quad\quad ( x\partial_x +\tfrac{3}{4} ) \partial_y H =0~.
\end{equation}
Thus $(M,g)$ is indeed generically of type II, and is of type D only if $H$ obeys \eqref{eq:Psi4}. If $(M,g)$ is of type D and Ricci-flat then $H$ is necessarily constant. 

Identification of the null tetrad one-forms as in \eqref{eq:NullTetradCoord} implies that the volume form \eqref{eq:NullTetradMetric} is $\varepsilon = x^{1/2}\, \rd u \wedge \rd v \wedge \rd x \wedge \rd y$. The action of $\nabla$ on \eqref{eq:NullTetradCoord} can be defined via the the triple of complex one-forms $(p,q,r)$ in \eqref{eq:NablaTetrad}, which are given by
\begin{equation}\label{eq:PQRNablaTetrad}
p = \tfrac{1}{4} x^{-1} ( \rd x - \tfrac{i}{2} \rd y ) \; , \quad q = -\tfrac{1}{2\sqrt{2}} x^{1/4} \rd u \; , \quad r = -\tfrac{1}{2\sqrt{2}} x^{-3/4} ( \rd v + ( \half  H + x ( \partial_x - i \partial_y ) H ) \rd u )~. 
\end{equation}
Comparison with \eqref{eq:CKVcondition} and \eqref{eq:rhoeps} confirms that $\xi$ is indeed a non-twisting Killing vector with $\rho_\xi =0$. 

Substituting \eqref{eq:NullTetradCoord} and \eqref{eq:PQRNablaTetrad} into \eqref{eq:CTSa} yields
\begin{equation}\label{eq:RicciFlatA}
A = \tfrac{3}{8} x^{-1} \rd y~.
\end{equation}
Whence, $F = -\tfrac{3}{8} x^{-2}  \rd x {\wedge} \rd y$. Furthermore, substituting \eqref{eq:NullTetradCoord} and \eqref{eq:PQRNablaTetrad} into \eqref{eq:DepsilonTorsion}, and using \eqref{eq:RicciFlatA}, implies the gauged connection in \eqref{eq:4dD} is given by 
\begin{equation}\label{eq:RicciFlatD}
D_\mu = \partial_\mu + \tfrac{1}{4} x^{-1/2} {\bm \Gamma}_\mu {\bm \Gamma}_x~.
\end{equation}
The identification $\xi = \xi_\epsilon = \partial_v$ implies ${\bm \Gamma}_v \epsilon_\pm =0$ while, from \eqref{eq:4dgammadual}, $i {\bm \Gamma}_{xy} \epsilon_\pm = \pm x^{-1/2} \epsilon_\pm$. Consequently, using ${\bm \chi} \epsilon_+ = \sqrt{2} \epsilon_-$ and ${\bm \chi}^* \epsilon_+ =0$, it follows that ${\bm \Gamma}_x \epsilon_\pm = x^{-1/4} \epsilon_\mp$. Moreover, $\epsilon$ is a charged twistor spinor with respect to \eqref{eq:RicciFlatD} only if $\epsilon$ is constant.   

All the non-generic cases, where $\xi = \partial_v$ is not the only Killing vector in $\fK(M,g)$, are summarised in table 24.1 of \cite{ExactSolutions}. The pertinent data is displayed in Table~\ref{tab:VacuumIID} below (the parameters $a,b \in \RR \backslash \{ 0\}$ and real functions $f,h$ are constrained such that $\partial_x ( x \partial_x H) + x \partial_y^2 H =0$). Every $X \in \fK(M,g)$ is of the form 
\begin{equation}\label{eq:RicciFlatKV}
X = \alpha ( u \partial_u - v \partial_v ) + \beta \partial_u + \gamma \partial_y + h(u) \partial_v~,
\end{equation}
for some choice of real numbers $\alpha$, $\beta$ and $\gamma$ and a real function $h$ of $u$. Thus, from \eqref{eq:RicciFlatA}, it follows that $\iota_X F = \tfrac{3}{8} \gamma x^{-2}  \rd x$. Whence, the condition \eqref{eq:iXF} is indeed satisfied for any $X \in \fK(M,g)$ of the form \eqref{eq:RicciFlatKV}, with $\rho_X = - \tfrac{3}{8} \gamma x^{-1}$, up to the addition of an arbitrary constant.   
\begin{table}
\begin{center}
\begin{tabular}{|c|c|c|c|c|}
  \hline
  Type & $H(u,x,y)$ & $\{ X \}$ & $\fK(M,g)$ & $\cS_\circ$ \\ 
  \hline\hline
  &&&& \\ [-.4cm]
  II &$f(u,x)$ & $\{ \partial_y , \partial_v \}$ & $\RR^2$ & $\RR \oplus \cS^\circ (\RR | \cR)$  \\ [.05cm]
  &&&& \\ [-.4cm]
  II & $f(ay - bu,x)$ & $\{ b \partial_y + a \partial_u , \partial_v \}$ & $\RR^2$ & $\RR \oplus \cS^\circ (\RR | \cR)$  \\ [.05cm]
  &&&& \\ [-.4cm]
  II & $f(u,x) - y \partial_u h(u)$ & $\{ \partial_y + h(u) \partial_v , \partial_v  \}$ & $\RR^2$ & $\RR \oplus \cS^\circ (\RR | \cR)$  \\ [.05cm]
  &&&& \\ [-.4cm]
  II & $u^{-2} f(y - a\! \ln (u) ,x)$ & $\{ a \partial_y + u \partial_u - v \partial_v , \partial_v \}$ & $\fa$ & $\cS^\lhd ( \fa | \cR)$ \\ [.15cm]
  \hline
  &&&& \\ [-.4cm]
  II & $f(x) \mathrm{e}^{-2ay}$ & $\{ \partial_y + a( u \partial_u - v \partial_v ) , \partial_u , \partial_v \}$ & $\fb$[VI$_0$] & $\cS^\lhd (\fb$[VI$_0$]$ | \cR)$ \\ [.05cm] 
  &&&& \\ [-.4cm]
  II & $f(x) +ay$ & $\{ \partial_y - au \partial_v , \partial_u , \partial_v \}$ & $\fb$[II] & $\cS^\circ (\fb$[II]$ | \cR)$ \\ [.05cm] 
  &&&& \\ [-.4cm]
  II & $f(x)$ & $\{ \partial_y , \partial_u , \partial_v \}$ & $\RR^3$ & $\RR^2 \oplus \cS^\circ (\RR | \cR)$ \\ [.15cm]
  \hline
  &&&& \\ [-.4cm]
  D & $0$ & $\{ \partial_y , u \partial_u - v \partial_v , \partial_u , \partial_v \}$ & $\RR \oplus \fb$[VI$_0$]  & $\RR \oplus \cS^\lhd (\fb$[VI$_0$]$ | \cR)$ \\ [.15cm]
  \hline  
\end{tabular} \vspace*{.2cm}
\caption{Data for Ricci-flat geometries with a null Killing vector and $F \neq 0$.}
\label{tab:VacuumIID}
\end{center}
\end{table}

Using this data, it is straightforward to compute the associated symmetry superalgebras $\cS_\circ$ which are displayed in the rightmost column of Table~\ref{tab:VacuumIID}, in terms of the notation introduced at the end of Appendix~\ref{sec:AppendixF}. In each case, $\cR = \fu(1)$ and the even-odd bracket in \eqref{eq:BFGauged} is of the form $[X , \epsilon ] = \half \alpha \epsilon$, for all $\epsilon \in \cF_\circ$ and $X \in \fK(M,g)$ of the form \eqref{eq:RicciFlatKV}. For some non-zero $R\in \cR$ and all $\epsilon \in \cF_\circ$, we can take $[R,\epsilon] = i {\bm \Gamma} \epsilon$ and $[\epsilon , \epsilon ] = c_\epsilon \xi$, for some non-zero real number $c_\epsilon$. 

The type D geometry $(M,g)$ from Table~\ref{tab:VacuumIID} has $\fC(M,[g]) = \fH(M,g)$. If $\vartheta \in \fH(M,g) / \fK(M,g)$ is normalised such that $[\vartheta,\xi]=\xi$ then
\begin{equation}\label{eq:HomothetyD}
\vartheta = - v \partial_v - 2x \partial_x - 2y \partial_y~.
\end{equation}
The five-dimensional Lie algebra $\fH(M,g)$ is thus defined by $\fK(M,g)$ together with the following non-trivial brackets 
\begin{equation}\label{eq:DBB}
[ \vartheta , \partial_v ]= \partial_v \; , \quad\quad [ \vartheta , \partial_y ]= 2 \partial_y~.
\end{equation}
From \eqref{eq:HomothetyD} and \eqref{eq:RicciFlatA}, it follows that $\vartheta$ obeys condition \eqref{eq:iXF} with $\rho_\vartheta = \tfrac{3}{4} x^{-1} y$. In addition to the $[\fK(M,g),\cF_\circ]$ brackets given above, \eqref{eq:BFGauged} also prescribes
\begin{equation}\label{eq:DBF}
[ \vartheta , \epsilon ]= \half \epsilon~,
\end{equation}
for all $\epsilon \in \cF_\circ = \cF$. Whence, the symmetry superalgebra $\cS_\circ$ together with \eqref{eq:DBB} and \eqref{eq:DBF} define the conformal symmetry superalgebra $\cS \cong \cS^\lhd ( \fH(M,g) | \cR)$ (in the notation of Appendix~\ref{sec:AppendixF}) for the conformal class of the unique Ricci-flat type D geometry in Table~\ref{tab:VacuumIID}. 


\appendix

\section{Coordinate basis conventions}
\label{sec:CoordinateBasisConventions}
Let $\{ \partial_\mu \, |\, \mu = 0,1,..., d-1 \}$ denote a local coordinate basis on $\fX ( M )$, where $d= {\mbox{dim}}\, M$. The volume form on $(M,g)$ is given by $\varepsilon = \pm \sqrt{|g|} \rd x^{0} \wedge \rd x^{1} \wedge ... \wedge \rd x^{d-1}$, in terms of the dual basis $\{ \rd x^\mu \, |\, \mu = 0,1,..., d-1 \}$ of differential forms on $M$. 

With respect to this basis,  the action of the Levi-Civit\`{a} connection is defined by $\nabla_\mu \partial_\nu = \Gamma_{\mu\nu}^\rho \partial_\rho$ in terms of the Christoffel symbols 
\begin{equation}\label{eq:ChristoffelSymbols}
\Gamma_{\mu\nu}^\rho = \half g^{\rho\sigma} ( \partial_\mu g_{\nu\sigma} + \partial_\nu g_{\mu\sigma} - \partial_\sigma g_{\mu\nu} )~.
\end{equation}
Components of the Riemann tensor are given by 
\begin{equation}\label{eq:RiemannTensor}
R^{\rho}{}_{\sigma\mu\nu} = \partial_\mu \Gamma_{\nu\sigma}^\rho - \partial_\nu \Gamma_{\mu\sigma}^\rho + \Gamma_{\mu\alpha}^\rho \Gamma_{\nu\sigma}^\alpha - \Gamma_{\nu\alpha}^\rho \Gamma_{\mu\sigma}^\alpha~,
\end{equation}
and let $R_{\rho\sigma\mu\nu} = g_{\rho\alpha} R^{\alpha}{}_{\sigma\mu\nu}$. The Ricci tensor has components $R_{\mu\nu} = R^{\rho}{}_{\mu\rho\nu}$ and the scalar curvature is $R= g^{\mu\nu} R_{\mu\nu}$. The Schouten tensor $K$ has components 
\begin{equation}\label{eq:SchoutenTensor}
K_{\mu\nu} = \tfrac{1}{d-2} \left( -R_{\mu\nu} + \tfrac{1}{2(d-1)} g_{\mu\nu} R \right)~.
\end{equation}
The Weyl tensor $W$ has components 
\begin{equation}\label{eq:WeylTensor}
W_{\mu\nu\rho\sigma} = R_{\mu\nu\rho\sigma} + g_{\mu\rho} K_{\nu\sigma} - g_{\nu\rho} K_{\mu\sigma} - g_{\mu\sigma} K_{\nu\rho} + g_{\nu\sigma} K_{\mu\rho}~,
\end{equation}
and we define $\rn{W}^2 = W_{\mu\nu\rho\sigma} W^{\mu\nu\rho\sigma}$.
The Cotton-York tensor $C$ has components 
\begin{equation}\label{eq:CottonTensor}
C_{\mu\nu\rho} = \nabla_\mu K_{\nu\rho} - \nabla_\nu K_{\mu\rho}~,
\end{equation}
and we define $\rn{C}^2 = C_{\mu\nu\rho} C^{\mu\nu\rho}$.

Let $\{ {\bm \Gamma}_{\mu_1 ... \mu_k} | k=0,1,...,d  \}$ denote a basis for sections of the Clifford bundle $\Cl (TM)$, such that
\begin{equation}\label{eq:ClBasis}
{\bm \Gamma}_{\mu_1 ... \mu_k} = {\bm \Gamma}_{[ \mu_1} ... {\bm \Gamma}_{\mu_k ]} \equiv \frac{1}{k!} \sum_{\sigma \in S_k} (-1)^{|\sigma|} {\bm \Gamma}_{\mu_{\sigma(1)}} ... {\bm \Gamma}_{\mu_{\sigma(k)}}~,
\end{equation}
for degree $k>0$ (i.e. unit weight skewsymmetrisation of $k$ distinct degree one basis elements) and the identity element ${\bf 1}$ for $k=0$. 

Let $\{ e_\mu^\alpha \}$ denote the components of a pseudo-orthonormal frame on $(M,g)$. By definition, $g_{\mu\nu} = e_\mu^\alpha e_\nu^\beta \eta_{\alpha\beta}$, in terms of the canonical Minkowskian metric $\eta$ on $\RR^{d-1,1}$. Components $\{ \omega_\mu^{\alpha\beta} \}$ of the associated spin connection are defined by the \lq no torsion' condition 
\begin{equation}\label{eq:NoTorsion}
\rd e^\alpha + \omega^\alpha{}_\beta \wedge e^\beta =0~.
\end{equation}
The action of $\nabla$ on any $\psi \in \fS(M)$ is defined by
\begin{equation}\label{eq:SpinConnection}
\nabla_\mu \psi = \partial_\mu \psi + \tfrac{1}{4} \omega_\mu^{\alpha\beta} {\bm \Gamma}_{\alpha\beta} \psi~.
\end{equation} 

\section{Conformal Killing vectors of pp-waves in $d=3$}
\label{sec:3dppwaves}

The general form of the three-dimensional pp-wave metric is
\begin{equation}\label{eq:ppwave}
g_{\mathrm{pp}} = 2 \rd u \rd v + H(u,x) {\rd u}^2 + {\rd x}^2~,
\end{equation}
in terms of Brinkmann coordinates $(u,v,x)$ on $M$, where $H$ is an arbitrary real function of $(u,x)$. In these coordinates, $\xi = \partial_v$ is a null Killing vector of $(M,g_{\mathrm{pp}})$. It is often convenient to adopt the notation ${}^\prime = \partial_u$.

Given a positive integrable real function $\Omega (u)$, it is useful to note that redefining
\begin{equation}\label{eq:ppwaveRedefinition} 
u \mapsto \int \! \rd u \, \Omega^2(u) \; , \quad v \mapsto v - \half x^2 \Omega^{-1} \partial_u \Omega \; , \quad x \mapsto \Omega x \; , \quad H \mapsto \Omega^{-2} ( H - x^2 \Omega \partial_u^2 \Omega^{-1}  )~, 
\end{equation}
induces the Weyl transformation $g_{\mathrm{pp}} \mapsto \Omega^2 g_{\mathrm{pp}}$ of the pp-wave metric.

The only non-trivial Christoffel symbols of $g_{\mathrm{pp}}$ are
\begin{equation}\label{eq:ppwaveChristoffelSymbols}
\Gamma_{ux}^v = - \Gamma_{uu}^x = \half \partial_x H \; , \quad\quad \Gamma_{uu}^v = \half H^\prime~.
\end{equation}
The only non-trivial component of the Riemann tensor of $g_{\mathrm{pp}}$ is $R_{uxux} = - \half \partial_x^2 H$. Whence, $(M,g_{\mathrm{pp}})$ is flat only if $H$ is a linear function of $x$. The only non-trivial component of the Ricci tensor of $g_{\mathrm{pp}}$ is $R_{uu} = R_{uxux}$ and the scalar curvature $R=0$. The only non-trivial component of the Cotton-York tensor of $g_{\mathrm{pp}}$ is $C_{uxu} = - \half \partial_x^3 H$ whose scalar norm-squared $\rn{C}^2 =0$. Thus, $(M,g_{\mathrm{pp}})$ is conformally flat only if $H$ is a quadratic function of $x$. Since we are concerned with geometries that are not conformally flat, we shall assume henceforth that $(M,g_{\mathrm{pp}})$ has $\partial_x^3 H \neq 0$.

From the data above, it follows that any $X \in \fC ( M,[g_{\mathrm{pp}}] )$ must be of the form
\begin{equation}\label{eq:ppwaveCKV}
X = \gamma_X \partial_u + ( \alpha_X - \beta_X^\prime x - \frac{\gamma_X^{\prime\prime}}{4} x^2 + 2 c_X v ) \partial_v + ( \beta_X + ( \frac{\gamma_X^\prime}{2} + c_X ) x ) \partial_x~,
\end{equation}
in terms of three real functions $(\alpha_X,\beta_X,\gamma_X)$ of $u$ and a real number $c_X$ which obey
\begin{equation}\label{eq:ppwaveCKVCondition}
2 ( \alpha_X - \beta_X^\prime x - \frac{\gamma_X^{\prime\prime}}{4} x^2 )^\prime + ( \gamma_X H )^\prime - 2 c_X H + ( \beta_X + ( \frac{\gamma_X^\prime}{2} + c_X ) x ) \partial_x H =0~.
\end{equation}
The conformal factor is
\begin{equation}\label{eq:ppwaveCKVConformalFactor}
\sigma_X = - \frac{\gamma_X^\prime}{2} - c_X~.
\end{equation}

The expression \eqref{eq:ppwaveCKV} shows that the Lie bracket of the generic null Killing vector $\xi$ with any $X \in \fC ( M,[g_{\mathrm{pp}}] )$ is given by 
\begin{equation}\label{eq:ppwaveXiCKVLieBracket}
[ \xi , X ] = 2 c_X \xi~. 
\end{equation}
Whence, the real line spanned by $\xi$ forms a one-dimensional ideal of $\fC ( M,[g_{\mathrm{pp}}] )$. Clearly $\xi$ is in the centre $Z( \fC ( M,[g_{\mathrm{pp}}] ) )$ of $\fC ( M,[g_{\mathrm{pp}}] )$ only if every $X \in \fC ( M,[g_{\mathrm{pp}}] )$ has $c_X =0$. If $\xi \notin Z( \fC ( M,[g_{\mathrm{pp}}] ) )$ then at least one $X \in \fC ( M,[g_{\mathrm{pp}}] )$ must have $c_X \neq 0$ and every other $Y \in \fC ( M,[g_{\mathrm{pp}}] )$ can be taken to have $c_Y =0$ (i.e. if $Y \in \fC ( M,[g_{\mathrm{pp}}] )$ has $c_Y \neq 0$ then ${\tilde Y} = Y - \frac{c_Y}{c_X} X \in \fC ( M,[g_{\mathrm{pp}}] )$ has $c_{\tilde Y}=0$).

The condition \eqref{eq:ppwaveCKVCondition} indicates that the existence of an extra conformal Killing vector (in addition to $\xi$) puts constraints on the function $H$. If ${\mbox{dim}}\, \fC ( M,[g_{\mathrm{pp}}] ) =2$, one can fix $\gamma_X =1$ in \eqref{eq:ppwaveCKV}, \eqref{eq:ppwaveCKVCondition} and \eqref{eq:ppwaveCKVConformalFactor} with respect to a conformally equivalent metric via \eqref{eq:ppwaveRedefinition} if it is possible to identify $\Omega^2 = \gamma_X^{-1}$. In this case, $X$ is homothetic with respect to $\gamma_X^{-1} g_{\mathrm{pp}}$, with conformal factor $-c_X$. Whence, from \eqref{eq:ppwaveXiCKVLieBracket}, $X$ is a Killing vector only if $[\xi,X]=0$, in which case \eqref{eq:ppwaveCKVCondition} fixes 
\begin{equation}\label{eq:ppwaveXKVH}
H = - 2\alpha_X - \beta_X^2 + 2\beta_X^\prime x + f(-x+\int\!\! \rd u \, \beta_X)~,
\end{equation}
in terms of any real function $f$ of one variable whose third derivative is not zero. A similar, but more complicated, expression for $H$ emerges when $c_X \neq 0$.  

Since $\rn{C}^2$ vanishes identically on $(M,g_{\mathrm{pp}})$, it cannot be used to define a conformal scalar. However, the fact that the Lie derivative of $C$ along any conformal Killing vector $X$ is zero implies that $\partial_X C_{uxu} = - ( 2 \partial_u X^u + \partial_x X^x ) C_{uxu}$ on $(M,g_{\mathrm{pp}})$. Using \eqref{eq:ppwaveCKVConformalFactor}, this gives $\partial_X C_{uxu} = ( 5\sigma_X + 4 c_X ) C_{uxu}$ for any $X \in \fC ( M,[g_{\mathrm{pp}}] )$. Consequently, if $\xi \in Z( \fC ( M,[g_{\mathrm{pp}}] ) )$, then $\phi = \partial_x^3 H$ is a proper conformal scalar with $p_\phi =5$ and $\fC ( M,[g_{\mathrm{pp}}] )$ is conformally isometric. If $\xi \notin Z( \fC ( M,[g_{\mathrm{pp}}] ) )$ then $(M,g_{\mathrm{pp}})$ admits a proper conformal gradient with $\varphi = \tfrac{1}{5} \ln \partial_x^3 H$ and $s_X = -\tfrac{4}{5} c_X$, in which case $(M,g_{\mathrm{pp}})$ is conformally homothetic. In both cases, the isometric/homothetic action of $\fC ( M,[g_{\mathrm{pp}}] )$ is with respect to the metric $( \partial_x^3 H )^{2/5} g_{\mathrm{pp}}$ on $M$ (which need not be locally isometric to a pp-wave).

\section{The Kru\v{c}kovi\v{c} classification}
\label{sec:TheKruckovicClassification}

The classification, up to local isometry, of all three-dimensional Lorentzian geometries with Killing vectors is due to Kru\v{c}kovi\v{c} \cite{Kru:1954} (see also section 5 in \cite{KruMat2013}). We shall adhere to the notation used in the \lq summary of results' section in \cite{Kru:1954}. Class IV contains eight geometries with ${\mbox{dim}}\, \fK(M,g) = 4$, three of which (IV.4, IV.5 and IV.6) are not conformally flat. Class III contains seven geometries with ${\mbox{dim}}\, \fK(M,g) = 3$ for which the action of $\fK(M,g)$ is transitive (in fact, the first two III.1$\cong$IV.4 and III.2$\cong$IV.5). Class II contains four geometries with ${\mbox{dim}}\, \fK(M,g) = 3$ for which the action of $\fK(M,g)$ is not transitive, but all of them are conformally flat. Finally, class I contains five geometries with ${\mbox{dim}}\, \fK(M,g) = 2$. The Lie algebra $\fK(M,g)$ is abelian for I.1 and I.2, and nonabelian for I.3, I.4 and I.5. 

Of the Lorentzian geometries $(M,g)$ in \cite{Kru:1954} which are not conformally flat, all have $\fK(M,g)$ with zero-dimensional centre $Z ( \fK(M,g) )$ except I.1, I.2, III.3, IV.4, IV.5 and IV.6. The geometries III.3, IV.4, IV.5 and IV.6 have ${\mbox{dim}}\, Z ( \fK(M,g) ) =1$ while I.1 and I.2 have ${\mbox{dim}}\, Z ( \fK(M,g) ) =2$. By explicit calculation, one finds that any Lorentzian three-manifold $(M,g)$ that is not conformally flat and admits a non-zero null $\xi \in Z ( \fK(M,g) )$ must have ${\mbox{dim}}\, \fK(M,g) < 3$. If ${\mbox{dim}}\, \fK(M,g) =2$ then $(M,g)$ must be in class I.1. In this case, one can choose local coordinates $(u,v,x)$ on $M$ such that $( \partial_u , \xi = \partial_v ) \in \fK(M,g) \cong \RR^2$ and, for some positive function $\Omega$ of $x$,  $g = \Omega(x)  g_{\mathrm{pp}}$ in terms of the pp-wave metric $g_{\mathrm{pp}}$ in \eqref{eq:ppwave} with $\partial_u H =0$ and $\partial_x^3 H \neq 0$. 

\section{Plane waves in $d>3$}
\label{sec:PlaneWaves}

Consider a $d$-dimensional pp-wave metric of the form
\begin{equation}\label{eq:ppwaveGen}
g_{\mathrm{pp}} = 2 \rd u \rd v + H(u,\vec{x}) {\rd u}^2 + {\rd \vec{x}}^2~,
\end{equation}
in terms of Brinkmann coordinates $(u,v,\vec{x} )$ on $M$, where $H$ is an arbitrary real function of $(u,\vec{x})$. It is convenient to write $\vec{x} = ( x^a )$, where $a=1,...,d-2$, and ${\rd \vec{x}}^2 = \rd x^a \rd x^a$ is the canonical Euclidean metric on $\RR^{d-2}$. In these coordinates, $\xi = \partial_v$ is a null Killing vector of $(M,g_{\mathrm{pp}})$.

The only non-trivial Christoffel symbols of \eqref{eq:ppwaveGen} are
\begin{equation}\label{eq:ppwaveChristoffelSymbolsGen}
\Gamma_{ua}^v = - \Gamma_{uu}^a = \half \partial_a H \; , \quad\quad \Gamma_{uu}^v = \half \partial_u H~.
\end{equation}
The only non-trivial components of the associated Riemann tensor are $R_{uaub} = - \half \partial_a \partial_b H$. Whence, $(M,g_{\mathrm{pp}})$ is flat only if $H$ is linear in $\vec{x}$. The only non-trivial component of the Ricci tensor is $R_{uu} = R_{uaua}$ and the scalar curvature $R=0$. Whence, the only non-trivial components of the Weyl tensor are $W_{uaub} = - \half (\partial_a \partial_b H - \tfrac{1}{d-2} \delta_{ab} \partial_c \partial_c H )$. We shall assume henceforth that $(M,g_{\mathrm{pp}})$ is not conformally flat.

With respect to the following pseudo-orthonormal frame on $(M,g_{\mathrm{pp}})$,
\begin{equation}\label{eq:ppwaveVielbeinGen}
e^- = \rd u \; , \quad\quad e^+ = \rd v + \half H \rd u \; , \quad\quad e^a = \rd x^a~,
\end{equation}
the only non-trivial component of the associated spin connection is $\omega_{u}^{+a} = \half \partial_a H$. Whence, from \eqref{eq:SpinConnection}, it follows that
\begin{equation}\label{eq:ppwaveNablaGen}
\nabla_u \epsilon = \partial_u \epsilon - \tfrac{1}{4} ( \partial_a H ) {\bm \Gamma}_a {\bm \xi} \epsilon \; , \quad\quad \nabla_v \epsilon = \partial_v \epsilon  \; , \quad\quad \nabla_a \epsilon = \partial_a \epsilon~,
\end{equation}
for any $\epsilon \in \fS(M)$. Any $\epsilon \in \fZ(M,[g_{\mathrm{pp}}])$ is actually $\nabla$-parallel. Moreover, using \eqref{eq:ppwaveNablaGen}, it follows that $\nabla_\mu \epsilon=0$ only if $\partial_\mu \epsilon =0$ with ${\bm \xi} \epsilon =0$. Consequently, ${\mbox{dim}} \, \fZ(M,[g_{\mathrm{pp}}]) = \half {\mbox{dim}} \, \fS(M)$.   

A special class of pp-waves with $H = H_{ab}(u) x^a x^b$ in \eqref{eq:ppwaveGen} are called {\emph{plane waves}} (see \cite{Blau:Notes} for a comprehensive review). 
\footnote{Note that a plane wave in $d=3$ is necessarily conformally flat.}
Let $g_{\mathrm{pw}}$ denote the generic plane wave metric in $d>3$. In addition to $\xi = \partial_v$, $\fK(M,g_{\mathrm{pw}})$ contains Killing vectors of the form
\begin{equation}\label{eq:planewaveKV}
k(\vec{f}) = f_a \partial_a - x_a ( \partial_u f_a ) \partial_v~,
\end{equation}
where
\begin{equation}\label{eq:planewaveKVcondition}
\partial_u^2 f_a = H_{ab} f_b~,
\end{equation}
and each component $f_a$ is a function only of $u$. Clearly $[ \xi , k(\vec{f}) ] =0$ while 
\begin{equation}\label{eq:planewaveKVbracket}
[k(\vec{f}),k(\vec{\tilde f}) ] = - ( f_a \partial_u {\tilde f}_a -  {\tilde f}_a \partial_u f_a ) \xi~.
\end{equation}
The condition \eqref{eq:planewaveKVcondition} for $\vec{f}$ and $\vec{\tilde f}$ implies that $f_a \partial_u {\tilde f}_a -  {\tilde f}_a \partial_u f_a$ is constant. In fact, the Lie subalgebra of $\fK(M,g_{\mathrm{pw}})$ spanned by $\xi$ and all linearly independent $k(\vec{f})$ as in \eqref{eq:planewaveKV} (with $\vec{f}$ solving \eqref{eq:planewaveKVcondition}) is isomorphic to the Heisenberg Lie algebra $\fheis_{d-2}$ of dimension $2d-3$. Furthermore, it follows that $\fZ(M,[g_{\mathrm{pw}}])$ is invariant under this $\fheis_{d-2}$. That is,  
\begin{equation}\label{eq:planewaveSpinorialLD}
\cL_\xi \epsilon = 0 \; , \quad\quad \cL_{k(\vec{f})} \epsilon = 0~,
\end{equation}
for all $\epsilon \in \fZ(M,[g_{\mathrm{pw}}])$ and $k(\vec{f})  \in \fK(M,g_{\mathrm{pw}})$, in terms of the spinorial Lie derivative \eqref{eq:SpinLieDer}.

A generic plane wave also admits a proper homothetic conformal Killing vector proportional to $2v \partial_v + x_a \partial_a$. In fact, $\vartheta = -\half ( 2v \partial_v + x_a \partial_a )$ obeys  
\begin{equation}\label{eq:planewaveTheta}
[\vartheta,\xi]=\xi \; , \quad\quad [\vartheta,k(\vec{f})]=\half k(\vec{f}) \; , \quad\quad  {\hat \cL}_{\vartheta} \epsilon = \half \epsilon~,
\end{equation}
for all $k(\vec{f}) \in \fK(M,g_{\mathrm{pw}})$ and $\epsilon \in \fZ(M,[g_{\mathrm{pw}}])$, in terms of the Kosmann-Schwarzbach Lie derivative \eqref{eq:KSLieDer}.

For certain choices of $H$, $(M,g_{\mathrm{pw}})$ may admit an additional Killing vector $l$. Let $h = ( H_{ab} )$ denote the symmetric $(d-2){\times}(d-2)$ matrix of $u$-dependent functions in $g_{\mathrm{pw}}$. The data for four classes of homogeneous plane waves with an extra Killing vector is displayed in Table~\ref{tab:PlaneWaveExtraKV}. Both $A$ and $B$ are in $\Mat_{d-2} ( \RR )$, with $A$ symmetric and $B$ skewsymmetric. Entries in the two rightmost columns of Table~\ref{tab:PlaneWaveExtraKV} apply to any $k(\vec{f})$ as in \eqref{eq:planewaveKV} and $\epsilon \in \fZ(M,[g_{\mathrm{pw}}])$. In all four cases, $[l,\vartheta]=0$. The first case, with constant $h=A$, is a symmetric space.   
\begin{table}
\begin{center}
\begin{tabular}{|c|c|c|c|c|}
  \hline
  &&&& \\ [-.4cm]
  $h$ & $l$ & $[ l ,\xi ]$ & $[l,k(\vec{f})]$  & ${\hat \cL}_l \epsilon$ \\ [.05cm] 
  \hline\hline
  &&&& \\ [-.4cm]
  $A$ & $\partial_u$ & $0$ & $k(\partial_u \vec{f})$ & $0$ \\ [.05cm] 
  \hline
  &&&& \\ [-.4cm]
  $u^{-2} A$ & $u\partial_u - v\partial_v$ & $\xi$ & $k(u\partial_u \vec{f})$ & $\half \epsilon$ \\ [.05cm] 
  \hline
  &&&& \\ [-.4cm]
  ${\mathrm{e}}^{uB} A\, {\mathrm{e}}^{-uB}$ & $\partial_u - B_{ab} x_a \partial_b$ & $0$ & $k(\partial_u \vec{f} - B\vec{f})$ & $-\half {\bm B} \epsilon$ \\ [.05cm]
  \hline 
  &&&& \\ [-.4cm]
  $u^{-2} {\mathrm{e}}^{(\ln u)B} A\, {\mathrm{e}}^{-(\ln u)B}$ & $u\partial_u - v\partial_v - B_{ab} x_a \partial_b$ & $\xi$ & $k(u\partial_u \vec{f} - B \vec{f})$ & $\half ( {\bf 1} - {\bm B} ) \epsilon$ \\ [.1cm] 
  \hline
  \end{tabular} \vspace*{.2cm}
\caption{Data for some homogeneous plane waves with an extra Killing vector.}
\label{tab:PlaneWaveExtraKV}
\end{center}
\end{table}

In $d=4$, the conformal symmetry superalgebra $\cS = \cB \oplus \cF$ associated with the conformal class of a plane wave $(M,g_{\mathrm{pw}})$ has $\cB = \fH(M,g_{\mathrm{pw}}) \oplus \cR$, $\cR = \fu(1)$ and $\cF = \fZ(M,[g_{\mathrm{pw}}])$. The twistor spinors in $\cF$ are uncharged with respect to $\cR$ so one can write $\cS = \cS_{\mathrm{pw}} \oplus \cR$, in terms of the Lie superalgebra $\cS_{\mathrm{pw}}$ with even part $\fH(M,g_{\mathrm{pw}})$ and odd part $\cF$. Generically, ${\mbox{dim}}\, \fH(M,g_{\mathrm{pw}}) =6$ and ${\mbox{dim}}\, \cF =2$. However, ${\mbox{dim}}\, \fH(M,g_{\mathrm{pw}}) =7$ if (and only if) $(M,g_{\mathrm{pw}})$ corresponds to one of the four homogeneous geometries in Table~\ref{tab:PlaneWaveExtraKV} (in the third and fourth cases, we can take $B = \rd x \wedge \rd y$ and ${\bm B} = -i {\bm \Gamma}$ in $d=4$). In $d=4$, each of the four homogeneous plane wave geometries in Table~\ref{tab:PlaneWaveExtraKV} has the function $H$ in \eqref{eq:ppwaveGen} of the form
\begin{equation}\label{eq:planewaveH}
H(u,x,y) = \alpha (u) (x^2 - y^2) + 2\beta (u) xy + \gamma(u) (x^2 + y^2)~,
\end{equation}
in terms of three real functions $\alpha$, $\beta$ and $\gamma$ of $u$, whose respective forms are displayed in Table~\ref{tab:PlaneWaveExtraKV4d}.  
\begin{table}
\begin{center}
\begin{tabular}{|c|c|c|}
  \hline
  && \\ [-.4cm]
  $\alpha$ & $\beta$ & $\gamma$ \\ [.05cm] 
  \hline\hline
  && \\ [-.4cm]
  $a$ & $b$ & $c$ \\ [.05cm] 
  \hline
  && \\ [-.4cm]
  $u^{-2} a$ & $u^{-2} b$ & $u^{-2} c$ \\ [.05cm] 
  \hline
  && \\ [-.4cm]
  $a \cos (2u) + b \sin (2u)$ & $b \cos (2u) - a \sin (2u)$ & $c$ \\ [.05cm]
  \hline 
  && \\ [-.4cm]
  $u^{-2} ( a \cos (2\ln u) + b \sin (2\ln u))$ & $u^{-2} ( b \cos (2\ln u) - a \sin (2\ln u))$ & $u^{-2} c$ \\ [.1cm] 
  \hline
  \end{tabular} \vspace*{.2cm}
\caption{Data for plane waves with ${\mbox{dim}}\, \fH(M,g_{\mathrm{pw}}) =7$ in $d=4$.}
\label{tab:PlaneWaveExtraKV4d}
\end{center}
\end{table}
In each case, $(M,g_{\mathrm{pw}})$ is not conformally flat provided the real numbers $a$ and $b$ are not both zero and is Ricci-flat only if the real number $c$ is zero. Furthermore, the first and second entries in Table~\ref{tab:PlaneWaveExtraKV4d} actually define conformally equivalent classes of plane wave metrics. The same applies to the third and fourth entries in Table~\ref{tab:PlaneWaveExtraKV4d}. In both cases, the explicit conformal isometry is induced by the coordinate transformation
\begin{equation}\label{eq:planewaveConformalIsometry}
(u,v,x,y) \mapsto ( {\mathrm{e}}^u , v - \tfrac{1}{4} ( x^2 + y^2 ), {\mathrm{e}}^{u/2} x , {\mathrm{e}}^{u/2} y )~.
\end{equation}
Whence, any $\cS_{\mathrm{pw}}$ with ${\mbox{dim}}\, \fH(M,g_{\mathrm{pw}}) =7$ can be assigned to the conformal class of a plane wave geometry defined by either the first or third entry in Table~\ref{tab:PlaneWaveExtraKV4d}.

The non-trivial $[\fH(M,g_{\mathrm{pw}}),\fH(M,g_{\mathrm{pw}})]$ and $[\fH(M,g_{\mathrm{pw}}),\cF]$ brackets for $\cS_{\mathrm{pw}}$ are prescribed by  \eqref{eq:planewaveKVbracket}, \eqref{eq:planewaveSpinorialLD} and \eqref{eq:planewaveTheta} (together with the data in Table~\ref{tab:PlaneWaveExtraKV} when $(M,g_{\mathrm{pw}})$ admits an extra Killing vector). The $[\cF,\cF]$ bracket for $\cS_{\mathrm{pw}}$ is of the form
\begin{equation}\label{eq:planewaveFF}
[\epsilon,\epsilon^\prime]= c_{\epsilon,\epsilon^\prime} \xi~,
\end{equation}
for every $\epsilon , \epsilon^\prime \in \cF$, in terms of some non-zero real number  $c_{\epsilon,\epsilon^\prime}$ (see section 7.1 of~\cite{deMedeiros:2013jja} for the proof). 

\section{Lie algebras in low dimension}
\label{sec:LALow}
Up to isomorphism, all real Lie algebras with dimension four or less have been classified \cite{Bia:1898,PSWZ:1976}. A Lie algebra $\fg$ is said to be {\emph{decomposable}} if it can be expressed as the direct sum of non-trivial ideals. Otherwise, $\fg$ is said to be {\emph{indecomposable}}. To classify all Lie algebras of a given dimension, it is clearly sufficient to classify all the indecomposable Lie algebras of less than or equal to that dimension. We list below in Table~\ref{tab:4dLA} all (isomorphism classes of) indecomposable real Lie algebras with dimension four or less. The symbol used to denote each class $\fg$ is prefixed by $\fa$, $\fb$, $\fc$ for ${\mbox{dim}}\, \fg =2,3,4$, respectively. For ${\mbox{dim}}\, \fg =3,4$, each class is denoted by a roman numeral in square brackets (we follow the classical notation of \cite{Bia:1898} for ${\mbox{dim}}\, \fg =3$). We specify each $\fg$ by its non-zero Lie brackets $[-,-]$ with respect to a basis $\{ \be_1 ,..., \be_{{\scriptsize \mbox{dim}} \fg} \}$. 
\begin{table}
\begin{center}
\hspace*{-.6cm}\begin{tabular}{|c|c|c|c|}
  \hline
  $\fg$ & Alias & Non-zero brackets & Real parameters  \\ [.05cm]
  \hline\hline
  &&& \\ [-.4cm]
  $\RR$ & - & - & - \\ [.05cm] 
  \hline
  &&& \\ [-.4cm]
  $\fa$ & non-abelian & $[ \be_1 , \be_2 ] = \be_1$ & - \\ [.05cm] 
  \hline
  &&& \\ [-.4cm]
 $\fb$[II] & ${\mathfrak{heis}}_1$ & $[ \be_2 , \be_3 ] = \be_1$ & - \\ [.05cm] 
  $\fb$[IV] & - & $[ \be_1 , \be_3 ] = \be_1$, $[ \be_2 , \be_3 ] = \be_2 - \be_1$ & - \\ [.05cm]
  $\fb$[V] & - & $[ \be_1 , \be_3 ] = \be_1$, $[ \be_2 , \be_3 ] = \be_2$ & - \\ [.05cm] 
  $\fb$[VI] & - & $[ \be_1 , \be_3 ] = \be_2 + a \be_1$, $[ \be_2 , \be_3 ] = \be_1 + a\be_2$ & $a{>}0$, $a {\neq} 1$ \\ [.05cm] 
  $\fb$[VI$_0$] & $\fso(1,1) {\ltimes} \RR^{1,1}$ & $[ \be_1 , \be_3 ] = \be_2$, $[ \be_2 , \be_3 ] = \be_1$ & - \\ [.05cm] 
  $\fb$[VII] & - & $[ \be_1 , \be_3 ] = \be_2 + a \be_1$, $[ \be_2 , \be_3 ] = a\be_2 - \be_1$ & $a{>}0$ \\ [.05cm] 
  $\fb$[VII$_0$] & $\fso(2) {\ltimes} \RR^{2}$ & $[ \be_1 , \be_3 ] = -\be_2$, $[ \be_2 , \be_3 ] = \be_1$ & - \\ [.05cm] 
 $\fb$[VIII] & $\fsl(2,\RR)$ & $[ \be_1 , \be_2 ] = -\be_3$, $[ \be_1 , \be_3 ] = -\be_2$, $[ \be_2 , \be_3 ] = \be_1$  & - \\ [.05cm] 
  $\fb$[IX] & $\fso(3)$ & $[ \be_1 , \be_2 ] = \be_3$, $[ \be_1 , \be_3 ] = -\be_2$, $[ \be_2 , \be_3 ] = \be_1$ & - \\ [.05cm] 
  \hline
  &&& \\ [-.4cm]
  $\fc$[I] & - & $[ \be_2 , \be_4 ] = \be_1$, $[ \be_3 , \be_4 ] = \be_2$ & - \\ [.05cm] 
  $\fc$[II] & - & $[ \be_1 , \be_4 ] = a \be_1$, $[ \be_2 , \be_4 ] = \be_2$, $[ \be_3 , \be_4 ] = \be_2 + \be_3$ & $a{\neq} 0$ \\ [.05cm]
  $\fc$[III] & - & $[ \be_1 , \be_4 ] = \be_1$, $[ \be_3 , \be_4 ] = \be_2$ & - \\ [.05cm] 
  $\fc$[IV] & - & $[ \be_1 , \be_4 ] = \be_1$, $[ \be_2 , \be_4 ] = \be_1 + \be_2$, $[ \be_3 , \be_4 ] = \be_2 + \be_3$ & - \\ [.05cm] 
  $\fc$[V] & - & $[ \be_1 , \be_4 ] = \be_1$, $[ \be_2 , \be_4 ] = a \be_2$, $[ \be_3 , \be_4 ] = b\be_3$ & $ab{\neq} 0$, $-1 {\leq} a {\leq} b {\leq} 1$ \\ [.05cm] 
  $\fc$[VI] & - & $[ \be_1 , \be_4 ] = a \be_1$, $[ \be_2 , \be_4 ] = b \be_2 - \be_3$, $[ \be_3 , \be_4 ] = \be_2 + b\be_3$ & $a{\neq} 0$, $b{\geq} 0$ \\ [.05cm]
 $\fc$[VII] & - & $[ \be_1 , \be_4 ] = 2\be_1$, $[ \be_2 , \be_4 ] = \be_2$, & - \\ [.05cm]
&& $[ \be_3 , \be_4 ] = \be_2 + \be_3$, $[ \be_2 , \be_3 ] = \be_1$ & \\ [.05cm] 
  $\fc$[VIII] & - & $[ \be_2 , \be_3 ] = \be_1$, $[ \be_2 , \be_4 ] = \be_2$, $[ \be_3 , \be_4 ] = -\be_3$ & - \\ [.05cm]
  $\fc$[IX] & - & $[ \be_1 , \be_4 ] = (a+1) \be_1$, $[ \be_2 , \be_3 ] = \be_1$, & $-1 {<} a {\leq} 1$ \\ [.05cm]
  && $[ \be_2 , \be_4 ] = \be_2$, $[ \be_3 , \be_4 ] = a\be_3$ & \\ [.05cm] 
  $\fc$[X] & - & $[ \be_2 , \be_3 ] = \be_1$, $[ \be_2 , \be_4 ] = - \be_3$, $[ \be_3 , \be_4 ] = \be_2$ & - \\ [.05cm]
  $\fc$[XI] & - & $[ \be_1 , \be_4 ] = 2a\be_1$, $[ \be_2 , \be_4 ] = a\be_2 - \be_3$, & $a{>} 0$ \\ [.05cm]
  && $[ \be_3 , \be_4 ] = \be_2 + a\be_3$, $[ \be_2 , \be_3 ] = \be_1$ & \\ [.15cm]
  \hline  
\end{tabular} \vspace*{.2cm}
\caption{Indecomposable real Lie algebras of dimension four or less.}
\label{tab:4dLA}
\end{center}
\end{table}

\section{Lie superalgebras in low dimension}
\label{sec:AppendixF}
A real Lie superalgebra consists of a $\ZZ_2$-graded real vector space $\cS$ (with even part $\cS_0$ and odd part $\cS_1$) that is equipped with the following additional structure. 

A real bilinear map $[-,-] : \cS \times \cS \rightarrow \cS$ which respects the $\ZZ_2$-grading such that  
\begin{equation}\label{eq:LSAbracket}
[ \cS_0 , \cS_0 ] \subset \cS_0 \; , \quad\quad [ \cS_0 , \cS_1 ] \subset \cS_1 \; , \quad\quad [ \cS_1 , \cS_1 ] \subset \cS_0~.
\end{equation}
For all $u,v \in \cS_0$ and $\alpha,\beta \in \cS_1$, 
\begin{equation}\label{eq:LSAbracket2}
[u,v] =-[v,u] \; , \quad\quad [u,\alpha] = -[\alpha,u] \; , \quad\quad [\alpha,\beta] = [\beta,\alpha]~.
\end{equation}
Since $[-,-]$ is symmetric bilinear on $\cS_1$, $[ \cS_1 , \cS_1 ]$ is defined by specifying $[\alpha , \alpha] \in \cS_0$ for all $\alpha \in \cS_1$ (i.e. via polarisation, any $[\alpha,\beta] = \half ( [\alpha+\beta,\alpha+\beta] - [\alpha , \alpha] - [\beta ,\beta] )$). 

Furthermore, $\cS$ is subject to a Jacobi identity which constrains 
\begin{align}\label{eq:LSAJacobi}
[[ u,v ],w] + [[v,w],u] + [[w,u],v] &= 0~, \nonumber \\
[[ u,v ],\alpha] + [[v,\alpha],u] + [[\alpha,u],v] &= 0~, \\
[[ u,\alpha ],\beta] + [[\alpha,\beta],u] - [[\beta,u],\alpha] &= 0~, \nonumber \\
[[\alpha,\beta],\gamma] + [[\beta,\gamma],\alpha] + [[\gamma,\alpha],\beta] &= 0~, \nonumber
\end{align}
for all $u,v,w \in \cS_0$ and $\alpha,\beta,\gamma \in \cS_1$. The first three conditions in \eqref{eq:LSAJacobi} have a simple conceptualisation. The first condition says that $\cS_0$ must be a real Lie algebra (i.e. it is the Jacobi identity for $\cS_0$). The second condition says that $\cS_1$ must be a real representation of $\cS_0$. The third condition says that the map $[-,-] : \cS_1 \times \cS_1 \rightarrow \cS_0$ must be $\cS_0$-equivariant. Notice that the fourth condition is symmetric trilinear on $\cS_1$ and therefore equivalent, via polarisation, to demanding $[[\alpha,\alpha] , \alpha] =0$, for all $\alpha \in \cS_1$. 

It is straightforward to extend the concept of an ideal of a Lie algebra to a superideal of a Lie superalgebra and $\cS$ is called decomposable if it can be expressed as the direct sum of non-trivial superideals. Otherwise $\cS$ is said to be indecomposable.

If $[\cS_1,\cS_1]=0$ then clearly the third and fourth conditions in \eqref{eq:LSAJacobi} are identically satisfied and a real Lie superalgebra $\cS$ is defined by any real Lie algebra $\cS_0$ with real $\cS_0$-module $\cS_1$. We shall call $\cS$ {\emph{proper}} if $[\cS_1,\cS_1]\neq 0$. If $[\cS_0,\cS_1]=0$ then the second and fourth conditions in \eqref{eq:LSAJacobi} are identically satisfied. Consequently, $\cS$ is then a proper real Lie superalgebra only if $[\cS_1,\cS_1] \subset Z( \cS_0 )$ for a given real Lie algebra $\cS_0$ with non-trivial centre $Z( \cS_0 )$. 

Consider now the class of proper real Lie superalgebras $\cS$ with ${\mbox{dim}}\, \cS_1 =1$. Up to a non-zero constant multiple, $\alpha \in \cS_1$ is unique and let us define a non-zero $\xi = [\alpha,\alpha] \in \cS_0$. Given a real Lie algebra $\cS_0 = \fg$, the $[\cS_0,\cS_1]$ bracket must be of the form
\begin{equation}\label{eq:BF1d}
[ u , \alpha ] = \varrho(u) \alpha~,
\end{equation}
in terms of a linear map $\varrho : \fg \rightarrow \RR$. From \eqref{eq:LSAJacobi}, it follows that $\cS$ is a real Lie superalgebra only if    
\begin{equation}\label{eq:BF1dJacobi}
\varrho ([u,v]) =0 \; , \quad\quad [ u , \xi ] = 2 \varrho(u) \xi \; , \quad\quad \varrho(\xi) =0~,
\end{equation}
for all $u,v \in \fg$. The first condition in \eqref{eq:BF1dJacobi} states that $[\fg,\fg] \subset \ker \varrho$, whence $\ker \varrho \lhd \fg$. The second and third conditions in \eqref{eq:BF1dJacobi} imply that $\xi \in Z( \ker \varrho )$. 

If $\varrho =0$ (i.e. $\ker \varrho = \fg$) then \eqref{eq:BF1dJacobi} are satisfied provided $\xi \in Z( \fg )$. Let $\cS^\circ (\fg)$ denote the proper real Lie superalgebra of this type defined by a real Lie algebra $\fg$ with non-trivial centre $Z( \fg )$. In this case, $\cS^\circ (\fg)$ is indecomposable as a Lie superalgebra only if $\fg$ is indecomposable as a Lie algebra. By identifying all the entries in Table~\ref{tab:4dLA} which have a non-trivial centre, one obtains all indecomposable $\cS^\circ (\fg)$ with ${\mbox{dim}}\, \fg \leq 4$, whose data is displayed in Table~\ref{tab:41LSACentered}. 
\begin{table}
\begin{center}
\begin{tabular}{|c||c|c|c|c|c|c|}
  \hline
  &&&&&& \\ [-.4cm]
  $\fg$ & $\RR$ & $\fb$[II] & $\fc$[I] & $\fc$[III] & $\fc$[VIII] & $\fc$[X] \\ [.05cm] 
  \hline
  &&&&&& \\ [-.4cm]
  $\xi$ & $\be_1$ & $\be_1$ & $\be_1$ & $\be_2$ & $\be_1$ & $\be_1$ \\ [.05cm] 
  \hline
  \end{tabular} \vspace*{.2cm}
\caption{Data for indecomposable $\cS^\circ (\fg)$ with ${\mbox{dim}}\, \fg \leq 4$.}
\label{tab:41LSACentered}
\end{center}
\end{table}

If $\varrho \neq 0$ then $\varrho(z) \neq 0$ for at least one $z \in \fg$. In this case, one can take every other $u \in \fg$ to be in $\ker \varrho$ (i.e. via the redefinition $u \mapsto u - \frac{\varrho(u)}{\varrho(z)} z$). Moreover, it is convenient to define $\vartheta = \frac{1}{2 \varrho(z)} z \in \fg / \ker \varrho$ so the second condition in \eqref{eq:BF1dJacobi} is equivalent to taking
\begin{equation}\label{eq:UncentredBrackets}
[\vartheta, \xi] = \xi~,
\end{equation}
with $\xi \in Z( \ker\varrho )$ (which also ensures that the third condition in \eqref{eq:BF1dJacobi} is satisfied). Since ${\mbox{dim}}\, \fg = 1+ {\mbox{dim}}\, \ker \varrho$, the first condition in \eqref{eq:BF1dJacobi} is equivalent to demanding $\ker \varrho \lhd \fg$. Let $\cS^\lhd (\fg)$ denote the proper real Lie superalgebra of this type defined by a real Lie algebra $\fg$ which contains a codimension one ideal $\ker \varrho$ with non-trivial centre $Z( \ker \varrho )$. By identifying all the entries in Table~\ref{tab:4dLA} which have a codimension one ideal with non-trivial centre, one obtains all indecomposable $\cS^\lhd (\fg)$ with ${\mbox{dim}}\, \fg \leq 4$, whose data is displayed in Tables~\ref{tab:41LSAUncentered},~\ref{tab:41LSAUncentered2},~\ref{tab:41LSAUncentered3}. In each case, $\ker \varrho$ is spanned by $\{ \be_1 ,..., \be_{{\scriptsize \mbox{dim}} \fg -1} \}$. For ${\mbox{dim}}\, \fg =4$, the classification described above was first obtained in \cite{Matiadou:2007gz}.
\begin{table}
\begin{center}
\begin{tabular}{|c||c|c|c|c|c|}
  \hline
  &&&&& \\ [-.4cm]
  $\fg$ & $\fa$ & $\fb$[IV] & $\fb$[V]  & $\fb$[VI] & $\fb$[VI$_0$] \\ [.05cm] 
  \hline
  &&&&& \\ [-.4cm]
  $\ker\varrho$ & $\RR$ & $\RR^2$ & $\RR^2$ & $\RR^2$ & $\RR^2$ \\ [.05cm] 
  \hline
  &&&&& \\ [-.4cm]
  $\xi$ & $\be_1$ & $\be_1$ & $\be_1$ & $\be_2 - \be_1$ & $\be_1 + \be_2$ \\ [.05cm] 
  \hline
  &&&&& \\ [-.4cm]
  $-\vartheta$ & $\be_2$ & $\be_3$ & $\be_3$ & $\frac{1}{a-1}\be_3$ & $\be_3$ \\ [.1cm] 
  \hline
  \end{tabular} \vspace*{.2cm}
\caption{Data for indecomposable $\cS^\lhd (\fg)$ with ${\mbox{dim}}\, \fg \leq 3$.}
\label{tab:41LSAUncentered}
\end{center}
\end{table}

\begin{table}
\begin{center}
\begin{tabular}{|c||c|c|c|c|c|c|c|c|c|c|c|}
  \hline
  &\multicolumn{2}{c|}{}&&&\multicolumn{3}{c|}{} &&&& \\ [-.4cm]
  $\fg$ & \multicolumn{2}{c|}{$\fc$[II]} & $\fc$[III]  & $\fc$[IV] & \multicolumn{3}{c|}{$\fc$[V]} & $\fc$[VI] & $\fc$[VII] & $\fc$[IX] & $\fc$[XI] \\ [.05cm] 
  \hline 
  &\multicolumn{2}{c|}{}&&&\multicolumn{3}{c|}{} &&&& \\ [-.4cm]
  $\ker\varrho$ & \multicolumn{2}{c|}{$\RR^3$} & $\RR^3$ & $\RR^3$ & \multicolumn{3}{c|}{$\RR^3$} & $\RR^3$ & $\fb$[II] & $\fb$[II] & $\fb$[II] \\ [.05cm] 
  \hline
  &&&&&&&&&&& \\ [-.4cm]
  $\xi$ & $\be_1$ & $\be_2$ & $\be_1$ & $\be_1$ & $\be_1$ & $\be_2$  & $\be_3$ & $\be_1$ & $\pm \be_1$ & $\be_1$ & $\pm \be_1$ \\ [.05cm] 
  \hline
  &&&&&&&&&&& \\ [-.4cm]
  $-\vartheta$ & $\frac{1}{a} \be_4$ & $\be_4$ & $\be_4$ & $\be_4$ & $\be_4$ & $\frac{1}{a} \be_4$ & $\frac{1}{b} \be_4$ & $\frac{1}{a} \be_4$ & $\half \be_4$ & $\frac{1}{b+1} \be_4$ & $\frac{1}{a} \be_4$ \\ [.1cm] 
  \hline
  &\multicolumn{2}{c|}{}&&&\multicolumn{3}{c|}{} &&&& \\ [-.4cm]
  Note & \multicolumn{2}{c|}{$a\neq 1$} & - & - & \multicolumn{3}{c|}{$a<b<1$} & - & - & - & - \\ [.05cm] 
 \hline
  \end{tabular} \vspace*{.2cm}
\caption{Data for generic indecomposable $\cS^\lhd (\fg)$ with ${\mbox{dim}}\, \fg = 4$.}
\label{tab:41LSAUncentered2}
\end{center}
\end{table}

\begin{table}
\begin{center}
\begin{tabular}{|c||c|c|c|c|c|c|}
  \hline
  &&\multicolumn{5}{c|}{}  \\ [-.4cm]
  $\fg$ & $\fc$[II] & \multicolumn{5}{c|}{$\fc$[V]} \\ [.05cm] 
  \hline 
  &&\multicolumn{5}{c|}{} \\ [-.4cm]
  $\ker\varrho$ & $\RR^3$ & \multicolumn{5}{c|}{$\RR^3$} \\ [.05cm] 
  \hline
  &&&&&& \\ [-.4cm]
  $\xi$ & $\cos \theta\, \be_1 {+} \sin \theta\, \be_2$ & $\be_1$ & $\be_3$ & $\be_2$ & $\be_3$  & $\be_3$ \\ [.05cm] 
  \hline
  &&&&&& \\ [-.4cm]
  $-\vartheta$ & $\be_4$ & $\be_4$ & $\be_4$ &  $\frac{1}{a}  \be_4$ & $\be_4$ & $\be_4$ \\ [.1cm] 
  \hline
  &&\multicolumn{2}{c|}{} & \multicolumn{2}{c|}{} & \\ [-.4cm]
  Note & $a=1$, $0 \leq \theta < 2\pi$ & \multicolumn{2}{c|}{$a=b<1$} & \multicolumn{2}{c|}{$a<b=1$} & $a=b=1$ \\ [.05cm] 
 \hline
  \end{tabular} \vspace*{.2cm}
\caption{Data for non-generic indecomposable $\cS^\lhd (\fg)$ with ${\mbox{dim}}\, \fg = 4$.}
\label{tab:41LSAUncentered3}
\end{center}
\end{table}

Finally, let us consider the class of proper real Lie superalgebras $\cS$ with ${\mbox{dim}}\, \cS_1 =2$. It is convenient to express any ${ \alpha \choose {\hat \alpha}} \in \cS_1$ as a complex element $\alpha_\CC = \alpha + i {\hat \alpha}$. Now let $\cS_0 = \fg \oplus \fr$, in terms of a real Lie algebra $\fg$ and a one-dimensional real Lie algebra $\fr$. Let us fix a non-zero element $r \in \fr$ such that $[r,u]=0$, for all $u \in \fg$. Furthermore, let  
\begin{equation}\label{eq:BF2dR}
[ r , \alpha_\CC ] = i \alpha_\CC~,
\end{equation}
for all $\alpha_\CC \in \cS_1$. The remaining brackets for $\cS$ are of the form
\begin{align}\label{eq:BF2dNonR}
[ u , \alpha_\CC ] &= \varrho_\CC (u) \alpha_\CC + \varpi_\CC (u) \alpha_\CC^*~,  \nonumber \\
[ \alpha_\CC , \alpha_\CC ] &= \Sigma + i \Pi \; , \quad\quad [ \alpha_\CC , \alpha_\CC^* ] = \Xi~, 
\end{align}
for all $u \in \fg$ and $\alpha_\CC \in \cS_1$, where $\varrho_\CC , \varpi_\CC : \fg \rightarrow \CC$ are linear maps and $\Sigma , \Pi , \Xi \in \cS_0$. The Jacobi identity \eqref{eq:LSAJacobi} refines the data in \eqref{eq:BF2dNonR}. In particular, the $[r u \alpha_\CC]$ component of the second condition in \eqref{eq:LSAJacobi} implies $\varpi_\CC =0$ while the $[r \alpha_\CC \alpha_\CC]$ component of the third condition in \eqref{eq:LSAJacobi} implies $\Sigma = 0 = \Pi$. Furthermore, using \eqref{eq:BF2dR}, the imaginary part of $\varrho_\CC$ can be set to zero by redefining $u \mapsto u - \Im ( \varrho_\CC (u) ) r$, for all $u\in \fg$. The remaining conditions from \eqref{eq:LSAJacobi} involve only $\Re ( \varrho_\CC )$ and $\Xi$. They are in fact precisely equivalent to those in \eqref{eq:BF1dJacobi}, identifying $\Re ( \varrho_\CC ) = \varrho$ and $\Xi = \xi$. 

Whence, the data for any proper real Lie superalgebra with one-dimensional odd part is sufficient to define another proper real Lie superalgebra $\cS$ of the type described above with $\cS_0 = \fg \oplus \fr$ and ${\mbox{dim}}\, \cS_1 =2$. We shall write $\cS = \cS^\circ (\fg|\fr)$ if $\cS$ is based on the data for $\cS^\circ (\fg)$ or $\cS = \cS^\lhd (\fg|\fr)$ if $\cS$ is based on the data for $\cS^\lhd (\fg)$. 

\bibliographystyle{utphys}
\bibliography{CurvedSCFT}

\providecommand{\href}[2]{#2}\begingroup\raggedright\begin{thebibliography}{10}

\bibitem{Festuccia:2011ws}
G.~Festuccia and N.~Seiberg, ``{Rigid Supersymmetric Theories in Curved
  Superspace},'' \href{http://dx.doi.org/10.1007/JHEP06(2011)114}{{\em JHEP}
  {\bf 1106} (2011)  114}, \href{http://arxiv.org/abs/1105.0689}{{\tt
  arXiv:1105.0689 [hep-th]}}.

\bibitem{Jia:2011hw}
B.~Jia and E.~Sharpe, ``{Rigidly Supersymmetric Gauge Theories on Curved
  Superspace},'' \href{http://dx.doi.org/10.1007/JHEP04(2012)139}{{\em JHEP}
  {\bf 1204} (2012)  139},
\href{http://arxiv.org/abs/1109.5421}{{\tt arXiv:1109.5421 [hep-th]}}.

\bibitem{Samtleben:2012gy}
H.~Samtleben and D.~Tsimpis, ``{Rigid supersymmetric theories in 4d Riemannian
  space},'' \href{http://dx.doi.org/10.1007/JHEP05(2012)132}{{\em JHEP} {\bf
  1205} (2012)  132},
\href{http://arxiv.org/abs/1203.3420}{{\tt arXiv:1203.3420 [hep-th]}}.

\bibitem{Klare:2012gn}
C.~Klare, A.~Tomasiello, and A.~Zaffaroni, ``{Supersymmetry on Curved Spaces
  and Holography},'' \href{http://dx.doi.org/10.1007/JHEP08(2012)061}{{\em
  JHEP} {\bf 1208} (2012)  061},
\href{http://arxiv.org/abs/1205.1062}{{\tt arXiv:1205.1062 [hep-th]}}.

\bibitem{Dumitrescu:2012he}
T.~T. Dumitrescu, G.~Festuccia, and N.~Seiberg, ``{Exploring Curved
  Superspace},'' \href{http://dx.doi.org/10.1007/JHEP08(2012)141}{{\em JHEP}
  {\bf 1208} (2012)  141},
\href{http://arxiv.org/abs/1205.1115}{{\tt arXiv:1205.1115 [hep-th]}}.

\bibitem{Cassani:2012ri}
D.~Cassani, C.~Klare, D.~Martelli, A.~Tomasiello, and A.~Zaffaroni,
  ``{Supersymmetry in Lorentzian Curved Spaces and Holography},''
  \href{http://dx.doi.org/10.1007/s00220-014-1983-3}{{\em Commun. Math. Phys.}
  {\bf 327} (2014)  577--602},
\href{http://arxiv.org/abs/1207.2181}{{\tt arXiv:1207.2181 [hep-th]}}.

\bibitem{Liu:2012bi}
J.~T. Liu, L.~A. Pando~Zayas, and D.~Reichmann, ``{Rigid Supersymmetric
  Backgrounds of Minimal Off-Shell Supergravity},''
  \href{http://dx.doi.org/10.1007/JHEP10(2012)034}{{\em JHEP} {\bf 1210} (2012)
   034},
\href{http://arxiv.org/abs/1207.2785}{{\tt arXiv:1207.2785 [hep-th]}}.

\bibitem{deMedeiros:2012sb}
P.~de~Medeiros, ``{Rigid supersymmetry, conformal coupling and twistor
  spinors},''
\href{http://arxiv.org/abs/1209.4043}{{\tt arXiv:1209.4043 [hep-th]}}.

\bibitem{Dumitrescu:2012at}
T.~T. Dumitrescu and G.~Festuccia, ``{Exploring Curved Superspace (II)},''
  \href{http://dx.doi.org/10.1007/JHEP01(2013)072}{{\em JHEP} {\bf 1301} (2013)
   072},
\href{http://arxiv.org/abs/1209.5408}{{\tt arXiv:1209.5408 [hep-th]}}.

\bibitem{Kehagias:2012fh}
A.~Kehagias and J.~G. Russo, ``{Global Supersymmetry on Curved Spaces in
  Various Dimensions},''
  \href{http://dx.doi.org/10.1016/j.nuclphysb.2013.04.010}{{\em Nucl. Phys.}
  {\bf B873} (2013)  116--136},
\href{http://arxiv.org/abs/1211.1367}{{\tt arXiv:1211.1367 [hep-th]}}.

\bibitem{Closset:2012ru}
C.~Closset, T.~T. Dumitrescu, G.~Festuccia, and Z.~Komargodski,
  ``{Supersymmetric Field Theories on Three-Manifolds},''
  \href{http://dx.doi.org/10.1007/JHEP05(2013)017}{{\em JHEP} {\bf 1305} (2013)
   017},
\href{http://arxiv.org/abs/1212.3388}{{\tt arXiv:1212.3388 [hep-th]}}.

\bibitem{Martelli:2012sz}
D.~Martelli, A.~Passias, and J.~Sparks, ``{The supersymmetric NUTs and bolts of
  holography},'' \href{http://dx.doi.org/10.1016/j.nuclphysb.2013.04.026}{{\em
  Nucl. Phys.} {\bf B876} (2013)  810--870},
\href{http://arxiv.org/abs/1212.4618}{{\tt arXiv:1212.4618 [hep-th]}}.

\bibitem{Samtleben:2012ua}
H.~Samtleben, E.~Sezgin, and D.~Tsimpis, ``{Rigid 6D supersymmetry and
  localization},'' \href{http://dx.doi.org/10.1007/JHEP03(2013)137}{{\em JHEP}
  {\bf 1303} (2013)  137},
\href{http://arxiv.org/abs/1212.4706}{{\tt arXiv:1212.4706 [hep-th]}}.

\bibitem{Kuzenko:2012vd}
S.~M. Kuzenko, ``{Symmetries of curved superspace},''
  \href{http://dx.doi.org/10.1007/JHEP03(2013)024}{{\em JHEP} {\bf 1303} (2013)
   024},
\href{http://arxiv.org/abs/1212.6179}{{\tt arXiv:1212.6179 [hep-th]}}.

\bibitem{Hristov:2013spa}
K.~Hristov, A.~Tomasiello, and A.~Zaffaroni, ``{Supersymmetry on
  Three-dimensional Lorentzian Curved Spaces and Black Hole Holography},''
  \href{http://dx.doi.org/10.1007/JHEP05(2013)057}{{\em JHEP} {\bf 1305} (2013)
   057},
\href{http://arxiv.org/abs/1302.5228}{{\tt arXiv:1302.5228 [hep-th]}}.

\bibitem{deMedeiros:2013jja}
P.~de~Medeiros and S.~Hollands, ``{Conformal symmetry superalgebras},'' {\em
  Class. Quant. Grav.} {\bf 30} (2013)  175016,
\href{http://arxiv.org/abs/1302.7269}{{\tt arXiv:1302.7269 [hep-th]}}.

\bibitem{deMedeiros:2013mca}
P.~de~Medeiros and S.~Hollands, ``{Superconformal quantum field theory in
  curved spacetime},'' {\em Class. Quant. Grav.} {\bf 30} (2013)  175015,
\href{http://arxiv.org/abs/1305.0499}{{\tt arXiv:1305.0499 [hep-th]}}.

\bibitem{Martelli:2013aqa}
D.~Martelli and A.~Passias, ``{The gravity dual of supersymmetric gauge
  theories on a two-parameter deformed three-sphere},''
  \href{http://dx.doi.org/10.1016/j.nuclphysb.2013.09.012}{{\em Nucl. Phys.}
  {\bf B877} (2013)  51--72},
\href{http://arxiv.org/abs/1306.3893}{{\tt arXiv:1306.3893 [hep-th]}}.

\bibitem{Cassani:2013dba}
D.~Cassani and D.~Martelli, ``{Supersymmetry on curved spaces and
  superconformal anomalies},''
  \href{http://dx.doi.org/10.1007/JHEP10(2013)025}{{\em JHEP} {\bf 1310} (2013)
   025},
\href{http://arxiv.org/abs/1307.6567}{{\tt arXiv:1307.6567 [hep-th]}}.

\bibitem{Alday:2013lba}
L.~F. Alday, D.~Martelli, P.~Richmond, and J.~Sparks, ``{Localization on
  Three-Manifolds},'' \href{http://dx.doi.org/10.1007/JHEP10(2013)095}{{\em
  JHEP} {\bf 1310} (2013)  095},
\href{http://arxiv.org/abs/1307.6848}{{\tt arXiv:1307.6848 [hep-th]}}.

\bibitem{Kuzenko:2013gva}
S.~M. Kuzenko, ``{Super-Weyl anomalies in N=2 supergravity and (non)local
  effective actions},'' \href{http://dx.doi.org/10.1007/JHEP10(2013)151}{{\em
  JHEP} {\bf 1310} (2013)  151},
\href{http://arxiv.org/abs/1307.7586}{{\tt arXiv:1307.7586}}.

\bibitem{Klare:2013dka}
C.~Klare and A.~Zaffaroni, ``{Extended Supersymmetry on Curved Spaces},''
  \href{http://dx.doi.org/10.1007/JHEP10(2013)218}{{\em JHEP} {\bf 1310} (2013)
   218},
\href{http://arxiv.org/abs/1308.1102}{{\tt arXiv:1308.1102 [hep-th]}}.

\bibitem{Pan:2013uoa}
Y.~Pan, ``{Rigid Supersymmetry on 5-dimensional Riemannian Manifolds and
  Contact Geometry},'' \href{http://dx.doi.org/10.1007/JHEP05(2014)041}{{\em
  JHEP} {\bf 1405} (2014)  041},
\href{http://arxiv.org/abs/1308.1567}{{\tt arXiv:1308.1567 [hep-th]}}.

\bibitem{Closset:2013vra}
C.~Closset, T.~T. Dumitrescu, G.~Festuccia, and Z.~Komargodski, ``{The Geometry
  of Supersymmetric Partition Functions},''
  \href{http://dx.doi.org/10.1007/JHEP01(2014)124}{{\em JHEP} {\bf 1401} (2014)
   124},
\href{http://arxiv.org/abs/1309.5876}{{\tt arXiv:1309.5876 [hep-th]}}.

\bibitem{Closset:2013sxa}
C.~Closset and I.~Shamir, ``{The $\mathcal{N}=1$ Chiral Multiplet on $T^2\times
  S^2$ and Supersymmetric Localization},''
  \href{http://dx.doi.org/10.1007/JHEP03(2014)040}{{\em JHEP} {\bf 1403} (2014)
   040},
\href{http://arxiv.org/abs/1311.2430}{{\tt arXiv:1311.2430 [hep-th]}}.

\bibitem{Deger:2013yla}
N.~S. Deger, A.~Kaya, H.~Samtleben, and E.~Sezgin, ``{Supersymmetric Warped AdS
  in Extended Topologically Massive Supergravity},''
  \href{http://dx.doi.org/10.1016/j.nuclphysb.2014.04.011}{{\em Nucl. Phys.}
  {\bf B884} (2014)  106--124},
\href{http://arxiv.org/abs/1311.4583}{{\tt arXiv:1311.4583 [hep-th]}}.

\bibitem{Kuzenko:2013uya}
S.~M. Kuzenko, U.~Lindstrom, M.~Rocek, I.~Sachs, and
  G.~Tartaglino-Mazzucchelli, ``{Three-dimensional N=2 supergravity theories:
  From superspace to components},''
  \href{http://dx.doi.org/10.1103/PhysRevD.89.085028}{{\em Phys. Rev.} {\bf
  D89} (2014)  085028},
\href{http://arxiv.org/abs/1312.4267}{{\tt arXiv:1312.4267 [hep-th]}}.

\bibitem{Cassani:2014zwa}
D.~Cassani and D.~Martelli, ``{The gravity dual of supersymmetric gauge
  theories on a squashed $S^1 \times S^3$},''
\href{http://arxiv.org/abs/1402.2278}{{\tt arXiv:1402.2278 [hep-th]}}.

\bibitem{DiPietro:2014moa}
L.~Di~Pietro, M.~Dine, and Z.~Komargodski, ``{(Non-)Decoupled Supersymmetric
  Field Theories},'' \href{http://dx.doi.org/10.1007/JHEP04(2014)073}{{\em
  JHEP} {\bf 1404} (2014)  073},
\href{http://arxiv.org/abs/1402.3385}{{\tt arXiv:1402.3385 [hep-th]}}.

\bibitem{Kuzenko:2014mva}
S.~M. Kuzenko and G.~Tartaglino-Mazzucchelli, ``{N = 4 supersymmetric
  Yang-Mills theories in $AdS_3$},''
  \href{http://dx.doi.org/10.1007/JHEP05(2014)018}{{\em JHEP} {\bf 1405} (2014)
   018},
\href{http://arxiv.org/abs/1402.3961}{{\tt arXiv:1402.3961 [hep-th]}}.

\bibitem{Anous:2014lia}
T.~Anous, D.~Z. Freedman, and A.~Maloney, ``{de Sitter Supersymmetry
  Revisited},''
\href{http://arxiv.org/abs/1403.5038}{{\tt arXiv:1403.5038 [hep-th]}}.

\bibitem{Imamura:2014ima}
Y.~Imamura and H.~Matsuno, ``{Supersymmetric backgrounds from 5d N=1
  supergravity},''
\href{http://arxiv.org/abs/1404.0210}{{\tt arXiv:1404.0210 [hep-th]}}.

\bibitem{Farquet:2014kma}
D.~Farquet, J.~Lorenzen, D.~Martelli, and J.~Sparks, ``{Gravity duals of
  supersymmetric gauge theories on three-manifolds},''
\href{http://arxiv.org/abs/1404.0268}{{\tt arXiv:1404.0268 [hep-th]}}.

\bibitem{Alday:2014rxa}
L.~F. Alday, M.~Fluder, P.~Richmond, and J.~Sparks, ``{The gravity dual of
  supersymmetric gauge theories on a squashed five-sphere},''
\href{http://arxiv.org/abs/1404.1925}{{\tt arXiv:1404.1925 [hep-th]}}.

\bibitem{Assel:2014paa}
B.~Assel, D.~Cassani, and D.~Martelli, ``{Localization on Hopf surfaces},''
\href{http://arxiv.org/abs/1405.5144}{{\tt arXiv:1405.5144 [hep-th]}}.

\bibitem{Alday:2014bta}
L.~F. Alday, M.~Fluder, C.~M. Gregory, P.~Richmond, and J.~Sparks,
  ``{Supersymmetric gauge theories on squashed five-spheres and their gravity
  duals},''
\href{http://arxiv.org/abs/1405.7194}{{\tt arXiv:1405.7194 [hep-th]}}.

\bibitem{Kuzenko:2014eqa}
S.~M. Kuzenko, J.~Novak, and G.~Tartaglino-Mazzucchelli, ``{Symmetries of
  curved superspace in five dimensions},''
\href{http://arxiv.org/abs/1406.0727}{{\tt arXiv:1406.0727 [hep-th]}}.

\bibitem{Farquet:2014bda}
D.~Farquet and J.~Sparks, ``{Wilson loops on three-manifolds and their M2-brane
  duals},''
\href{http://arxiv.org/abs/1406.2493}{{\tt arXiv:1406.2493 [hep-th]}}.

\bibitem{Kaku:1978nz}
M.~Kaku, P.~Townsend, and P.~van Nieuwenhuizen, ``{Properties of Conformal
  Supergravity},''
\href{http://dx.doi.org/10.1103/PhysRevD.17.3179}{{\em Phys. Rev.} {\bf D17}
  (1978)  3179}.

\bibitem{Fradkin:1985am}
E.~Fradkin and A.~A. Tseytlin, ``{Conformal Supergravity},''
\href{http://dx.doi.org/10.1016/0370-1573(85)90138-3}{{\em Phys. Rept.} {\bf
  119} (1985)  233--362}.

\bibitem{vanNieuwenhuizen:1985cx}
P.~van Nieuwenhuizen, ``{$D=3$ Conformal Supergravity and Chern-Simons
  Terms},''
\href{http://dx.doi.org/10.1103/PhysRevD.32.872}{{\em Phys. Rev.} {\bf D32}
  (1985)  872}.

\bibitem{Rocek:1985bk}
M.~Rocek and P.~van Nieuwenhuizen, ``{$N\geq 2$ Supersymmetric Chern-Simons
  Terms As d=3 Extended Conformal Supergravity},''
\href{http://dx.doi.org/10.1088/0264-9381/3/1/007}{{\em Class. Quant. Grav.}
  {\bf 3} (1986)  43}.

\bibitem{FigueroaO'Farrill:2004mx}
J.~M. Figueroa-O'Farrill, P.~Meessen, and S.~Philip, ``{Supersymmetry and
  homogeneity of M-theory backgrounds},''
  \href{http://dx.doi.org/10.1088/0264-9381/22/1/014}{{\em Class. Quant. Grav.}
  {\bf 22} (2005)  207--226},
\href{http://arxiv.org/abs/hep-th/0409170}{{\tt arXiv:hep-th/0409170
  [hep-th]}}.

\bibitem{FigueroaO'Farrill:2007ar}
J.~M. Figueroa-O'Farrill, E.~Hackett-Jones, and G.~Moutsopoulos, ``{The Killing
  superalgebra of ten-dimensional supergravity backgrounds},''
  \href{http://dx.doi.org/10.1088/0264-9381/24/13/010}{{\em Class. Quant.
  Grav.} {\bf 24} (2007)  3291--3308},
\href{http://arxiv.org/abs/hep-th/0703192}{{\tt arXiv:hep-th/0703192
  [hep-th]}}.

\bibitem{FOF:2007F4E8}
J.~Figueroa-O'Farrill, ``{A geometric construction of the exceptional Lie
  algebras F4 and E8},'' {\em Commun. Math. Phys.} {\bf 283} (2008)  663--674,
  \href{http://arxiv.org/abs/0706.2829}{{\tt arXiv:0706.2829 [math.DG]}}.

\bibitem{FigueroaO'Farrill:2008ka}
J.~Figueroa-O'Farrill, E.~Hackett-Jones, G.~Moutsopoulos, and J.~Simon, ``{On
  the maximal superalgebras of supersymmetric backgrounds},''
  \href{http://dx.doi.org/10.1088/0264-9381/26/3/035016}{{\em Class. Quant.
  Grav.} {\bf 26} (2009)  035016},
\href{http://arxiv.org/abs/0809.5034}{{\tt arXiv:0809.5034 [hep-th]}}.

\bibitem{FigueroaO'Farrill:2008if}
J.~M. Figueroa-O'Farrill, ``{The Homogeneity conjecture for supergravity
  backgrounds},'' \href{http://dx.doi.org/10.1088/1742-6596/175/1/012002}{{\em
  J. Phys. Conf. Ser.} {\bf 175} (2009)  012002},
\href{http://arxiv.org/abs/0812.1258}{{\tt arXiv:0812.1258 [hep-th]}}.

\bibitem{FigueroaO'Farrill:2012fp}
J.~Figueroa-O'Farrill and N.~Hustler, ``{The homogeneity theorem for
  supergravity backgrounds},''
  \href{http://dx.doi.org/10.1007/JHEP10(2012)014}{{\em JHEP} {\bf 1210} (2012)
   014},
\href{http://arxiv.org/abs/1208.0553}{{\tt arXiv:1208.0553 [hep-th]}}.

\bibitem{Figueroa-O'Farrill:2013aca}
J.~Figueroa-O'Farrill and N.~Hustler, ``{The homogeneity theorem for
  supergravity backgrounds II: the six-dimensional theories},''
\href{http://arxiv.org/abs/1312.7509}{{\tt arXiv:1312.7509 [hep-th]}}.

\bibitem{Hab:1990}
K.~Habermann, ``{The twistor equation on Riemannian manifolds},'' {\em J. Geom.
  Phys.} {\bf 7} (1990) no.~4, 469--488.

\bibitem{Duval:1993hs}
C.~Duval and P.~Horv\'{a}thy, ``{On Schr\"{o}dinger superalgebras},''
  \href{http://dx.doi.org/10.1063/1.530521}{{\em J. Math. Phys.} {\bf 35}
  (1994)  2516--2538},
\href{http://arxiv.org/abs/hep-th/0508079}{{\tt arXiv:hep-th/0508079
  [hep-th]}}.

\bibitem{Klinker:2005}
F.~Klinker, ``{Supersymmetric Killing Structures},'' {\em Commun. Math. Phys.}
  {\bf 255} (2005) no.~2, 419--467.

\bibitem{Rajaniemi:2006}
H.~Rajaniemi, ``{Conformal Killing spinors in supergravity and related aspects
  of spin geometry},'' {\em PhD thesis, University of Edinburgh} (2006)  .

\bibitem{Baum:2002}
H.~Baum, ``{Conformal Killing spinors and special geometric structures in
  Lorentzian geometry - a survey},''
  \href{http://arxiv.org/abs/math/0202008}{{\tt arXiv:math/0202008}}.

\bibitem{BL:2003}
H.~Baum and F.~Leitner, ``{The twistor equation in Lorentzian spin geometry},''
  \href{http://arxiv.org/abs/math/0305063}{{\tt arXiv:math/0305063}}.

\bibitem{Baum:2012}
H.~Baum, ``{Holonomy groups of Lorentzian manifolds - a status report},'' {\em
  Global Differential Geometry, eds. C.B{\"{a}}r, J. Lohkamp and M. Schwarz,
  Springer Proceedings in Mathematics, Springer-Verlag} {\bf 17} (2012)
  163--200.

\bibitem{Leitner:2005}
F.~Leitner, ``{Conformal Killing forms with normalization condition},'' {\em
  Rend. Circ. Mat. Palermo, suppl. Ser II} {\bf 75} (2005)  279--292.

\bibitem{Baum:2008}
H.~Baum, ``{Conformal Killing spinors and the holonomy problem in Lorentzian
  geometry - a survey of new results},'' {\em Symmetries and Overdetermined
  Systems of Partial Differential Equations, eds. M. Eastwood and W. Miller,
  IMA Volumes in Mathematics, Springer} (2008)  251--264.

\bibitem{Nahm:1977tg}
W.~Nahm, ``{Supersymmetries and their Representations},''
\href{http://dx.doi.org/10.1016/0550-3213(78)90218-3}{{\em Nucl. Phys.} {\bf
  B135} (1978)  149}.

\bibitem{KruThe2013}
B.~{Kruglikov} and D.~{The}, ``{The gap phenomenon in parabolic geometries},''
  \href{http://arxiv.org/abs/1303.1307}{{\tt arXiv:1303.1307 [math.DG]}}.

\bibitem{KruMat2013}
B.~{Kruglikov} and V.~{Matveev}, ``{Submaximal metric projective and metric
  affine structures},'' \href{http://arxiv.org/abs/1304.4426}{{\tt
  arXiv:1304.4426 [math.DG]}}.

\bibitem{DouThe2013}
B.~{Doubrov} and D.~{The}, ``{Maximally degenerate Weyl tensors in Riemannian
  and Lorentzian signatures},'' \href{http://arxiv.org/abs/1305.3499}{{\tt
  arXiv:1305.3499 [math.DG]}}.

\bibitem{Kru:1954}
G.~Kru\v{c}kovi\v{c}, ``{Classification of three-dimensional Riemannian spaces
  according to groups of motions},'' {\em Uspehi Matem. Nauk (N.S.)} {\bf 9}
  (1954)  no. 1 (59) 3--40.

\bibitem{DC:1975}
L.~Defrise-Carter, ``{Conformal groups and conformally equivalent isometry
  groups},'' {\em Commun. Math. Phys.} {\bf 40} (1975)  273--282.

\bibitem{ExactSolutions}
H.~Stephani, D.~Kramer, M.~MacCallum, C.~Hoenselaers, and E.~Herlt, {\em {Exact
  Solutions of Einstein's Field Equations}}.
\newblock Cambridge University Press, 2nd edition, 2009.

\bibitem{AlekCort1995math}
{{Alekseevsky}, D.~V. and {Cort{\'e}s}, V.}, ``{Classification of
  $N$-(super)-extended Poincar{\'e} algebras and bilinear invariants of the
  spinor representation of $Spin(p,q)$},''
  \href{http://arxiv.org/abs/{arXiv:math/9511215 [math.RT]}}{{\tt
  {arXiv:math/9511215 [math.RT]}}}.

\bibitem{Alekseevsky:2003vw}
D.~V. Alekseevsky, V.~Cort\'{e}s, C.~Devchand, and A.~Van~Proeyen,
  ``{Polyvector superPoincare algebras},''
  \href{http://dx.doi.org/10.1007/s00220-004-1155-y}{{\em Commun. Math. Phys.}
  {\bf 253} (2004)  385--422},
\href{http://arxiv.org/abs/hep-th/0311107}{{\tt arXiv:hep-th/0311107
  [hep-th]}}.

\bibitem{Lic:1963}
A.~Lichnerowicz, ``{Spineurs harmoniques},'' {\em C. R. Acad. Sci. Paris} {\bf
  257} (1963)  7--9.

\bibitem{Kosmann:1972}
Y.~Kosmann, ``{D\'{e}riv\'{e}es de Lie des spineurs},'' {\em Ann. Mat. Pura
  Appl.} {\bf 91} (1972)  317--395.

\bibitem{BG:1992}
J.-P. Bourguignon and P.~Gauduchon, ``{Spineurs, op\'{e}rateurs de Dirac et
  variations de m\'{e}triques},'' {\em Commun. Math. Phys.} {\bf 144} (1992)
  581--599.

\bibitem{Habermann:1996}
K.~Habermann, ``{The graded algebra and the lie derivative of spinor fields
  related to the twistor equation},'' {\em Journal of Geometry and Physics}
  {\bf 18} (1996)  131--146.

\bibitem{Lewandowski:1991bx}
J.~Lewandowski, ``{Twistor equation in a curved space-time},''
\href{http://dx.doi.org/10.1088/0264-9381/8/1/003}{{\em Class.Quant.Grav.} {\bf
  8} (1991)  L11--L18}.

\bibitem{Feff:1976}
C.~Fefferman, ``{Monge-Amp\`{e}re equations, the Bergman kernel, and geometry
  of pseudoconvex domains},'' {\em Ann. Math.} {\bf 103} (1976)  395--416.

\bibitem{Keane:2013fta}
A.~J. Keane and B.~O.~J. Tupper, ``{Conformal symmetry classes for pp-wave
  spacetimes},'' \href{http://dx.doi.org/10.1088/0264-9381/21/8/009}{{\em
  Class. Quant. Grav. 21,} {\bf 2037} (2004)  },
\href{http://arxiv.org/abs/1308.1683}{{\tt arXiv:1308.1683 [gr-qc]}}.

\bibitem{Blau:Notes}
M.~Blau, ``{Plane Waves and Penrose Limits - Gravity and String Theory
  Group},''
  \href{http://arxiv.org/abs/http://www.blau.itp.unibe.ch/lecturesPP.pdf}{{\tt
  http://www.blau.itp.unibe.ch/lecturesPP.pdf}}.

\bibitem{Bia:1898}
L.~Bianchi, ``{Sugli spazii a tre dimensioni che ammettono un gruppo continuo
  di movimenti},'' {\em Soc. Ital. Sci. Mem. di Mat.} {\bf 11} (1898)  267.

\bibitem{PSWZ:1976}
J.~Patera, R.~Sharp, P.~Winternitz, and H.~Zassenhaus, ``{Invariants of real
  low dimension Lie algebras},'' {\em J. Math. Phys.} {\bf 17} (1976)  no. 6,
  986--994.

\bibitem{Matiadou:2007gz}
N.~Matiadou and A.~Fellouris, ``{Classification of the five-dimensional Lie
  superalgebras over the real numbers},''
\href{http://dx.doi.org/10.1007/s10773-006-9055-x}{{\em Int. J. Theor. Phys.}
  {\bf 46} (2007)  451--470}.

\end{thebibliography}\endgroup

\end{document}